\documentclass[14pt,a4paper]{report}

\usepackage{epsfig}
\title{}

\begin{document}
\baselineskip=21pt
\begin{titlepage}
\begin{center}
Lebedev Physical Institute of Russian Academy of Sciences \\
Astro Space Center \\
Pushchino Radio Astronomy Observatory
\end{center}
\vspace{1.0 cm}
\begin{flushright}
	Copyrighted as a manuscript
\end{flushright}

\vspace{2.5 cm}
\begin{center}
{\bf\Large Rodin Alexander Evgenievich} \\
\vspace{5mm}
{\bf\Large Pulsar Astrometry in the Presence of Low Frequency Noise}\\
\vspace{1.5 cm}
Speciality 01.03.02 -- Astrophysics, radioastronomy\\
\vspace{1.5 cm}
{\large Dissertation for an academic degree of\\
candidate of physical and mathematical sciences}
\end{center}
\vspace{2cm}

\begin{flushright}
	Scientific supervisors:\\	
	doctor of physical and mathematical sciences Yu.P.Ilyasov\\
	doctor of physical and mathematical sciences S.M.Kopeikin\\
\end{flushright}

\vspace{3.5 cm}
\begin{center}
{Moscow -- 2000} \\
\end{center}
\end{titlepage}

\pagestyle{empty}
\tableofcontents
\listoffigures
\listoftables

\newcommand{\ch}{\,{\rm ch}\,}
\newcommand{\sh}{\,{\rm sh}\,}
\newcommand{\tg}{\,{\rm tg}\,}

\baselineskip=20pt

\setcounter{figure}{0}
\setcounter{table}{0}

\chapter*{Introduction}
\addcontentsline{toc}{chapter}{Introduction}

Pulsar astrometry is a relatively recent part of astrometry, which, as the name implies, measures
the space-time coordinates of pulsars. Since the vast majority of pulsars are observed in the radio range, methods developed
in radio astronomy are used. Two main methods are considered and used in this work : very long baseline interferometry (VLBI) and
timing. VLBI is used to observe not only pulsars, but all radio sources. Timing, however, due to the specificity of the
method is used only for observations of objects emitting periodic pulses. Both methods allow you to determine the coordinates
of pulsars in a geometric way: VLBI - knowing the geometry of the location of the base in space, timing - based on the geometry of the Earth orbit. In addition to measuring coordinates, pulsar timing allows the construction of an independent time scale, which is based
only on observations of a group of highly stable pulsars. 

Low-frequency noise is such a noise, an autocorrelation function which is different from the delta function. Other names of low-frequency
noise -- correlated, colored, red noise. All these names will be used in this work

High-precision astrometric observations of pulsars are a good tool for solving various problems of astrometry, astrophysics,
cosmology and fundamental metrology. Namely, on the basis of such observations, it becomes possible to establish an inertial
coordinate system in the sky, realized kinematically, i.e. based on the precise coordinates and proper motions of pulsars; establishing a link between
different coordinate systems (quasar and dynamic); constructing pulsar time scales based on both the periodicity of their own rotation of the pulsar, and on the periodicity of the pulsar's movement around the barycenter of the binary system; assessment of the background of gravitational radiation; the study of noise spectra in the residual deviations of the times of arrival of pulses and, as a result, the study of physical processes inside and near the pulsar, in particular, the study of gravitational fields and mass distribution in globular clusters.

Since modern observations are characterized by very high accuracy, the problem arises of adequately accounting for the noise present in the observed
quantities. At the same time, a number of basic principles that were followed when the low level of observation accuracy (uncorrelation of random errors,
constancy of the parameters of the mathematical model used to reduce observations) is no longer fulfilled. First of all, this applies to
the independence of observation errors. In both VLBI and timing, the influence of correlated (or low-frequency) noises is great, which are very
 significantly distort estimates of model parameters, and over time even lead to an increase in the variance of estimates. Proper understanding of the influence
of low-frequency noise in the reduction of observations, as well as their correct accounting are therefore extremely important in data processing.

One of the main tasks of astrometry is the construction of an inertial coordinate system (ISC). It will be appropriate to agree on
terminology right away. In astrometry, several terms are used that are similar in meaning: reference frame, coordinate system, reference system. At the same time, some of the authors (Kovalevsky, 1991) distinguish several levels of the hierarchy of reference systems:

\begin{enumerate}\item The ideal frame of reference is the theoretical principle on which the final reference system is based. \item Frame of
reference -- defines the physical system on the basis of which the definition of an ideal frame of reference is applied. \item Conventional
reference frame -- in addition to paragraphs 1, 2, the parameters describing the physical system are assigned certain values (and therefore this system
it becomes conventional). \item Conventional reference system - set the starting points along with their coordinates, which materialize
the conventional frame of reference. \end{enumerate} The purpose of introducing celestial reference frames is to determine the only 
means of assigning coordinates to celestial bodies, either observed instrumentally or deduced from some theory.

Currently, the required accuracy of this system is able to provide VLBI,  and in recent years, the timing of highly stable pulsars has also been added to it. Now we can already talk about the accuracy of determining the coordinates of radio sources better than 0.001 seconds of arc, as about realistically achievable by modern methods. The construction of an inertial coordinate system is required for various tasks: astrophysical, astrometric and geophysical.

The inertial coordinate system can be implemented in three ways (Abalakin, 1979, Gubanov et al., 1983). At the same time, it would be more correct to
talk about a quasi-inertial coordinate system:

\begin{enumerate}
\item Geometrically, i.e. when the reference objects have practically no visible angular movements in the sky.  
Quasars and compact details of galaxies can serve as such objects. In this case, the implementation of the ISC is reduced to measuring the arcs between the reference radio sources. The stability of such a system is ensured up to the stability of the reference radio sources.

\item Kinematically. In this method, it is assumed that the reference objects move uniformly and rectilinearly.  In optics, this role has always
been performed by stars.  In recent years, radio sources with well-measured proper motions, in particular pulsars, have been added to the stars (Fedorov, 1986). A coordinate system based on such sources will not be rotating only if proper movements are determined by the absolute method. It can be seen that
pulsars can significantly help in this case if their proper motions are determined in relation to practically stationary extragalactic sources.

\item Dynamically.  Here, celestial bodies moving in a gravitational field serve as reference objects. A classic example of such objects are the bodies of the Solar system and artificial satellites of the Earth.  To create an inertial coordinates system in this method need to know the theory of motion
of the reference bodies . Along with purely gravitational interactions, non-gravitational forces also act on bodies, which are much worse
to account for (this applies primarily to artificial satellites).  Thus, this method does not provide good accuracy, because
it requires additional information, largely arbitrary.
\end{enumerate}

Thus, at present, the first method provides the simplest way to build a claim.  However, there are disadvantages to this method: if we
want to absolutize the coordinates of the reference radio sources, then we will encounter uncertainties in the rotational motion of the Earth, 
plus, the right ascents of radio sources are determined with accuracy to an arbitrary constant, which also makes it difficult
to determine the constant of precession. This paper sets out below  a method that allows you to link two coordinate systems 
dynamic, based on the annual movement of the Earth around the Sun, and quasar, based on the positions of remote, and therefore almost
stationary quasars and radio galaxies, and thereby find the position of the point of the vernal equinox, which traditionally is
the zero point of coordinate systems in astronomy, and the inclination of the ecliptic to the equator.

The traditional and most straightforward and simple way to connect two coordinate systems is to compare the positions of celestial
sources observed in both one and the other system  coordinates (Murray, 1986).  Among such sources, pulsars can be distinguished (Fedorov, 1986), which have a noticeable advantage in the accuracy of determining their coordinates compared to other objects.  The position of pulsars is determined by variations in the arrival times of pulses throughout the year due to the movement of the Earth around the Sun. For this reason the coordinates of pulsars derived from timing, associated with ephemerides, which describe the orbital parameters of the Earth. Positions of pulsars in the quasar coordinate system
are tied to distant quasars, since they are the ones that currently best define the orientation of this coordinate system.  Fairly
accurate observations were made by Bartel {\it et al.} (1985) and Gwinn{\it et al.} (1986).  They observed pulsars with an accuracy of about 4
milliarcseconds (mas). The work (Bartel {\it et al.}, 1985) is a demonstration character in order to show the capabilities
of the Mark III registration system for pulsar observations. In the work (Gwinn {\it et al.}, 1986), the task of determining parallaxes was also set.  The parallaxes were obtained for two relatively strong and close pulsars with accuracy 0.6 - 0.8 mas.  The technique of differential VLBI observations was used.
Many authors of the VLBI make observations of pulsars in order to determine their proper motions. Proper motions are needed to
determine their spatial velocities, which, in turn, allows us to establish the place of their formation and connection with supernova remnants. At work
(Lyne {\it et al.}, 1982) proper motions were determined for 26 pulsars with an average accuracy of 1-10 mas/year. It was used non-traditional technique of differential VLBI observations: pulsars were observed at a relatively low frequency of 408 MHz in the same diagram beam
with reference sources. This made it possible to almost completely eliminate the influence of the atmosphere and ionosphere. This technique is described
in detail in the work (Peckham, 1973).

Precise timing positions have been obtained by many authors (see works: (Rawley {\it et al.}, 1988), (Kaspi {\it et al.}, 1994),
(Matsakis, Foster, 1995)). V.M.Kaspi {\it et al.} in 1994 reviews the results of timing of pulsars PSR B1855+09,
B1937+21 for 7 and 8 years respectively. The accuracy of determining coordinates and proper motions is $\le 0.12$mas and $\le 0.06$mas/year, respectively, in the DE200 planetary ephemeris system. The extraordinary stability of the orbital period in the PSR B1855+09 system allows the authors to put
a limit on the secular change in the Newtonian gravitational constant $\dot G/G = (-9\pm 18)\times 10^{-12}\;{\rm year}^{-1}$.  Later in this
dissertation, it will be shown that the stability of the orbital period can be used to maintain a new independent time scale. In more detail
the parameters of the binary system  PSR B1855+09 are understood in operation (Ryba {\it et al.}, 1991). The masses of the pulsar and companion were determined, which turned out to be in good agreement with theoretical predictions based on the physics of neutron stars and the evolutionary model B1855+09.

D.N.Matsakis {\it et al.} (1996) consider the possibility of using millisecond pulsars to establish a long-term scale and
a quasi-inertial coordinate system. These authors conclude that so far the contribution to the terrestrial time scales of the two longest-observed
millisecond pulsars PSR B1937+21 and B1855+09 is apparently minimal, although they may be useful for maintaining
an independent time scale over long time intervals and for identifying the source of errors in atomic scales that are otherwise it is hard
to identify because of the finite lifetime of atomic standards.

In addition to the inertial coordinate system, the needs of modern science require the most accurate and stable time scale. In all
theories of the motion of celestial bodies, ephemeris time (ET) is present as an argument. It is clear that this time is an ideal
construction, and it is necessary to have a practical implementation of the ET timeline. Until the advent of atomic clocks in the late 50s of the twentieth century, the only time scale used for recording observations  for long periods was the average solar (universal) time
UT, based on the daily rotation of the Earth. The exact implementation of UT requires knowledge of the disturbing external moment caused by the Moon and
Sun, as well as knowledge of the Earth's orientation parameters (EOP). The EOP are determined quite well, and there
are no fundamental difficulties here. Variability of the UT time scale relative to   the scale used to calculate the ephemerides of the Solar system bodies in
accordance with Newton's theory of gravity was noticed at the end of the XIX century. It was finally established in the first half of the twentieth century . This
variability is caused by tidal friction in the Earth-Moon system and leads to a secular slowdown in the axial rotation of the Earth and the average
angular motion of the Moon. Thus, the variability of UT led to the establishment of an ephemeris time scale, precisely determined through the parameters
of the Earth's orbital motion.

Although universal time is not used as a time scale in astronomy, knowledge of it remains necessary, since it determines the instantaneous orientation of the Earth in space, and   astrometric (and, in particular, VLBI) observations are carried out from the Earth.  Currently, the study of variations in the rotation of the Earth
is of direct interest to geophysics.

For astronomical purposes, Universal Time (UT) has been replaced by International Atomic Time (TAI), which is easily accessible to users via
radio and TV channels. When TAI was introduced, its zero point was chosen so as to obtain the best agreement with the universal
time corrected for seasonal fluctuations for the epoch 1958, January, 1. Atomic time scale is currently being established by the International Bureau of Time and Measurements (BIPM) in Paris by comparing a group of cesium clocks at the disposal of organizations located around the globe.
The fundamental unit of this scale is the SI second at sea level. Although the TAI scale was officially introduced in 1972, it has actually
existed since 1955, when BIPM began comparing world time with the atomic scale. Currently, ephemeris time (ET)
and TAI can be considered equivalent, except for the constant difference, which was found from observations and for the epoch of 1958, January, 1 was 32.184
seconds. Currently, the question of possible discrepancies between the ET and TAI scales is open.

In this dissertation, the idea of a pulsar time scale is developed, but already based on the motion of a pulsar in a binary system. 
The pulsar's natural frequency here plays to some extent the role of a "carrier frequency" (to use terms from radiophysics), and the orbital frequency
acts as a reference. This idea was outlined in the works (Rodin, Kopeilin, Ilyasov, 1997; Ilyasov, Kopeikin, Rodin, 1998).
These works consider the real case of determining parameters against the background of correlated noise, and the main attention is paid to the behavior of the variances of the estimated parameters depending on the observation time interval.

Correlated noises can have completely different origins. This may be the stochastic background of gravitational waves formed at
an early stage of the Universe, variations in electron density along the visual ray in the interplanetary medium and the ionosphere of the Earth, precession
of the pulsar, the planetary system around the pulsar, etc. In the last chapter of this paper, the role of low-frequency correlated noises is performed
by gravitational disturbances in the quasi-uniform motion of the pulsar. It is shown that the variations in the residual deviations of the TOAs, interpreted
how perturbations in pulsar motion are well explained within the framework of gravitational perturbations.

The object of research in this work are pulsars characterized by a set of parameters interesting from the point of view
of astrometry, metrology and cosmology. First of all, such parameters are coordinates and proper motion, as well
as the pulsar's natural rotation frequency and orbital period if the pulsar is binary.

The subject of the study in this paper is observational data in the form of geometric delays and interference frequencies (in
VLBI observations), the time of arrival of pulses (TOA are used in the chapter on the BPT scale) and in the form of residual deviations of TOA (these
data are used in the chapter on gravitational disturbances as a source of low-frequency noise).

The main purpose of the work is to analyze observational data in the presence of low-frequency noise, namely:
\begin{itemize}
\item the use of more computationally advanced algorithms that allow you to obtain more correct estimates of parameters not
subject to the distorting effect of correlated noise;
\item analysis of the behavior of the dispersions of the rotational and orbital parameters of the pulsar depending on the observation interval by the least
squares method; A theoretical explanation of the long-term variations in TOAs observed in a number of pulsars due to the deviation of the pulsar's motion from
quasi-uniform and rectilinear, caused, in turn, by gravitational perturbations of massive bodies.
\end{itemize}

The theoretical basis and the basic method presented in this research is the theory of statistical inference for various
probabilistic models described by a finite number of parameters. The initial probabilistic model may include a deterministic part and a random
component forming a stationary random process. Among all statistical methods, regression analysis (least squares method, LSQ) is primarily used. Since the conditions for the use of classical LSQ are often not fulfilled, a modified LSQ is used, taking into account violations of the initial assumptions about
the properties of the random component.

All the results presented in this paper were reported at the following scientific events:

\begin{enumerate}

\item of the ASC (Astro Space Center) Reporting sessions in 1996, 1997, 1998 and 1999.

\item XXVI Radio Astronomy Conference in St. Petersburg in 1995.

\item IAU Colloquium No. 160 "Pulsars: problems and progress", Sydney, Australia, 1996.

\item International Conference "Modern problems and methods of astrometry and geodynamics", St. Petersburg, 1996.

\item International Workshop "Asia Pacific Telescope and Asia Pacific Space Geodynamics", Kashima, Japan, 1996.

\item XXVII Radio Astronomy Conference in St. Petersburg in 1997

\item XXX Conference of Young European Radio Astronomers, Krakow, Poland, 1997.

\item School-seminar of young radio astronomers "Radio Astronomy in space", Pushchino, April 14-16, 1998.

\item Colloquium MAC 164, San Francisco, USA, 1998.

\item EVN/JIVE Symposium, Holland, 1998.

\item School-seminar of young radio astronomers "Ultrahigh angular resolution in radio astronomy", Pushchino, June 9-11, 1998.

\item IAU Colloquium No. 177 "Pulsar Astronomy - 2000 and beyond: ", Bonn, Germany, 1999.

\end{enumerate}

List of the author's publications on the topic of this dissertation:
\begin{enumerate}

\item M.Sekido, M.Imae, Yu.Hanado, Yu.P.Ilyasov, V.V.Oreshko, A.E.Rodin, S.Hama, J. Nakajima, E. Kawai, Y. Koyama, T. Kondo, N. Kurihara and M.
Hosokawa, "Astrometric VLBI observation PSR0329+54". 1999, PASJ, {\bf 51}, No. 5, pp.595-601.

\item  M.Sekido, M. Imae, S. Hama, Yu. Koyama, T. Kondo, J. Nakajima, E. Kawai, N. Kurihara, Yu. P. Ilyasov, V.V. Oreshko, A.E. Rodin,
B.A.Pererechenko, "The VLBI experiment with the Kashima pulsar (Japan) - The basic level in Kalyazin (Russia)", New Astronomy Review, 1999, {\bf 43}/8-10,
pp. 599-602.

\item A.~E.~Rodin. Gravitational disturbances as a source of timing noise, Proceedings of the colloquium. IAU 177, August 31 - September 3, 1999, Bonn,
Germany.

\item A.~E.~Rodin, Yu.~P.~Ilyasov, V.~V.~Oreshko, M.~Sekido. Timing noise as a source of discrepancy between the timing position of the pulsar and
the position of the VLBI. Proceedings of the colloquium. IAU 177, August 31 - September 3, 1999, Bonn, Germany.

\item Yu.~P.~Ilyasov, V.~A.~Potapov, A.~E.~Rodin.  Time noise spectra of pulsars 0834+06,1237+25, 1919+21, 2016+28
observations for 1978-1999. Proceedings of the colloquium. IAU 177, August 31 - September 3. 1999, Bonn, Germany.

\item A.~E.~Rodin. Gravitational disturbances as a source of pulsar timing noise. Abstracts of the reports of the school-seminar of young
radio astronomers "Ultrahigh angular resolution in radio astronomy", June 9-11, 1999, Pushchino, pp. 19-20.

\item by M. Sekido, A. E. Rodin, Y. P. Ilyasov, M. Imae, V. V. Oreshko, S. Hama. The exact coordinates and proper motion of the pulsar PSR 0329+54
according to Kashima-Kalyazin's VLBI. Accepted in Astron. J. 1999.

\item M.~Sekido, S.~Hama, H.~Kiuchi, M.~Imae, Yu.~Hanado, Yu.~ Takahashi, A.~ E.~Rodin, Yu.~P.~Ilyasov. 1998, in the proceedings of the colloquium MAS 164, ed. J.~A.~Zensus, G.~B.~Taylor, J.~B.~Vorobel, A.S.P.~Conf.~Ser. Volume 105, (BookVrafter, San Francisco), p.~403.

\item Yu. P. Ilyasov, S. M. Kopeikin, A. E. Rodin, Astronomical time scale based on the orbital motion of a pulsar in a binary system, 1998, Astron.~Lett., No. 4, pp. 275-284.

\item A. E. Rodin, M. Sekido, VLBI - observation of pulsar B0329+54, Abstracts of the report of the young  radio astronomers seminar "Radio astronomy
in space" April 14-16, 1998, Pushchino, pp. 8-10.

\item by A. E. Rodin, S. M. Kopeikin, Yu. P. Ilyasov, Astronomical A time scale based on the orbital motion of a pulsar in a binary system, 1997,
Acta cosmologica, FASCICULUS XXIII-2, pp. 163-166.

\item Yu. P. Ilyasov, S. M. Kopeikin, A. E. Rodin, Artistkicals  ephemeris time based on the orbital motion of a double
pulsar. 1997, In the collection "Problems of modern radio astronomy", p. -  St. Petersburg, vol. 2, p. 189.

\item A. E. Rodin. The effect of the flyby of a massive body on the appearance of residual
deviations of the pulsar TOAs. 1997, In the collection "Problems of modern  radio astronomy", St. Petersburg, vol. 2, p. 193.

\item R. Akhmetov, S. Khama, Y. Ilyasov, A. Rodin, M. Sekido. Reference catalog of radio sources for VLBI observations of pulsars, 1997, Baltic
Astronomy, vol. 6, No. 4, p. 347.

\item M. Sekido, S. Hama, H. Kiuchi, M. Imae, Yu. Hanado, Yu. Takahashi, A. E. Rodin, V. V. Oreshko, Yu. P. Ilyasov, B. A. Poperechenko.
Development of the K4 correlator for the Japanese-Russian pulsar VLBI, 1996, Proceedings of TWAA, Kashima, Japan, pp. 183-187.

\item by A. E. Rodin, Y. P. Ilyasov, V. V. Oreshko, A. E. Avramenko, B. A. Pererechenko, M. Sekido, M. Imae, Y. Hanado. VLBI pulsar on Kalyazin
(Russia) -- Kashima (Japan) baseline.  1996, TWAA Materials, Kashima, Japan, pp. 265-268.

\item Yu. P. Ilyasov, M. Imae, S. M. Kopeikin, A. E. Rodin, T. Fukushima. Double pulsars as a high-precision astronomical clocks. Proceedings
of the conference "Modern problems and methods of astrometry and geodynamics". St. Petersburg, 1996.

\item A. E. Avramenko, M. Imae, E. P. Ilyasov, B. A. Pererechenko, V. V. Oreshko, A. E. Rodin, M. Sekido, Yu. Hanado. VLBI-pulsar observations
based on Kalyazin (Russia) -- Kashima (Japan) baseline. Proceedings of the conference "Modern problems and methods of astrometry and geodynamics". St. Petersburg, 1996.

\item Yu. P. Ilyasov, A. E. Rodin, A. E. Avramenko, V. V. Oreshko, etc. VLBI Pulsar Experiment in Kashima (Japan) - Kalyazin (Russia).
IAU 160 Colloquium -- Pulsars: Problems and Progress, 1996.

\item M. Sekido, Yu. Hanado, M. Imae, Yu. Takahashi, Yu. Koyama, Yu. Koyama Jr. Ilyasov, A.Rodin, A. Avramenko, V. Oreshko, B. Pererechenko. Kashima
(Japan) -- Kalyazin (Russia) experiment on VLBI with pulsar in 1995. TDC News in CRL, No. 7, October 1995, p. 17. 

\item A.Avramenko, M.Imae, Yu.Ilyasov, Ya.Koyama, V.Oreshko, B.Pererechenko, A.Rodin, M.Sekido, Yu.Takahashi and Yu.Hanado. 
"VLBI-observations of pulsars based on Kalyazin-Kasim at a frequency of 1.4 GHz. Pulsar timeline program".  XXVI Radio Astronomy Conference.  Abstracts
of the reports. p.235 (1995).

\item A. Rodin, M. Sekido, V. Oreshko, Yu. Hanado, V. Potapov. "Modernization of the SKED software package for the Russian-Japanese
VLBI observation program "Pulsar Time Scale". XXVI Radio Astronomy Conference.  Abstracts of reports. p.303. (1995).

\item A. Avramenko, O. Doroshenko, Y. Ilyasov, V. Potapov, A. Rodin, G.Khechinashvili.  "Automation of investments and information
support for pulsar synchronization". XXVI Radio Astronomy Conference. Abstracts of the reports. p.309. (1995).

\end{enumerate}

{\bf The following results are submitted for defense:}
\begin{enumerate}
\item Precise measurements of coordinates and proper motion of the pulsar PSR 0329+54 using the VLBI method.
\item Establishing of the reason for the discrepancy between the coordinates of pulsars measured by VLBI and timing methods, which comes down to the influence of low-frequency noise at the time of pulse arrivals (TOAs) from the pulsar.
\item A special method for processing timing observations that allows you to correct TOA - coordinates of pulsars.
\item Theoretical dependences of the behavior of dispersions of pulsar parameter depending on the observation interval and the type of correlated
noise, on the basis of which it became possible to propose a new time scale BPT based on the orbital motion of pulsar in binary system stable over long periods of time (more than 10 years).
\item A theory that explains spontaneous changes in the rotational frequency of pulsars through their interaction with the passing gravitating mass and the theoretical power spectrum of low-frequency fluctuations of the pulsar rotational phase caused by gravitational disturbances from the passage of bodies near the pulsar.
\end{enumerate}

\newpage

\chapter{Pulsar radio interferometry with very long baseline}\label{1}

In March 1995, joint Russian-Japanese VLBI observations of pulsars began (Rodin {\it et al.}, 1996; Sekido {\it et al.}, 1998; Sekido
{\it et al.}, 1999). The purpose of these observations is to determine the positions of pulsars with high accuracy, which will allow: 
\begin{itemize} 
\item to determine the parallaxes of pulsars, 
\item to measure their proper motions,
\item to link celestial coordinate systems: dynamic, based on the ephemerides of the bodies of the Solar system, and quasar, based on the positions
of extragalactic radio sources determined from the Earth rotating around its axis. 
\end{itemize}

\section{Algorithm}
\subsection{Geometric delay}

A theoretical expression for the geometric delay, i.e., for the difference in the moments of arrival of the wavefront at the first and second antennas
of the interferometer, was obtained by several authors (Kopeikin, 1990; Doroshenko et al., 1990). There is also an International Earth Rotation Service (IERS) standard, which prescribes how to calculate the geometric delay for full compatibility of VLBI data.

The geometric delay is represented by the following formula (Kopeikin, 1990; Petrov, 1995)
\begin{equation}
\begin{array}{rcl}
\tau_{geom}&=&\displaystyle\frac{1-(\displaystyle\frac 52 V^2+{\bf V}{\bf
v}_2)}{c} \cdot\displaystyle\frac{{\bf b}{\bf s}-\displaystyle\frac
 12\frac{({\bf b}{\bf V})({\bf V}{\bf s})}{c^2}+ \tau_{gr,pl}+\tau_{gr} }{
1+\displaystyle\frac 1c ({\bf V}+{\bf v}_2){\bf s} }+\\
&&\displaystyle\frac{1}{c^2}{\bf V}(t_1){\bf b}(t_{1_g}) + \frac{1}{c^2}
(g_{1,loc}h_{1,ort}-g_{2,loc}h_{2,ort})(t_1-t_{syn}),
\end{array}
\end{equation}
\begin{equation}
\tau_{gr,pl}=\sum\limits_k\displaystyle\frac{2fM_k}{c^3}
\ln\left|1+\displaystyle\frac 1{|G_{2k}|}\frac{{\bf b}(g_{2k}-{\bf s})}
{1-{\bf g}_{2k}{\bf s}}\right|,
\end{equation}
\begin{equation}
\tau_{gr}= \displaystyle\frac{2fM_\oplus}{c^3}
\ln\left|\displaystyle\frac{1+\sin E_1}{1+\sin E_2}\right|.
\end{equation}
Here ${\bf V}$ is the vector of the barycentric velocity of the geocenter, ${\bf v}_2$ is the vector of the geocentric velocity of the station $\sharp$2, ${\bf b}$ is
the geocentric vector of the base, ${\bf s}$ is a single vector directed from the barycenter of the Solar system to the source, $g_{i,loc}$ is the local
acceleration of gravity at the $i$ station, $h_{1,ort}$ is the orthometric height of the $i$ station, $k=1$ (Sun) and $k=2$ (Jupiter), ${\bf
G}_{2k}$ is the topocentric vector of the $k$th body, corrected for planetary aberration, $g_{2k}={\bf G}_{2k}/|{\bf G}_{2k}|$, $E_i$ -
the height of the observed source above the horizon of the $i$ station. The barycentric coordinates of the vectors are set in a barycentric
coordinate system, and the geocentric vectors are set in a geocentric coordinate system that uses TT (Time Terrestrial) time like
your proper time. All vectors are calculated at the time $t_{1g}$, which corresponds to the arrival of the wavefront at the phase center of the first antenna.

\subsection{Features of pulsar VLBI}

The primary processing of VLBI observations of pulsars has a feature that distinguishes it from the processing of other radio sources. The signals from pulsars
are pulsed, i.e., during primary processing (correlation), only the part of the magnetic recording corresponding to the time when there
is a pulse contributes to the amplitude of the correlation. The same part of the data, when there is no pulse, only worsens the signal-to-noise ratio. Thus,
a natural way of processing pulsar VLBI data suggests itself, when only the part of them where the pulse is recorded is correlated, and the part
records where there is no pulse are not correlated. The signal-to-noise ratio is improved by $\sqrt{\frac{P}{W}}$ times, where $P$ is the pulsar period,
$W$ is the pulse width (Sekido {\it et al.}, 1992). A detailed description of the methods of pre-calculation of the pulsar period and the necessary accuracy is presented in the work (Rodin,  Sekido, 1998). Here we will briefly repeat the reasoning from the work (Rodin and Sekido, 1998) on what restrictions are imposed on
the time interval for VLBI observations of pulsars.

In correlation processing, the term "characteristic period" is $T_{par}$. During this period, the correlation parameters
(the quadratic term of the interference frequency, the bit shift rate, the gating period, etc.) remain constant. The number of pulses during this
period is $T_{\rm par}/P$, where $P$ is the pulsar period. To keep the pulse inside the gating window for the time $T_{\rm par}$, the pulse shift
in this window must be less than the width of the window $W$, i.e. $\Delta P\,T_{\rm par}/P <W$, where $\Delta P$ is the error of the calculated period.
Thus, it is possible to put a limit on the relative accuracy of the period calculation
\begin{equation}\label{acc}
\frac{\Delta P}{P}<\frac{W}{T_{\rm par}}.
\end{equation}
As an example, we can consider the pulsar PSR 1937+21 with a period of $P$=1.5 ms. Suppose $W\simeq 0.1 P=1.5\cdot 10^{-4}$s, $T_{\rm par}\simeq 3$s.
Then $\Delta P/P<5\cdot 10^{-5}$.

The shift between the pulse and the strobe can occur not only due to the error of calculating the period, but also due to a change in the period due to the accelerated
movement of the observer on Earth, as well as the acceleration of the pulsar in a binary system. Under these circumstances, how long can you count
is the pulsar period constant? The change in the pulsar period $\delta P(t)$ over time $\delta t$ can be calculated as
\begin{equation}
\frac{\delta P(t)}{P} \sim \frac{v}{c} = \frac{\alpha\, t}{c},
\end{equation}
here $\alpha$ is acceleration, $c$ is the speed of light. The pulsar phase change during $T_{\rm par}$ should be less than $W$. 
The following condition can be deduced from this 
\begin{equation} \frac{1}{P}\int^{T_{\rm par}}_0 \delta P {\rm d} t = \frac{\alpha}{c}\int_0^{T_{\rm par}}t\,{\rm d}t<W,\quad T_{\rm
par}< \sqrt{\frac{2Wc}{\alpha}}.
\end{equation}
The magnitude of the change in the pulsar period over time $T_{\rm par}$ due to acceleration
\begin{equation}
\frac{\Delta P}{P} \sim \frac{\alpha}{c} T_{\rm par} \leq \sqrt{\frac{2W\alpha}{c}}.
\end{equation}
The magnitude of the acceleration due to the rotation of the Earth, the movement of the Earth around the Sun and the motion of the pulsar in a binary system (the orbital period is assumed 0.1 days) is equal to $3 \cdot 10^{-2}, 6 \cdot 10^{-3}$, and $3 \cdot 10^2$ m/s$^2$, respectively. Table 1.1 shows the maximum
parametric period and the magnitude of the period change at these accelerations $\alpha$, where W=0.15 ms is assumed.

\begin{table}[t]\centering
\begin{tabular}{|l|r|r|l|}
\hline $\alpha$ (m/s$^2$) & $T_{\rm par}$ (s) & $\Delta P/P$ & Comment
\\[5pt]
\hline $6\cdot 10^{-3}$ & 3900 & $3.9\cdot 10^{-8}$ & Earth rotation \\[5pt]
$3\cdot 10^{-2}$ & 1700 & $8.6 \cdot 10^{-8}$ &Earth orbital movement  \\[5pt]
$3\cdot 10^2$ & 17 & $8.6 \cdot 10^{-6}$ & Movement in binary system \\[5pt]
\hline
\end{tabular}
\label{tab1}
\caption{The limit value of the parametric period $T_{\rm par}$ and the relative change in the period $\Delta P/P$ caused by acceleration
$\alpha$ calculated for a pulsar with a period of 1.5 ms at the gate width $0.1$.}
\end{table}

During the VLBI session, the observation time of one scan is usually less than 10$^3$ s. The parametric period of $T_{\rm par}$ correlation processing
is usually less than 8 s. Thus, it can be concluded that the period of a single pulsar can be considered constant during the observational scan, and the parametric period of $T_{\rm par}$ is not limited by a change in the pulsar period for any pulsar.

\subsection{The influence of the troposphere}

As you know, the lower few tens of kilometers of the Earth's atmosphere are called the troposphere. With a good degree of accuracy, the troposphere can be
consider it electrically neutral. The radio signal, passing through this layer of the atmosphere, acquires a delay, curvature and attenuation relative
to the equivalent path in a vacuum. The additional delay at the zenith is equal to $\simeq$2 m and increases to $\simeq$20 m at an angle of $6^{\circ}$ above
the horizon. Thus, accurate VLBI models must take into account the delay in the troposphere.

Here is an expression for the tropospheric delay $\tau_{tr}$
\begin{equation}\label{trop}
\tau_{tr}=(\rho_{zd}+\rho_{zw})R(E),
\end{equation}
where $\rho_{zd}$, $\rho_{zw}$ are tropospheric delays in the zenith for the dry and wet components, respectively, $E$ is the height of the source
above the horizon. The expression (\ref{trop}) used a common $\rho_{zd},\;\rho_{zw}$ mapping function.

To analyze the tropospheric parameters in our observations, the mapping function CfA (Center for Astrophysics) (Sovers, Jacobs,1996) was used.
\begin{equation}
R(E)=\frac{1}{\displaystyle\sin E+\frac{a}{\displaystyle\tg E+
\frac{b}{\sin E+c}}},
\end{equation}
$a,\,b,\,c$ are certain parameters depending on temperature, pressure, humidity of the atmosphere. Their exact appearance can be found, for example, in the work
(Petrov, 1995; Sovers, Jacobs 1996). The height of the source above the horizon $E$ should be calculated taking into account the annual aberration (daily can
be ignored) and the flatness of the Earth.  The delay at the zenith and the rate of its change were estimated in our studies at 6-hour intervals,
which can be considered typical for weather changes.

\subsection{The influence of the ionosphere}
The ionospheric delay is usually determined by two-frequency observations at widely spaced frequencies (for example, in the S (13 cm) and X (3.5 cm)
ranges) and is usually excluded from the data before conducting a secondary processing. Here is an expression for the ionospheric delay in the X band (Petrov, 1995):
\begin{equation}
\tau_{ion,x}=\frac{\nu_s^2}{\nu_s^2-\nu_x^2}(\tau_s-\tau_x),
\end{equation}
where $\nu_s,\;\nu_x$ is the frequency of the S and X bands, $\tau_s,\;\tau_x$ are the group delays measured in the S and X bands.

If two-frequency observations are not carried out, it is possible to improve the model for group delay by modeling the ionosphere and measuring
the total electron content in the direction to the zenith at a single site. The delay at the zenith is then recalculated according to the known
mapping function to the desired height.

The ionospheric delay in a form similar to the tropospheric delay is (Bartel, 1990)
\begin{equation}
\tau_{ion}=\frac{kcr_0}{2\pi}\frac{1}{\nu^2}I_1(t)f(E_1),
\end{equation}
where $k=-1$ for phase delay and $k=1$ for group delay, $r_0$ is the classical electron radius, $c$ is the speed of light in vacuum, $\nu$ is
the observational frequency, $I_1(t)$ is the total electron content in the section of a unit area in the direction at Zenith at station 1,
$f(E_1)$ is a mapping function. $I_1(t)$ - varies significantly during the day, as well as with changes in solar activity.

The mapping function can be calculated by modeling the geometry of the ionosphere as a sphere with an inner radius of $r+h_i$ and an outer radius of $r+h_o$,
$r$ is the radius of the Earth:
\begin{equation}
f(E)=\frac{1}{h_o-h_i}\left[(r^2\sin^2E+2rh_o+h_o^2)^{1/2}-
(r^2\sin^2E+2rh_i+h_i^2)^{1/2}\right].
\end{equation}

For an interferometer with stations 1 and 2, the ionospheric delay will be:
\begin{equation}
\tau_{ion}=\frac{kcr_0}{2\pi}\frac{1}{\nu^2}[I_2(t)f(E_2)-I_1(t)f(E_1)].
\end{equation}

\subsection{Catalogs of reference radio sources}

Catalogs of compact radio sources observed using VLBI technology began to be actively created in the 70s. Now there are several hundred
radio sources that have been observed a total of several thousand times and have an accuracy of determining coordinates at the level of tens and hundreds
of angular microseconds. The catalog of radio sources published in the 1994 IERS Annual Report provides access to the International Coordinate Reference System (ICRS). It currently includes only 608 objects. In the future, based on new observations, it is planned to conduct
monitoring the stability of the coordinates of radio sources with appropriate warnings and coordinate updates.  Sources from the IERS catalog were chosen as reference objects for pulsar VLBI observations. Unfortunately, several hundred sources are still not enough to make it easy to choose the one closest to any predetermined pulsar. Therefore, the problem of choice remains quite serious. During the observations, the catalog of the Green Bank observatory
at 1400 MHz was additionally used, since the IERS catalog does not contain information about the flux of radio sources.

\section{VLBI registration equipment}
The K-4 registration system was developed in Japan as a system following
K-3 generations (Kiuchi, 1991). The K-4 system consists of the following parts:
\begin{enumerate}
\item Frequency Synthesizer
\item Video Converter
\item Input interface
\item Output interface
\item Video recorder.
\end{enumerate}

The frequency standard synthesizes a signal of a given frequency for the video converter.

The video converter converts the window in the IF signal input (100-500 MHz) into a video signal (0-2 MHz).

The input interface is used to collect data and write it to VLBI-site. It makes a 1-bit digitization of the 2 MHz (4 MHz) video signal and
together with the 4 MHz (8 MHz) time signal, which is obtained from an external frequency standard, outputs a data stream at a speed of 64 (128) Mbit/s.

The output interface is used for correlation processing. It converts the signal from the VCR into the format required for the correlator. The output interface format is compatible with the Mark-III format.

The Sony DIR-1000 industrial broadcast tape recorder is used as a video recorder. Commercial video cassettes of the D-1 type are used.

\section{Observations and data analysis}

The observations discussed in this chapter were carried out in 1995, 1996 and 1998. The 1997 data are not included in this dissertation, because
the pulsar PSR 0329+54 was not observed enough times for a full-fledged analysis this year. RT-64 radio telescopes
in Kalyazin, Tver region, Russia and RT-34 in Kashima, Ibaraki Prefecture, Japan were used. Observations of PSR B0329+54 alternated with observations of reference
radio sources. Table 1.3 shows which sources were used in the 1995, 1996 and 1998 sessions. One pulsar observation (scan) had a duration of 900 s in 1995 and 300 s in 1996 and 1998, and observations of reference sources had a duration of 300 s in 1995 and 240 s in 1996 and 1998. In the 1995 experiment, the observation band
was 1392-1432 MHz.  8 channels of 2 MHz each of the upper side frequency band (USB) with a 5 MHz distance were used.  In the 1996 experiment, the channel frequency arrangement was not equidistant and did not obey any dependence, but was dictated by minimizing external interference on
the radio telescope in Kashima. Interference situation on the radio telescope in Kalyazine was relatively favorable. The observations were carried out at
frequencies of 1392-1436 MHz in 15 channels of 2 MHz each. In May 1998 , observations were made in the S band (2.2 GHz) in the band 2200 - 2287 MHz in
15 channels of 2 MHz each. The Japanese registration system was used K4. The observation cycle was organized as follows: pulsar - quasar~1 - pulsar - quasar~2.

The Kashima Space Research Center K3 correlator was used for primary data processing. This correlator has a gating function, which was not used during processing, and which could improve the signal-to-noise ratio for the pulsar by 2-3 times. Weather data during
the observation session (temperature, pressure, humidity), as well as the partial derivatives of the group delay and the interference frequency
for the parameters of interest at the time of each observation scan were added to the correlated data. All this was recorded in DBH (Database handler) format, designed for processing by the CALC/SOLVE program.

The signal-to-noise ratio was monitored for each scan. Scans with a poor signal-to-noise ratio were excluded from subsequent processing because
noticeably distorted the final result. The average signal-to-noise value for a given integration time for the PSR B0329+54 pulsar and reference radio sources are shown in Table 1.2.

\begin{table}\centering
\begin{tabular}{|r|r|c|}
\hline
&&\\
IAU name & SNR $\qquad$ & Integration         \\
        &              & time, s \\
\hline
&&\\
&   14 March 1995&\\
&&\\
PSR 0329+54 & 63.2  & 394 \\
0300+470    & 27.3  & 396 \\
0316+413    & 459.0 & 276 \\
0331+545    & 9.1   & 270 \\
&&\\
&   12 May 1996&\\
&&\\
PSR 0329+54 & 8.6   & 270 \\
0300+470    & 38.4  & 210 \\
0316+413    & 263.1 & 153 \\
0333+321    & 69.0  & 210 \\
0429+415    & 311.7 & 210 \\
&&\\
&   25 May 1998&\\
&&\\
PSR 0329+54 & 7.7   & 270 \\
0300+470    & 110.7 & 210 \\
0316+413    & 188.5 & 153 \\
\hline
\end{tabular}
\caption{The signal-to-noise ratio during observations of the pulsar PSR 0329+54 together with reference sources.}
\label{snoise}
\end{table}

Among the fitted parameters were: the shift and the mutual course of the time scales at the observation sites, the tropospheric delay at the zenith at both sites, the coordinates of the antenna in Kalyazin, the coordinates of the pulsar PSR B0329+54. Since the observations were carried out at the single frequency, no estimates of the ionosphere parameters were carried out. Also, no ionospheric models were used that could predict the delay caused by it during
each scan of observations. Also, due to the single frequency and, consequently, the limited accuracy of the observations, an assessment was not carried out
coordinates of the instantaneous pole of the Earth's rotation, corrections to the universal time UT1 and nutation values. These values were taken ready-made from
the IERS (International Earth Rotation Service) bulletins. In order to take into account the influence of the ionosphere on the group delay, the entire observation session was divided into several subintervals, in each of which an independent assessment of the clock parameters was carried out. In a short time interval, the behavior of the ionospheric delay can be described by a linear function of time and, thus, it can be included in the clock parameters in this interval. 
In this way, the contribution of the ionosphere redefined the clock parameters.  It should be recalled that the clock parameters and tropospheric delay at the zenith
are determined at the very first stage of secondary processing VLBI-observations. After that, other parameters are added.

In the observation sessions in March 1995, May 1996 and May 1998, the pulsar PSR B0329+54 was observed together with other radio sources. They were selected
from the ICRF (IERS Celestial Reference Frame) catalog, which has a very good coordinate determination accuracy for today. In the ICRF catalog, radio sources are divided into three accuracy classes.  Sources of the first and second accuracy classes were selected for our observations.
Their coordinates were not adjusted, but were considered set. Thus, all subsequent parameter adjustment was reduced to minimizing
residual deviations from these radio sources. I.e., in other words, we can say that the parameters of the Kalyazin-Kashima radio interferometer
were adjusted to the quasar coordinate system determined by the ICRF catalog. At the last stage, coordinates were included in the number of customized parameters pulsar PSR B0329+54. The reference sources and their parameters are given in Table \ref{opora}.

\begin{table}\centering
\begin{tabular}{|c|l|llrl|}
\hline 
&&&&&\\
IAU name & Alt.name & Right asc.& Declination & Ang.dist.& Flux \\
& & $h\quad m\quad s$& ${}^{\circ}\quad '\quad "$ & ${}^{\circ}\quad$ & Jy \\
\hline
&&&&&\\
& & 14 March 1995&&&\\
&&&&&\\
0300+470 &$\emptyset$ & 03 03 35.24216 &+47 16 16.2732 &  8.64&  1.80   \\
0316+413 &3C84        & 03 19 48.16012 &+41 30 42.1028 & 13.25&  14.50  \\
0331+545 &$\emptyset$ & 03 34 55.1256  &+54 37 24.6080 &  0.28&  0.508  \\
&&&&&\\
& &  12 May 1996&&&\\
&&&&&\\
0300+470 &$\emptyset$ & 03 03 35.24199 &+47 16 16.2696 & 8.64 &  1.80   \\
0316+413 &3C84        & 03 19 48.16012 &+41 30 42.1028 & 13.25&  14.50  \\
0333+321 &NRAO140     & 03 36 30.10738 &+32 18 29.3367 & 22.28&  3.14   \\
0429+415 &3C119       & 04 32 36.503   &+41 38 28.43   & 16.25&  8.60   \\
&&&&&\\
& &  25 May 1998&&&\\
&&&&&\\
0300+470 &$\emptyset$ & 03 03 35.24199 &+47 16 16.2696 & 8.64 &  1.80   \\
0316+413 &3C84        & 03 19 48.16012 &+41 30 42.1028 & 13.25&  14.50  \\
\hline
\end{tabular}
\caption{Reference sources used for observations of pulsar PSR B0329+54.}
\label{opora}
\end{table}

To control the correctness of the procedure for estimating the coordinates of the pulsar, the coordinates of one of the reference sources, the quasar 0300+470, were determined, the coordinates of which are considered to be known very accurately.  Table \ref{0300} shows the coordinates obtained, their corrections and errors for 0300+470.

\begin{table}
\centering
\begin{tabular}{|l|ccl|}
\hline
&&&\\
$\qquad$ Date & Coordinates & Coord. correction & Coord. error\\
&(J2000.0)&&\\
\hline
&&&\\
14 March 1995 & $03^h\, 03^m\, 35^s\!.2409$ & $-0^s\!.0013$
& $0^s\!.0014$ \\
& $47^o\, 16'\, 16"\!.315$ & $0"\!.040$ & $0"\!.019$ \\
&&&\\
12 May 1996& $03^h\, 03^m\, 35^s\!.24225$ & $0^s\!.00031$
& $0^s\!.00120$ \\
& $47^o\, 16'\, 16"\!.287263$ & $0"\!.012$ & $0"\!.015$ \\
&&&\\
25 May 1998 & $03^h\, 03^m\, 35^s\!.24261$ & $0^s\!.00038$
& $0^s\!.00018$ \\
 & $47^o\, 16'\, 16"\!.2830$ & $0"\!.0075$ & $0"\!.0026$ \\
&&&\\              
\hline
\end{tabular}
\caption{The coordinates of the radio source 0300+470, obtained from VLBI-observations}
\label{0300}
\end{table}

The correction of coordinates 0300+470 can be regarded as the total effect of unaccounted fluctuations of the ionosphere and a methodological error in
the processing of observations. Therefore, it is necessary to include an amendment coordinates 0300+470 in the pulsar coordinate error.

The antenna coordinates were adjusted only for the Kalyazin station, since the coordinates of the antenna in Kashima had already been determined with high accuracy earlier during numerous geodetic experiments. Before the 1995 experiment, the coordinates of the radio telescope in Kalyazin were determined using
GPS technology in the WGS-84 system (Yunoshev, 1995). Accuracy was guaranteed 0.5 m, which was enough to start astrometric
experiments. The adjustment of the coordinates of the RT-64 radio telescope showed in an experiment on May 12, 1996, that the coordinate corrections along the X, Y and Z axes are comparable or even less than the standard error of the corresponding corrections. The exclusion of the coordinates of the Kalyazin station from the number of adjusted parameters did not lead to any significant change in the magnitude of the residual deviations of the group delay. In this way, you can do
the conclusion is that the corrections of the RT-64 coordinates were not significant and may not be taken into account without prejudice to the final result.
Apparently, it will be possible to really improve the coordinates of RT-64 in Kalyazin only through standard two-frequency observations, which
are usually used in geodesy. Below are the coordinate corrections and their root-mean-square errors according to the experiment data on May 12, 1996.

$$\Delta X=-0.70 {\rm m}, \qquad \sigma_{\Delta X}=0.47 {\rm m}, $$
$$\Delta Y=\;\, 0.15 {\rm m}, \qquad \sigma_{\Delta Y}=0.73 {\rm m}, $$
$$\Delta Z=1.05 {\rm m}, \qquad \sigma_{\Delta Z}=1.00 {\rm m}, $$

\noindent and according to the experiment data on May 25, 1998

$$\Delta X=\;\,0.14 {\rm m}, \qquad \sigma_{\Delta X}=0.20 {\rm m}, $$
$$\Delta Y=   -0.62 {\rm m}, \qquad \sigma_{\Delta Y}=0.19 {\rm m}, $$
$$\Delta Z=  -2.96 {\rm m}, \qquad \sigma_{\Delta Z}=0.32 {\rm m}, $$

\section{VLBI coordinates of PSR 0329+54 and their comparison with coordinates obtained by timing}

\begin{figure}
\centering
\includegraphics[width=15cm]{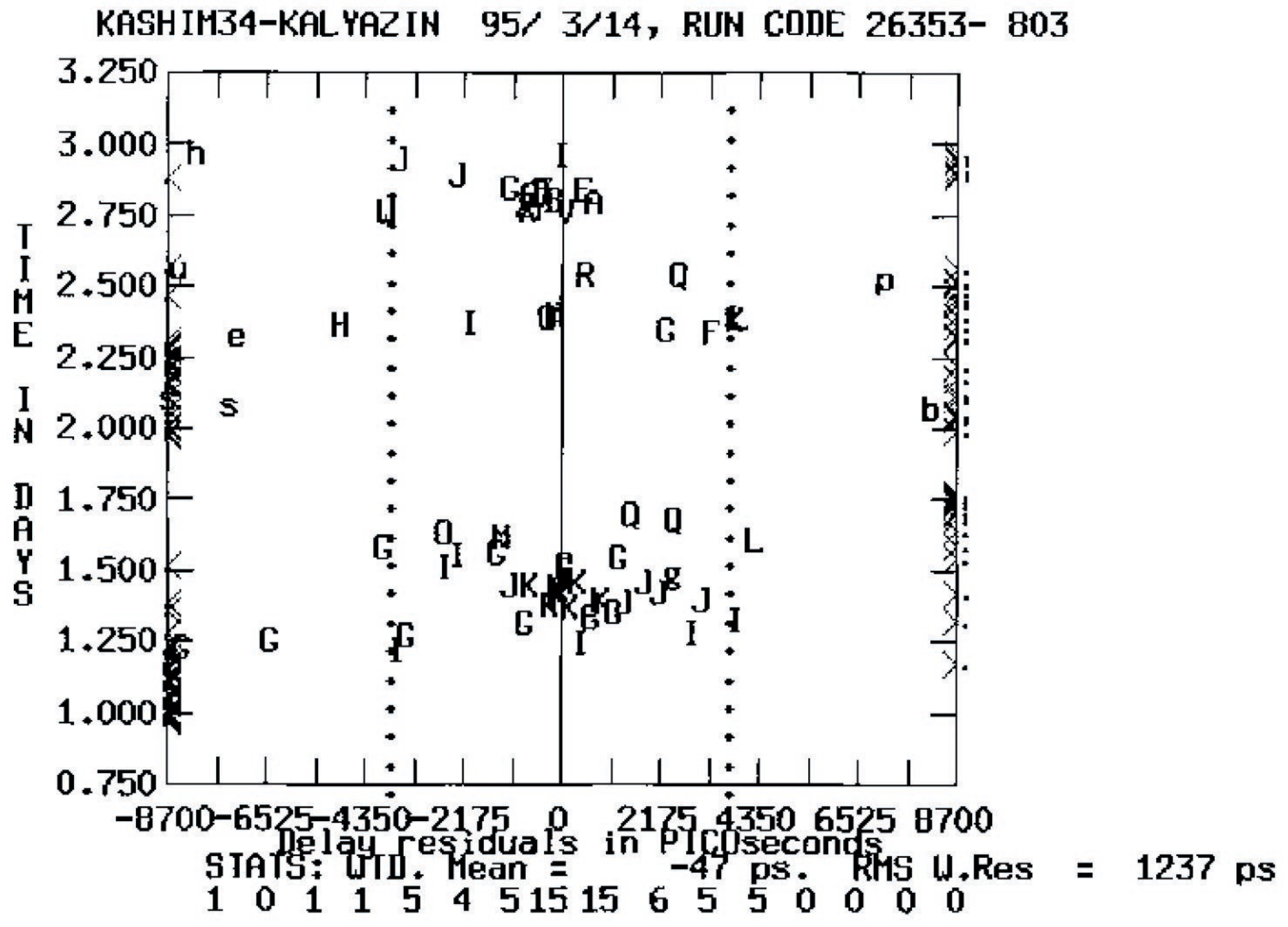}
\caption{Residuals  of group delay in pulsar observations PSR 0329+54 with reference sources, made in March 1995. On
the horizontal axis - residual deviations in picoseconds of time, on the vertical axis - time in days. Different
sources are marked with different letters.}
\label{fig11}
\end{figure}

\begin{figure}
\centering
\includegraphics[width=15cm]{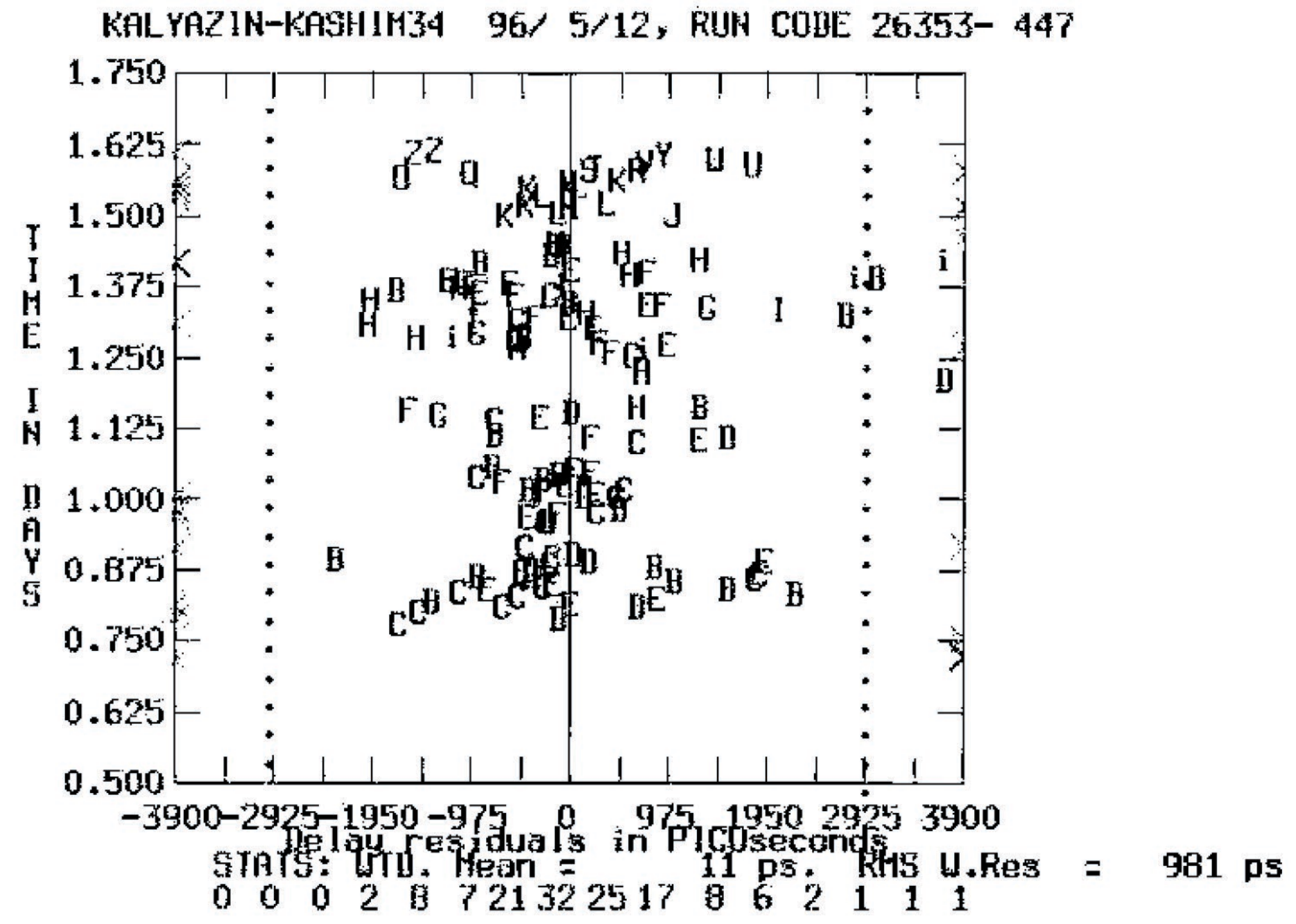}
\caption{Residuals of group delay in pulsar observations PSR 0329+54 with reference sources, made in May 1996.}
\label{fig12}
\end{figure}

\begin{figure}
\centering
\includegraphics[width=15cm]{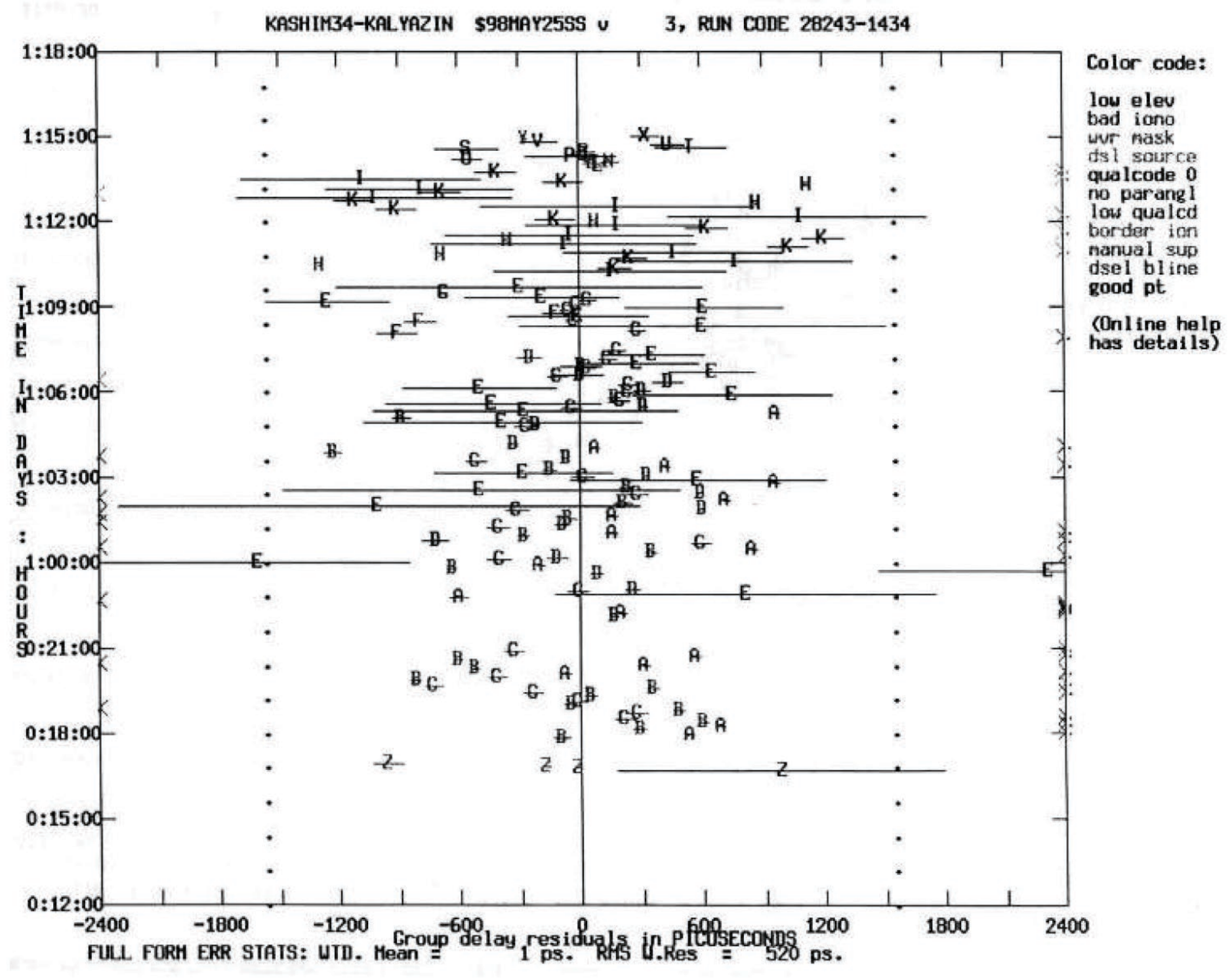}
\caption{Residuals of group delay in pulsar observations PSR 0329+54 with reference sources, made in May 1998.}
\label{fig13}
\end{figure}

The measured VLBI coordinates of PSR B0329+54 are shown in Table 1.5. The RMS error of the time delay is 1.24, 0.980 and 0.520 ns, the fringe rates are $0.177\cdot 10^{-12}$, $1.814 \cdot 10^{-12}$ and $0.604\cdot 10^{-12}$ s/s in the 1995, 1996 and 1998 sessions, respectively. It is most likely that these residual deviations are caused by ionospheric fluctuations, which could not be excluded from our observations, as well as by a limited signal-to-noise ratio. Residual
deviations of the group delay are shown in Fig. \ref{fig11}, \ref{fig12}, \ref{fig13}.

\begin{table}\centering
\label{position}
\begin{tabular}{|r|cc|l|}
\hline
&&&\\
Epoch & Right ascension & Declination & Comment \\
      &      (J2000)       & (J2000)   & \\
\hline
&&&\\
1995.20 &$3^h32^m59^s\!.3738\pm 0^s\!.0020$& $54^\circ 34'43"\!.487\pm0"\!.022$
& 14 March 1995 \\
1996.36 &$3^h32^m59^s\!.3812\pm 0^s\!.0006$& $54^\circ 34'43"\!.501\pm0"\!.006$
& 12 May 1996\\
1998.39 &$3^h32^m59^s\!.3825\pm 0^s\!.0002$& $54^\circ 34'43"\!.474\pm0"\!.003$
& 25 May 1998 \\
1981.21 &$3^h32^m59^s\!.3484\pm 0^s\!.0005$& $54^\circ 34'43"\!.663\pm0"\!.005$
& N.Bartel, 1985\\ \hline
\end{tabular}
\caption{The coordinates of the pulsar PSR B0329+54, obtained from VLBI-observations. For comparison, the coordinates taken from the work have been added
((Bartel, 1985). Only the formal coordinate error obtained during least squares processing is indicated, without taking into account ionospheric fluctuations and methodological errors.}
\end{table}

\begin{figure}
\centering
\includegraphics[width=15cm]{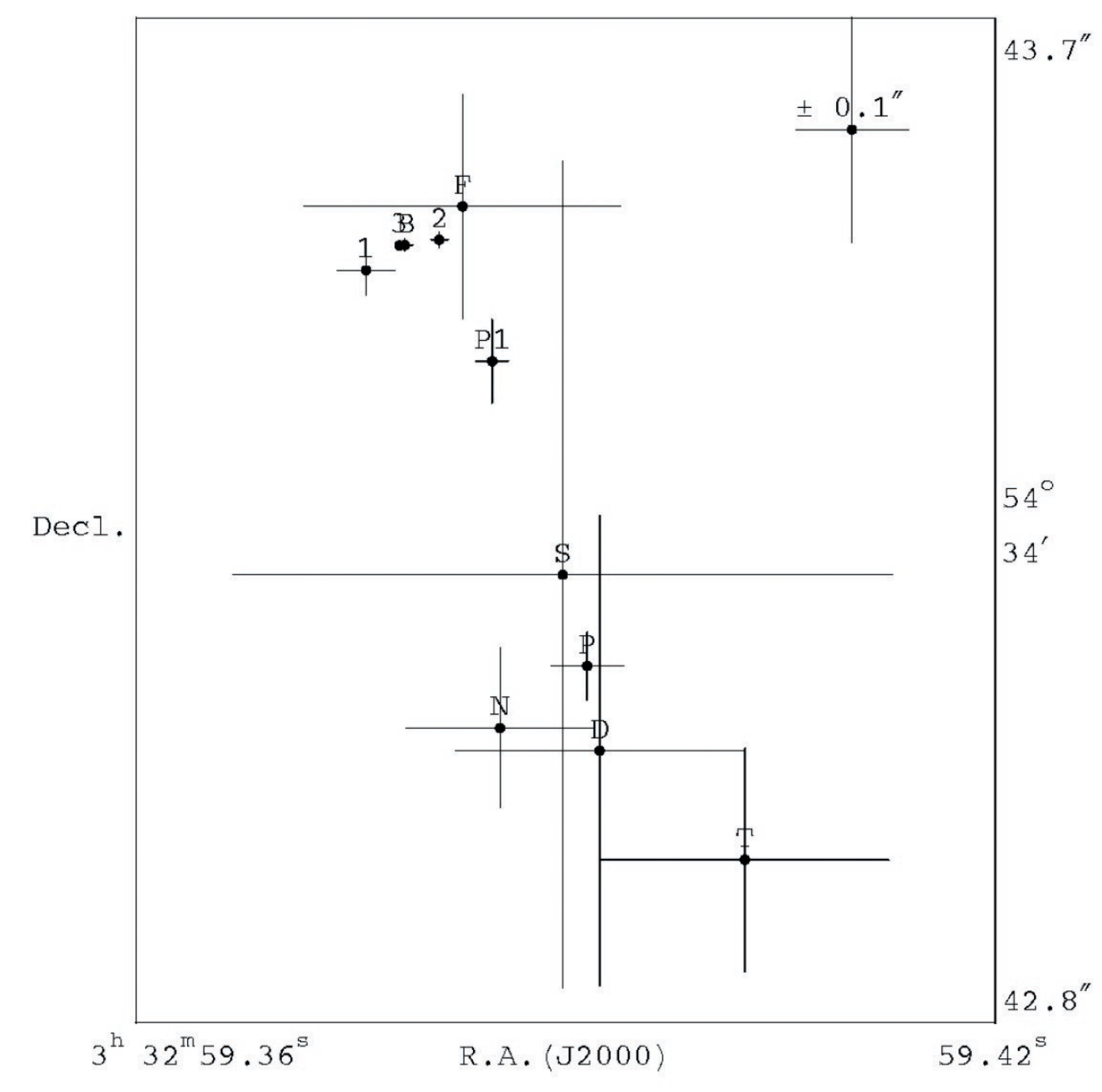}
\caption{Astrometric positions of PSR B0329+54 according to the observations of various authors, given for the epoch 1996.36. 
The error $\pm 0.1$" is shown in the upper right corner for the scale. See the notation in the text.}
\label{koord}
\end{figure}

Figure \ref{koord} shows the astrometric positions of the pulsar B0329+54 according to the observations of different authors, summarized for the epoch 1996.36. All
observations are indicated by different symbols:
\begin{itemize}
\item 1 - VLBI, Kalyazin-Kashima, 1995
\item 2 - VLBI, Kalyazin-Kashima, 1996
\item 3 - VLBI, Kalyazin-Kashima, 1998
\item B - VLBI, Bartel et al, 1983
\item V - VLA, Fomalont et al, 1983
\item R - TOA, Reichly, Downs, 1983
\item S - Backer, Shramek, 35 km interferometer, 1981
\item T - Taylor, catalog
\item N - TOA, TIMAPR, "no planet"
\item P - TOA, TIMAPR, "with 1 planet"
\item P1 - TOA, Rodin, this work.
\end{itemize}

Figure \ref{koord} shows the VLBI measurements and measurements obtained by the timing method. It can be seen that there is a significant variation in
coordinates. It can also be noted that there are two separate groups of positions: VLBI and VLA measurements are located in one place, and
timing is in another (the difference is $\approx 0"\!.45$). This discrepancy is consistent with observations made on VLA (Fomalont et
al., 1984). There, the coordinate discrepancy was also observed for other pulsars. On the other hand, no significant discrepancies were found
between TOA and VLA coordinates for pulsars B1913+16 and B1937+21, except at the level of accuracy obtained at VLA ($0"\!.2,\;0"\!.05$
, respectively) (Backer {\it et al.}, 1985). The difference in the weighted average VLBI-the position obtained in the works (Bartel {\it et al.}, 1996, Dewey,
{\it et al.}, 1996) and the position based on the ephemerides of DE200 and corrected for the rotation of DE200 (Folkner {\it et al.}, 1994), even less
- 5.4 milliseconds of arc. All this may speak in favor of the fact that the difference in coordinate systems, which is usually mentioned in the first place,
cannot fully explain the significant discrepancy in $0"\!.$45 between TOA and VLBI coordinates observed at pulsar B0329+54.

The positions indicated by N and P are obtained by processing using the program TIMAPR (Doroshenko, Kopeikin, 1990) observations conducted in the 
Jet Propulsion Laboratory (JPL, USA). This pulsar has significant noise in the residual deviations of the moments of arrival of pulses, which allows
some researchers to suspect the presence of a planet around this pulsar, possibly more than one (Shabanova, 1995). The position marked
"N" corresponds to processing without planets. The "P" position corresponds to the processing, which included the presence of one planet in PSR B0329+54.

Among other reasons that could lead to such a significant discrepancy in the pulsar positions observed by different methods, it is worth
mentioning the fluctuation effect of the ionosphere on the time delay. Fluctuation is due to the fact that the diurnal behavior of the ionospheric delay
is described by a fairly smooth function and can be included in the correction and the course of the clock. As an example, we can give a graph of the total
electron content in the ionosphere on April 2-7, 1998.

\begin{figure}
\centering
\includegraphics[width=15cm]{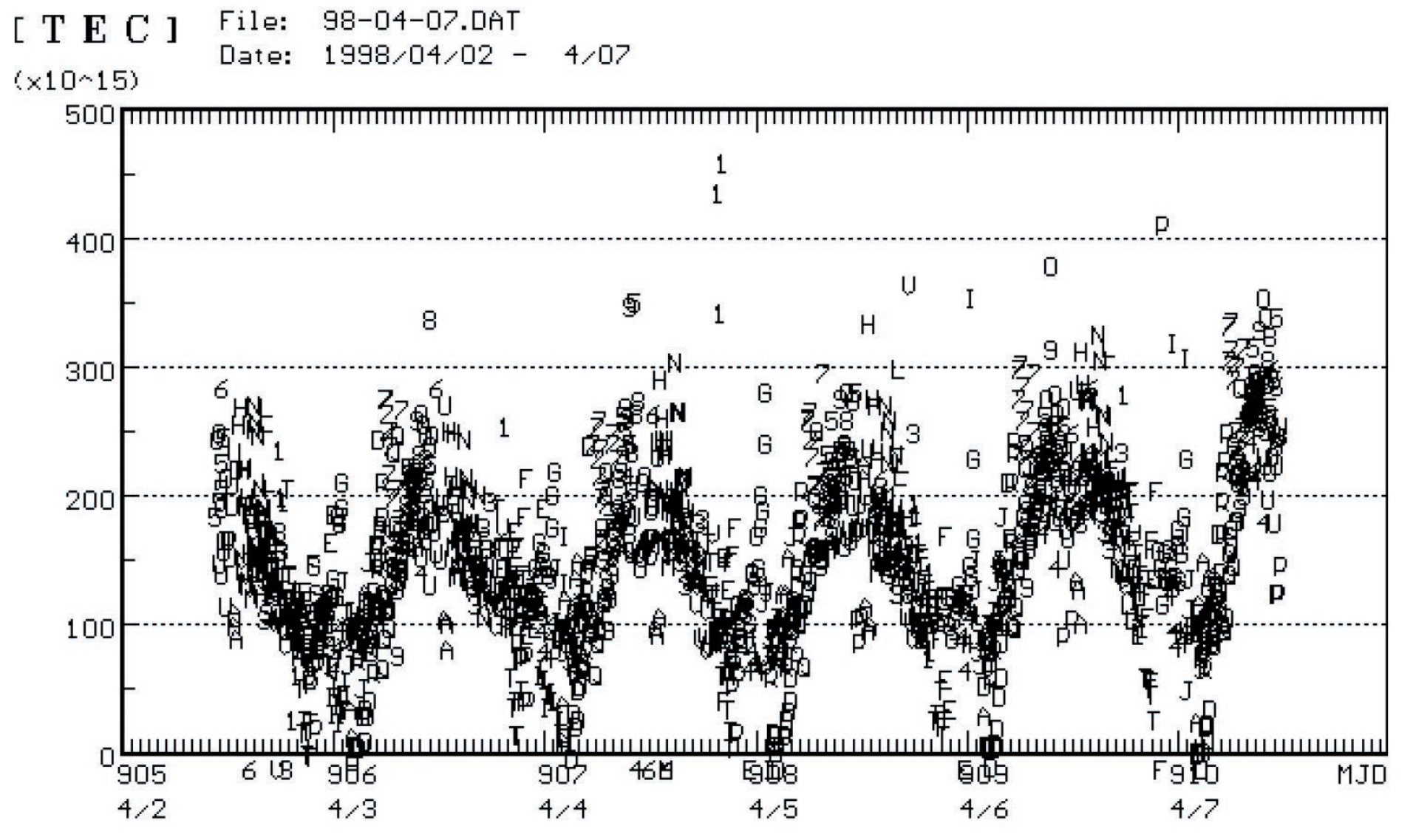}
\caption{A graph of the total electron content in the Earth's ionosphere in the direction of the zenith. The measurements were made from April 2 to April 7, 1998 at
the RT-22 radio telescope in Pushchino.}
\label{ionosph}
\end{figure}

Figure \ref{ionosph} shows a graph of the electron content in a column of section 1 m$^2$ in the direction of zenith. The daily frequency
of the electron content is clearly visible. Incorrect input of coordinates of radio sources lead to modulation of residual deviations of the group delay also with
the daily period and, thus, strongly correlate with the behavior of the ionospheric delay. The value of $I(t)=2\cdot 10^{17}\;{\rm m}^{-2}$
corresponds to a delay at a frequency of 1.4 GHz of 13.7 ns, and at a frequency of 2.2 GHz -- 5.6 ns. Immediately note that adding a sine wave to the time delay with
a period of one day and an amplitude of 1 ns will lead to a shift in the coordinates of the radio source by an amount of about 0.01" on a base with a length of 7000 km.  Figure 2.5 shows that along with the pronounced daily component , there are significant random fluctuations that have a range of approximately
$\pm$ 30 \% of the value of the periodic component. From this it can be concluded that the accuracy of measuring the geometric delay is limited
to about 4 ns in the 1.4 GHz range and 2 ns in the 2.2 GHz range.

Another reason for the limited accuracy is the insufficient ratio signal/noise, which is usually the case for all pulsars. 
The uncertainty of the time delay can be estimated by the formula (Gubanov et al., 1983) $\sigma_{\tau}=1/(2\pi B\,s/w)$, where $B$ is the effective receiver band, ${s/w}$ is the signal-to-noise ratio. At $B=40$ MHz and ${ s/w}=5$, the geometric delay error is $\sigma_{\tau}\approx 0.8$ ns.

Based on VLBI observations carried out in 1996 - 1998 at the base Kalyazin - Kashima one can determine the proper motion of the pulsar
PSR~0329+54. The magnitudes of the proper motion are obtained as follows $$\mu_\alpha=\;15\pm 4\; {\rm mas/year},$$
$$\mu_\delta=-8\pm 6\; {\rm mas/year}.$$ 
If we additionally use the coordinates PSR 0329+54 taken from the work
(Bartel {\it et al.}, 1985), then the accuracy of proper motion estimates increases significantly:
$$\mu_\alpha=\;17.4\pm 0.6\,{\rm mas\, yr}^{-1},$$
$$\mu_\delta=-11.0\pm 0.2\,{\rm mas\, yr}^{-1}.$$

Coordinates of the pulsar PSR 0329+54, determined from Russian -- Japanese observations and observations (Bartel {\it et al.}, 1985) and reduced to the epoch
1998.0 are as follows: 
$$\alpha=3^h32^m59^s\!.38211\pm 0^s\!.00088,$$
$$\delta=54^\circ 34'43"\!.4785\pm0"\!.0029.$$

\section{Another possible reason for the discrepancy in coordinates}

It is necessary to note another possible reason for pulsar coordinate discrepancies, which is usually not mentioned and which deserves to be included in
a separate section, it is the noise nature of the moments of pulse arrivals (TOA) from a pulsar (Rodin {\it et al.}, 1999a). This possible reason
no longer affects VLBI coordinates, but TOA coordinates. Depending on the nature of the noise present in the TOA, parameter estimates of
pulsar, including coordinate estimates determined during fitting using the least squares method, turn out to be either biased or
unbiased (Draper and Smith, 1973; Kostylev et al., 1991; Gubanov, 1997). If the noise in the TOA is pure white, then the parameter estimates are
unbiased, but if the noise is correlated, i.e. having a power spectrum of the form $S(f)=1/f^s$, where $s$ is a positive integer, then the estimates
parameters are biased. This can be explained by the following way: correlated noise, by its definition, does not change as much
in a random and fast way like white noise and, in principle, maybe represented as a finite Fourier series or a linear combination
orthogonal polynomials. Thus, in order to describe the noise it is necessary to explicitly introduce into the mathematical model additional
parameters. If this is not done (assuming the noise is pure white), then we have an incomplete data model that does not adequately describe observations and
leading to biased estimates.

\begin{figure}
\includegraphics[width=15cm]{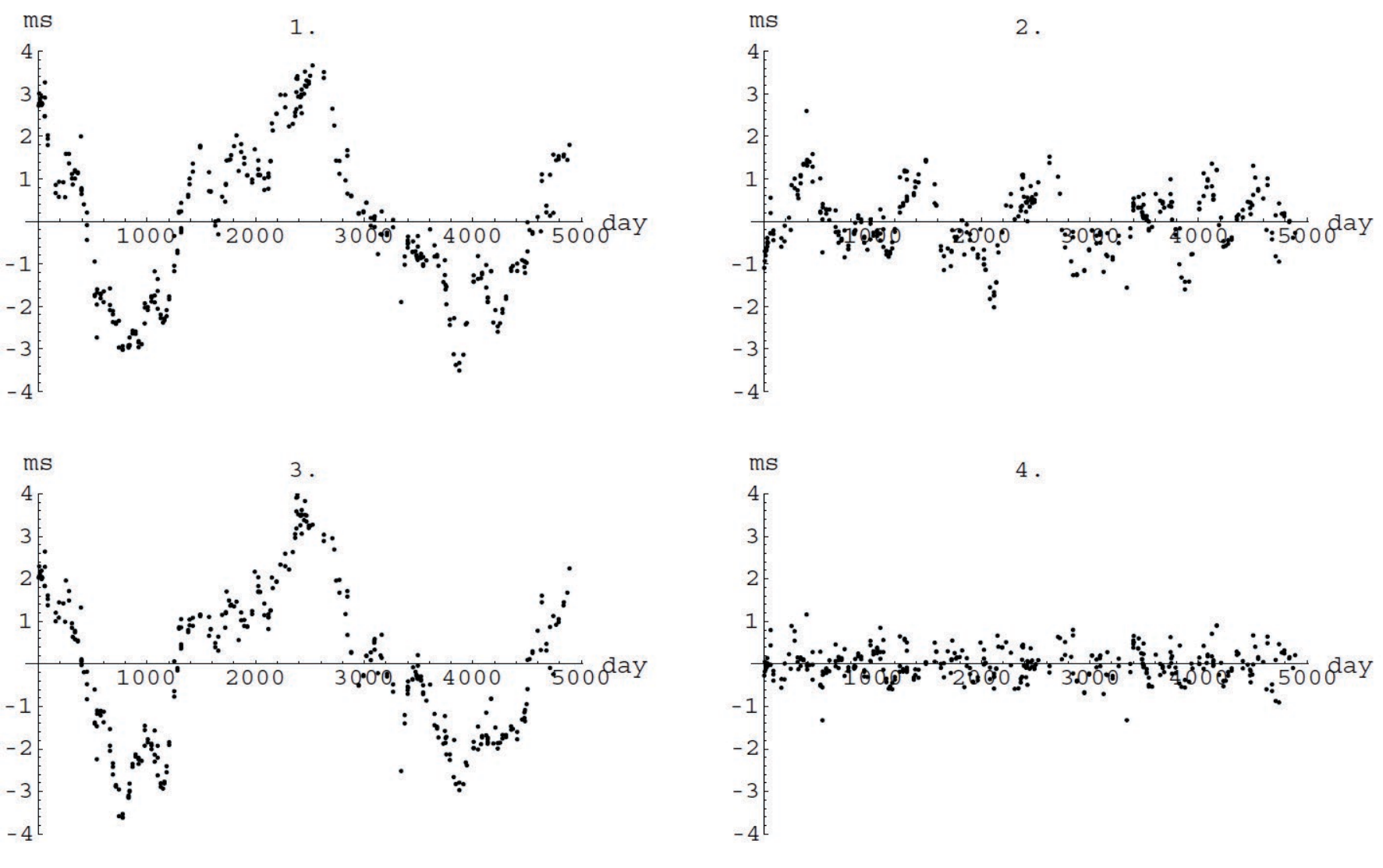}
\caption{Residual deviations of the pulsar PSR B0329+54 arising after:
1) fitting with a time polynomial of the 2nd degree;
2) fitting with a time polynomial of the 2nd degree and a periodic function arising due to the possible
the movements of one planet around the pulsar;
3) fitting with a 2nd degree time polynomial using fixed
VLBI coordinates of the pulsar;
4) Fitting the 1st graph using Fourier components, including the component
with a period of 1 year.}
\label{ukl}
\end{figure}

In order to illustrate the above, we show in Fig.\ref{ukl} graphs of residual deviations of the TOA of the pulsar PSR B0329+54, obtained by processing TOA with the TIMAPR program (Doroshenko, Kopeikin, 1990). TOAs are taken from observations carried out in the Jet Propulsion Laboratory (USA) (Downs, Reichly, 1983). Graph 1) corresponds to the case when the rotational phase of the pulsar included quadratic polynomial, 2) fit into a polynomial of the 2nd degree and were included in
fitting the orbital parameters of one planet, 3) fitting quadratic polynomial at fixed coordinates taken from VLBI observations, 4) spectral analysis of residual deviations was carried out graph 1 and subtracted the sinusoids with the found periods and the sinusoid with annual period.

Let us present the parameters of the annual sinusoid $a\cos\omega t + b\sin\omega t$, $\omega=2\pi$ years$^{-1}$, found from the residual deviations of plot 1
of Figure \ref{ukl}.

\begin{center}
\begin{tabular}{|c|c|c|c|}
\hline
     & Estimate & Standard deviation & Confidence Interval \\
     & ms & ms & 95 \%, ms \\
\hline
$a$ & 0.379 & 0.047 & $0.314 \div 0.443 $ \\
$b$ & 0.057 & 0.045 & $-0.008 \div 0.121$ \\
\hline
\end{tabular}
\end{center}

Coordinate corrections can be easily found by going to the ecliptic coordinate system. Let us write the scalar product of the unit vector
$\bf k$ directed towards the pulsar, and the Earth's radius vector $\bf r$. We will measure $\bf r$ in light seconds

\begin{equation}\label{kdotr}
{\bf k\cdot r}=\tau_A(\sin\beta_1\sin\beta_2+\cos\beta_1\cos\beta_2
\cos(\lambda_2-\lambda_1)),
\end{equation}
where $\lambda_{1,2}$ $\beta_{1,2}$ - ecliptic longitude and latitude of pulsar (index 1) and Earth (index 2), respectively,
$\tau_A=499.004784$ s -- the time it takes light to travel a distance of one astronomical unit.

Let us accept, with sufficient accuracy for our purposes, that the Earth moves along circular orbit. Then the ecliptic latitude of the Earth is $\beta_2\approx 0$ and
the equation (\ref{kdotr}) is transformed to the form
\begin{equation}\label{kdr}
{\bf k\cdot r}=\tau_A\cos\beta_1\cos(\lambda_2-\lambda_1).
\end{equation}

\noindent Let's take variations $\delta$ of the equation (\ref{kdr})
\begin{equation}
\delta{\bf k\cdot r}=\tau_A(-\delta\beta_1\sin\beta_1
\cos(\lambda_2-\lambda_1)-\delta\lambda_1\cos\beta_1
\sin(\lambda_2-\lambda_1)).
\end{equation}

\noindent From here we can derive corrections to the ecliptic coordinates of the pulsar through the parameters $a$ and $b$ of the annual sinusoid (replacing the variations $\delta$ with finite differences $\Delta$)
\begin{equation}\label{correc}
\Delta\beta_1 =\displaystyle\frac{a}{\tau_A\sin\beta_1},\quad
\Delta\lambda_1=\displaystyle\frac{b}{\tau_A\cos\beta_1}.
\end{equation}

\noindent After substituting the values of $a$ and $b$ into the equations
(\ref{correc}) we get
\begin{equation}
\begin{array}{llll}
\Delta\lambda_1= & \; 0"\!.0285\pm0"\!.024, &
\Delta\beta_1= & 0"\!.278\pm 0"\!.033, \\
\Delta\alpha\;=- & 0.0066\pm 0.0011 s, &
\Delta\delta\;= & 0"\!.273\pm0"\!.037,
\end{array}
\end{equation}

Figure 1.7 shows the power spectra of residual deviations 1 and 4 of Figure 1.6. The power spectrum was calculated as a sum of squares
Fourier transform coefficients of the time series taken as a function frequency and frequency smoothed by the Chebyshev window. Average value
periodograms can be considered as an estimate of the power spectrum time series (Bendat and Pearsol, 1974). Fourier transform of a function
$f(t)$, determined at discrete times $t_l$, $l=1,2,\ldots,n$, was taken in the form:
\begin{equation}
F(k)=\frac{1}{\sqrt{n}}\sum_{l=1}^n f(t_l)
\exp\{\frac{2\pi i(l-1)(k-1)}{n}\}.
\end{equation}
The periodogram was calculated using the formula
\begin{equation}
P(k)=\Delta t\left[(\Re F(k))^2+(\Im F(k))^2\right].
\end{equation}
Here $\Re$ and $\Im$ denote the real and imaginary parts, and $\Delta t$ - sampling interval. The power spectrum was calculated using
using the convolution of the periodogram $P(k)$ with the Chebyshev window (Marple Jr., 1990), which is characterized by having a constant level of side lobes, set manually.

\begin{equation}
S(f_k)=\frac{1}{2m+1}\sum_{i=-m}^m P(k+i).
\end{equation}

Before performing the Fourier transform, the residual series deviations were reduced to uniform, with a step between readings of 10 days.
Uniform series were obtained by spline approximation of the original series and then taking readings at the required nodes. This operation distorts
high-frequency part of the time series (at frequencies $\sim \Delta t^{-1}$), and the low-frequency part of the spectrum, which interests us,
leaves unchanged. In passing, let us note one practical computational moment: for comparison, uniform series were also constructed
by simple linear interpolation between samples. As shown comparison of two periodograms, they practically do not differ from each other, because their difference, as mentioned above, affects only high frequency component.

However, the given estimates of spectral density cannot be treated as unbiased. According to the results of the work (Deshpande
{\it et al.}, 1996) to estimate the power spectra of non-uniform time series (namely, such series are considered in astronomy)
the "CLEAN" algorithm must be used (Roberts {\it et al.}, 1987). This leads to a significant improvement in the dynamic range of the spectrum
and, as a consequence, a change in the value of the spectral index. Nevertheless, even such a simple spectral analysis as a regular periodogram
allows us to conclude that residual deviations have a power spectrum, inherent specifically in red noise.

\begin{figure} 
\centerline{\includegraphics[width=12cm,height=10cm]{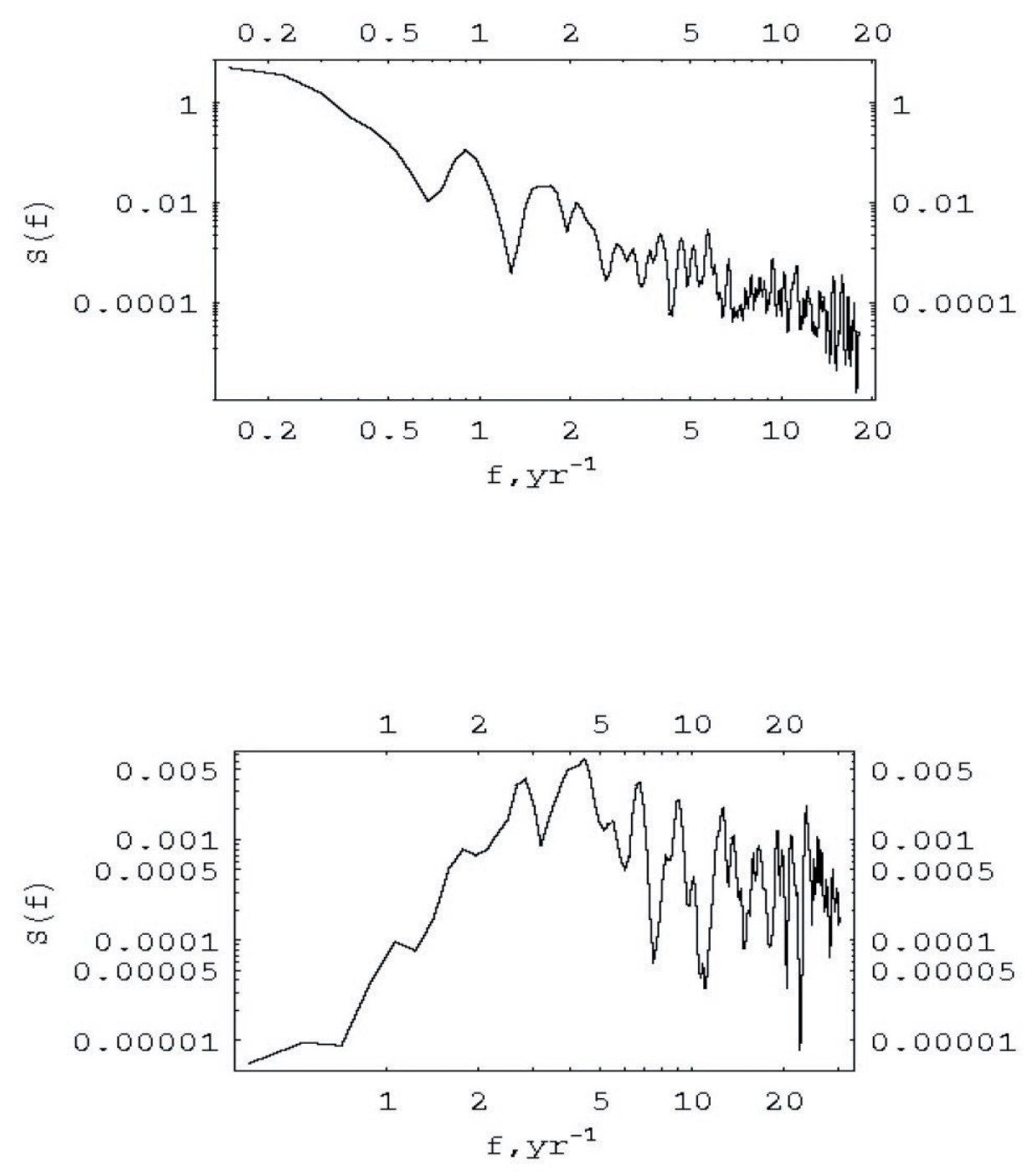}}
\caption{Power spectra of residual deviations of the pulsar PSR B0329+54. The upper graph corresponds to the power spectrum of residual TOA deviations fitted by a time polynomial of the 2nd degree. Bottom graph corresponds to the power spectrum of residual deviations after additional harmonic series fittings. It can be seen that after the frequency $\sim 3\,{\rm year}^{-1}$ the spectrum of residual deviations in the lower graph becomes character of white noise. It is also clear that all low-frequency components ($f<3$ year$^{-1}$) completely removed.}
\label{spectra}
\end{figure}

Since the power spectrum of residual deviations PSR B0329+54 even after subtraction of the main periodicities is not reduced to the spectrum of white
noise, we can conclude that the obtained estimates of the Fourier coefficients of the series 1 in Figure 1.6 are still biased relative to the real ones
values. This means that the coordinate correction PSR B0329+54 based on the parameters of the annual sinusoid, although it shifted the position
the pulsar PSR B0329+54 closer to the VLBI coordinates (which is not bad), cannot be considered final.

\section{Conclusions to Chapter 1}

Pulsar radio interferometry makes it possible to solve a number of important problems in field of fundamental astrometry. First of all, this concerns
reference systems in the sky: quasar and dynamic. Pulsars like objects observed by two independent methods (VLBI and timing), which have coordinates determined in different reference systems, have an undoubted advantage over other astronomical objects, such as radio stars, planets or asteroids. Radio interferometry of pulsars also makes it possible to determine their proper motions and parallaxes, which, in turn, gives the ability to measure distances and tangential velocities of pulsars. To obtain consistent results in pulsar VLBI it is required placing high demands on hardware and software. Thus, it is necessary to take into account the contribution of the ionosphere to the group delay. If dual-frequency observations, which allow one to directly exclude the influence
ionosphere are impossible for one reason or another, then you need to use data on the total electron content obtained by other methods, for example, using GPS satellites or using maps containing necessary information and available on the internet. To improve  signal/noise ratio during correlation processing of pulsar VLBI data it is highly desirable to use gated correlation pulsar pulses. The problem here is modification of standard software, i.e. use along with standard software for VLBI and software used in processing timing data, which allows you to pre-calculate the rotational phase pulsar.

So, from the first chapter we can draw the following conclusions:
\begin{enumerate}
\item Three sessions of VLBI observations of the pulsar PSR 0329+54 were carried out, which were planned as efficiently as possible, which made it possible to completely realize the potential accuracy of the Kalyazin-Kashima interferometer.
\item The coordinates and proper motion of pulsar PSR 0329+54 were accurately measured using the VLBI method.
\item It has been established that the reason for the discrepancy in coordinates is PSR 0329+54, measured by the VLBI method and timing, lies in the presence
low-frequency correlated noise in the pulsar TOAs.
\item A special method for processing observations is proposed, based on harmonic analysis, which eliminates low-frequency
noise component from residual TOA deviations, correct coordinates of pulsars obtained by the timing method, and significantly
reduce the discrepancy between VLBI and TOA coordinates.
\end{enumerate}

\newpage
\chapter{Dynamic pulsar time scale} \label{dvs}

The formation and keeping of time scales is one of the most important problems of modern astronomy. Progress made in creating
ultra-stable quantum frequency standards, allows you to keep single time intervals with fractional accuracy better than $10^{-14}$ and
stability is better than $10^{-15}$ at intervals of $10^3\div 10^4$ s. On the other hand, the discovery of natural very stable astronomical
clocks -- pulsars, provides the ability to recreate the scale ephemeris time at a new level.

\section{A brief overview of astronomical time scales}
\subsection{Universal Time}
Universal time UT is defined as the angle of rotation of the Earth around its axis, counted from a certain era. Until recently, UT was considered as the most accurate implementation of time. This belief was questioned by Newcomb in the process of analyzing observations of the Moon, made in the 18th-19th centuries. The observed effect was visible irregular fluctuations in the mean longitude of the Moon relative to values predicted by theory, reached $\pm 15"$ during decades. Further studies by Brown, De Sitter and Spencer Jones established the reality of such fluctuations in mean longitudes of other bodies of the Solar system, which turned out to be proportional to their average movements. With the advent of quartz and atomic clocks UT as the uniform time scale was rejected. However, although it turned out that UT is not uniform, its measurements still represent interest for geodynamics and geophysics.

\subsection{Ephemeris time}

Ephemeris time ET is an independent argument in differential equations underlying gravitational theories of body motion in Solar System (Abalakin, 1979). The definition of ET is based on movement of the Earth around the Sun. Equation for the mean longitude of the Sun $L(t_E)$ given by Newcomb and approved by the IAU in 1952.
\begin{equation}
L(t_E)=L_0+L_1t_E+L_2t_E^2,
\end{equation}
where $t_E$ is ephemeris time, $L_0$, $L_1$, $L_2$ are constants, which are derived from the theory of motion of bodies in the Solar system. Since longitude of
the Sun is determined from observations made during the daytime, when thermal deformations are strong, and since the solar disk has quite
large visible dimensions, the geometric center is determined with a rather poor accuracy of the order of 0.5". Thus, derived from
Solar observations ephemeris time has a relatively low accuracy. To improve the accuracy of determining ET, the movement of the Moon around
Earth was used. The angular motion of the Moon occurs 13 times faster than that of the Sun. Unfortunately, the theory of the motion of the Moon cannot be considered purely gravitational, since tidal acceleration in the movement of the Moon cannot be accurately determined quantitative accounting within the framework of this theory and is not completely determined only by gravitational forces. In 1950, a new time scale called "ephemeris time" was introduced at the initiative of
American astronomer Clemens.

\subsection{Atomic time}

Progress in quantum radiophysics and electronics in the 1950s made it possible create new frequency standards based on natural, repeating
with a high degree of accuracy of the oscillatory process occurring during resonant transitions of atoms from one energy level to another.
The atomic time (AT) system has a very high uniformity over over long periods of time and does not depend on the rotation of the Earth,
nor from the theory of motion of the celestial bodies of the Solar system.

The unit of measurement of time in the AT system is the atomic second, determined in accordance with the resolution of the XIII Conference of the International
Bureau of Weights and Measures as the period of time during which $9\,192\,631\,770$ oscillations are performed corresponding to the frequency
radiation absorbed by a cesium atom Cs$^{133}$ during a resonant transition between energy levels of the hyperfine structure of the ground state
in the absence of disturbances from external magnetic fields.

This definition of the atomic second is based on the results of experiment conducted by the Naval Observatory (USNO, Washington, USA) and the National Physical Laboratory (Teddington, England) for determination of the nominal frequency of the cesium standard based on observations of the Moon. This frequency for the epoch 1957.0 is determined to be $9\,192\,631\,770\pm 20$ oscillations per ephemeris second.

\subsection{Pulsar time}

Search for objects that can serve as highly stable frequency standards, led to the fact that within a short time after the discovery of pulsars it was suggested that the stable period of rotation of some of them can be used to establish a new pulsar time scale (PT). The practical implementation of the pulsar scale was developed in the works of Russian scientists (Shabanova et al., 1979; Ilyin, Ilyasov, 1985; Il'in {\it et al.}, 1986; Ilyasov et al, 1989).

The pulsar time scale is constructed in a barycentric reference frame of the Solar system as a sequence of discrete intervals between radio pulses from pulsars. It is assumed that the rotational frequency of pulsar and its derivatives are known exactly, which makes it possible to precalculate the number of the recorded pulse at any given point in time. In practice, this idealized situation does not hold true, and forward precomputation can only be carried out to a limited extent of
time interval, after which the rotational pulsar parameters are refined. Thus, the use of a single pulsar does not allow us to establish a time scale completely independent of terrestrial standards. A solution can be found using a group pulsar time scale based on several pulsars (at least three) (Ilyin, Ilyasov, 1985; Foster, Backer, 1990). Then variations in the phase of any of the pulsars can be detected and excluded by comparison with variations in the phase of other pulsars. It is assumed that it is unlikely that several pulsars will have the same variations.

By analogy with the conventional pulsar scale PT, the dynamical (binary) pulsar time scale BPT is introduced (Ilyasov et al., 1996, 1998; Kopeikin, 1997a; Rodin {\it et al.}, 1997;), which is based on the motion of the pulsar around the barycenter of the binary system. In this case, the number of revolutions around the barycenter is calculated, and it is also assumed that the period of revolution and its derivatives are known exactly. The BPT construction algorithm is described in more detail in the next section.

\section{Orbital parameters and timing algorithm for binaries pulsars}

In this section, we briefly consider the model for measuring the pulsar pulse arrival time from a pulsar located in a binary system (Kopeikin, 1997a; Rodin {\it et al.},
1997; Ilyasov et al., 1998). Let us assume that the pulsar orbits in an elliptical orbit with eccentricity $e\neq 0$ around a common center of mass of the binary system. At the same time, the pulsar rotates around its axis with a frequency $\nu$, which decreases due to energy losses on electromagnetic radiation, which is characterized by the derivative of frequency $\dot\nu$. Let us denote by $n$ the frequency of rotation of the pulsar in its orbit ($n=2\pi/P_b$, where $P_b$ is the orbital period), and through $a_r$ is the semimajor axis of the orbit.

The rotational phase of a pulsar is expressed by the formula
\begin{equation}\label{NotT}
{\cal N}(T)={\cal N}_0+\nu T + \frac 12 \dot\nu
T^2+\frac 16 \ddot\nu T^3,
\end{equation}
where $T$ is the proper emission time of the ${\cal N}$th pulse, $\ddot\nu$ is the second derivative of the frequency. Barycentric time $\tau_a$ of arrival
${\cal N}$-th pulse to the barycenter of the Solar System in the absence of the gravitational field of the Solar System and interstellar dispersion is associated with
time $T$ by formula (Damour, Deruelle, 1986)

\begin{equation}\label{tDT}
t=D\tau_a=T+\Delta_R(T)+\Delta_E(T)+\Delta_S(T)+\Delta_A(T)+
O\left(\frac{v^4}{c^4}\right),
\end{equation}
\noindent where $\Delta_R$ is the Roemer correction, which is expressed by the formula
\begin{equation}
\Delta_R(T)=\frac{a_r\sin i}{c}\{\sin\omega[\cos u-e]+(1-e^2)^{1/2}
\cos\omega\sin u\},
\end{equation}
Einstein's correction $\Delta_E$ is expressed by the formula
\begin{equation}
\Delta_E=\gamma\sin u,
\end{equation}
Shapiro correction $\Delta_S$ is expressed by the formula
\begin{equation}
\begin{array}{ll}
\Delta_S(T)=\displaystyle
-\frac{2Gm_c}{c^3}\ln\{1-e\cos u-\sin i\!\! &[\sin\omega(\cos u-e)\\
& +(1-e^2)^{1/2}\cos\omega\sin u]\},
\end{array}
\end{equation}
the aberration correction $\Delta_A$ is expressed by the formula
\begin{equation}
\Delta_A(T)=A\{\sin(\omega+A_e(u))+e\sin\omega\}+
B\{\cos(\omega+A_e(u))+e\cos\omega \},
\end{equation}
where $a_r$ is the semimajor axis of the pulsar orbit, $e$ is the eccentricity of
orbit, $i$ is the angle between the perpendicular to the orbit and the line of sight,
$\omega = \omega_0 + k A_e(u)$ - slowly precessing argument of periapsis,
\begin{equation}
A_e(u)=2{\rm arctan}
\left[\left(\frac{1+e}{1-e}\right)^{1/2}{\rm tg\frac u2} \right],
\end{equation}
$u$ is the eccentric anomaly, which is determined from the Kepler equation
\begin{equation}
u-e\cos u=n(t-T_0),
\end{equation}
$k=\dot\omega/n$, $D=\frac{1+V_R/c}{\sqrt{1-\frac{V^2}{c^2}}}$ --
Doppler factor caused by the motion of the pulsar, $V_R$ - radial velocity of the barycenter, and $V$ is the total velocity of the barycenter of the binary system
relative to the barycenter of the Solar system. It is usually assumed that $D=1$ and changes slowly, therefore in the final formulas this value is not
appears. In general, the motion of a binary system relative to the barycenter of Solar system can significantly affect orbital parameters of the binary system (Kopeikin, 1996).

In order to use the formula (\ref{NotT}), you must express time $T$ in terms of barycentric time $t$, i.e. invert formula (\ref{tDT}). Let's redesignate
\begin{equation}
\Delta(T)=\Delta_R(T)+\Delta_E(T)+\Delta_S(T)+\Delta_A(T),
\end{equation}
and let, as mentioned earlier, $D=1$. Then the solution to the equation $t=T+\Delta(T)$ relative to $t$ will be $T=t-\bar{\bar\Delta}(t)$, where
\begin{equation}\label{invDelta}
\bar{\bar\Delta}(t)=\Delta(t-\Delta(t-\Delta(t)))+
O\left(\frac{1}{c^4}\right).
\end{equation}
The formula (\ref{invDelta}) is obtained by an iterative method, which suitable for computer calculation. In the work (Damour {\it et al.}, 1986) also provides an explicit formula for $\bar{\bar\Delta}(t)$.

This section provides an algorithm for calculating the phase of a binary pulsar by an observer located at the barycenter of the Solar System. Amendments,
related to the movement of the observer were not considered, because available works that present this problem in sufficient detail.

\section{Timing noises. Correlated noises}

Any measurement process is subject to errors, which are divided into several types: random, systematic (including errors of observer) and methodological. Random errors have a white character noise, i.e. their autocorrelation function is equal to the delta function $R(t_i,t_j)=\delta(t_i-t_j)$, and the power spectrum is constant value $S(f)=const$. Systematic errors are no longer characterized by random behavior, but long-term fluctuations that depend or from external conditions, for example, temperature deformations of (radio) telescope, or from the imperfection of the observation instrument itself (a classic example is errors in the orientation of the mount telescope, non-perpendicularity of its axes, etc.). This also includes observer errors, which, as a rule, depend on his experience. All these systematic errors are studied in a special way and are largely eliminated. Currently in radio astronomy observations registration occurs automatically without direct human participation. Its role comes down to control and decision-making in a particular emergency situation, as well as to the correct organization of observations. Here we can already talk about methodological errors during the experiment and subsequent data processing. As an example is the observation of polarized radiation from pulsars in one polarization using a radio telescope mounted on an azimuthal mount. As the hour angle of the pulsar changes, the position angle between the plane of the polarized wave and the plane receiver polarization changes also. This leads to the fact that the shape of the pulsar pulse changes, and, as a consequence, the moment of registration of the center of gravity of the pulse shifts. In this example, the role of the observer is reduced to the correct choice of observation moments or to the study of additional delay caused by a change in the pulse shape so that a correction can then be introduced. Methodological ones can also include the error of the algorithm used in the subsequent processing of observational data. For example, converting the terrestrial time scale TT to the barycentric scale time TB is produced with different numbers of terms depending on the required accuracy. Failure to take into account high-order terms also leads to systematic error.

Let's take a closer look at long-term errors. Their very name is already suggests that the power spectrum of such errors rises in the region low frequencies. The power spectrum of the form $S(f)=h_s/f^s$ is often considered, where $f$ is frequency, $s$ is spectral index ($s=0,1,2,\ldots,$ in literature on pulsars is usually limited to $s\le 6$), $h_s$ - noise intensity with spectral index $s$. $s=0$ matches white noise. If we talk about timing errors, then here this means white {\it phase} noise. Further in the text, without specifying each time, noise will be considered during timing pulsars. Noises with $s\ge 1$ are called correlated, colored or
just red. In the radiophysical literature there is no established terminology on this matter. In this work we will use all terms. Spectral indices $s=2,4,6$ correspond to random walk in the phase, frequency and derivative of the pulsar rotation frequency respectively. Spectral indices $s=1,3,5$ correspond to flicker noise of phase, frequency and derivative frequency, respectively. Both the random walk noise, and flicker noise are obtained within the framework of shot noise, i.e. noise that is formed as a result of the superposition of a large number pulses of a certain shape, amplitude and duration of which are random variables, and the moments of disturbing impulses are Poisson subsequence. In the specific case of measurements of the TOA of pulsars, noise from spectral index $s=3$ is formed as a result of density fluctuations of interstellar medium through which the impulse propagates (Blandford {\it et al.}, 1984), and noise with $s=5$ is obtained due to the presence of background gravitational radiation left over from the origin of the Universe (Bertotti {\it et al.}, 1985; Kopeikin, 1997b).

\section{Estimating pulsar parameters using the least squares method}

Today, millisecond and binary pulsars are the most stable frequency standards created by nature. Their high stability can be applied in various fields of science from
relativistic astrophysics to fundamental metrology: search stochastic background of gravitational waves, testing the general theory
relativity (GR), establishment and keeping of new time scales. This chapter is devoted specifically to this area of application of double pulsars --
ephemeris time scale, and, to a lesser extent, the establishment of an upper limit to the background of gravitational waves that arose in the early Universe. Any
of the above problems requires precise knowledge of rotational and orbital parameters of the pulsar. In practice, these parameters are never
are known absolutely precisely, only numerical estimates of the parameters are known, obtained by one method or another (least squares method, method
maximum likelihood, maximum entropy method, etc.), with varying degrees of accuracy. Any researcher in his work strives to
so that these accuracies are as high as possible. Accuracy is characterized by the magnitude of the dispersion and strongly depends on the type of noise present in the observational data and having completely different physical origin. This chapter covers both white noise and correlated (red, colored) low frequency noise having a spectrum power of the form $h_s/f^s$, where $h_s$ is the noise intensity, $f$ is the frequency, $s=0,1,2,\ldots,6$ - spectral index.

The essence of our research comes down to the study of functional dependence of the parameters of a binary pulsar on the time interval of
observations. To compare the stability of the pulsar time scale PT, which is based on the rotation of the pulsar around its own axis, and the scale
BPT (short for Binary Pulsar Time scale), based on the orbital the motion of the pulsar around the center of mass of the binary system, are used
dispersion of rotational frequency $\sigma_\nu$ and orbital mean motion $\sigma_n$.

Let us consider a simplified model of a double pulsar (Kopeikin, 1997a; Rodin {\it et al.}, 1997; Ilyasov et al., 1998). The simplification will be
that we are considering a pulsar that moves in a circular orbit, which has only one derivative of the rotational frequency caused by
radiation of electromagnetic waves, derivatives of orbital frequency $n$ and projections of the semimajor axis $x$, the existence of which is due to
radiation of gravitational waves or accelerated motion of a binary system, neglected due to their smallness.

The moment of emission of the $\cal N$-th pulse $T$ is related to its moment of arrival $t$ equation
\begin{equation}\label{emis}
D[T+x\sin n(T-T_0)]=t+\epsilon(t).
\end{equation}
The equation (\ref{emis}) uses the following notation: $T$-time in pulsar time scale, $t$-time at the barycenter of the Solar system,
$x$-projection of the semimajor axis of the pulsar orbit onto the line of sight, $n$-orbital frequency (mean motion) of the pulsar, $T_0$-- passage momentum
 of the periapsis, $\epsilon(t)$ -- additive noise, $D=\frac{1+V_R/c}{\sqrt{1-\frac{V^2}{c^2}}}$ -- Doppler factor caused by
motion of the pulsar, $V_R$ is the radial velocity of the pulsar, $V$ is its full speed, $c$ is the speed of light.

The rotational phase of a pulsar is given by the formula
\begin{equation}
{\cal N}=\nu_p T+\frac 12\dot\nu_p T^2+\frac 16 \ddot\nu_p T^3+\ldots,
\end{equation}
where $\nu_p$, $\dot\nu_p$, etc. is the rotational frequency of the pulsar, its derivative, etc. in the pulsar reference frame at epoch $T=0$

The pulsar phase is written as
\begin{equation}\label{1.7}
{\cal N}(t)={\cal N}_0+\nu t+\frac 12\dot\nu t^2+\nu x\sin[n(T-T_0)]
\end{equation}

Let us assume for simplicity that all observations of the binary pulsar carried out with the same accuracy. Let us further define the residual deviations of
pulsar phase $r(t)$ as the difference between the observed phase ${\cal N}^{obs}$ and precomputed ${\cal N} (t,\theta )$ based on the best
estimates of the parameters of the pulsar model.
\begin{equation}
r(t,\theta )={{\cal N}^{obs}-{\cal N}(t,\theta )},
\label{1.8a}
\end{equation}
where $\theta =\{\theta _{a},a=1,2,...k\}$ denotes the set of defined parameters ($k=5$ in the model (\ref{1.7})).

If the numerical value of the $\theta$ parameters coincides with their physical value $\hat\theta$, then the residual deviations will represent
actual noise, i.e.
\begin{equation}
r(t,\hat\theta )=\epsilon(t).
\label{1.8b}
\end{equation}
In practice, the true physical values of the $\theta$ parameters are never are known absolutely precisely, only their estimates are known, obtained by
least squares method, so the residual deviations are represented by the expression (Kopeikin, 1999)

\begin{equation}
r(t,\theta ^{*})=\epsilon (t)-{\displaystyle \sum_{a=1}^{5}}\beta _{a}\psi_{a}
^{*}+O(\beta _{a}^{2}), \label{1.9}
\end{equation}
where $\beta_a=\delta\theta_a=\theta_a^*-\hat\theta_a$ corrections to estimates parameters $\theta$, functions $\psi_a(t,\theta^*)=\left.\frac{\partial\cal
N}{\partial\theta_a}\right|_{\theta=\theta^*}$ and are written below in explicit form
\begin{equation}
\begin{array}{ccl}
\psi_1=\frac{\partial{\cal N}}{\partial \nu} &=& t-x\sin nt, \\\\
\psi_2=\frac{\partial{\cal N}}{\partial n} &=& -tx\nu\cos nt,\\\\
\psi_3=\frac{\partial{\cal N}}{\partial t_0} &=& -\nu, \\\\
\psi_4=\frac{\partial{\cal N}}{\partial T_0} &=& nx\nu\cos nt, \\\\
\psi_5=\frac{\partial{\cal N}}{\partial x} &=& -\nu\sin nt.
\end{array}
\end{equation}

Let us further assume that during one orbital revolution we make $m$ measurements at regular intervals and a total of $N$ observed
orbital revolutions. Then we will have $mN$ residual deviations $r_i=r(t_i),\; i=1,2,\ldots,mN$. Standard LSQ estimation procedure
gives the best estimates for parameter corrections (or simply parameters)
$\beta_a$:
\begin{equation}
\beta _{a}(T)={\displaystyle \sum_{b=1}^{5}}
\sum_{i=1}^{mN}L_{ab}^{-1}\psi(t_{i})\epsilon (t_{i}),
\qquad a=1,...,5,
\label{1.11}
\end{equation}
where $T=NP_b$ is the full observation interval, information matrix $L_{ab}$ is presented in the form
\begin{equation} L_{ab}(T)={\displaystyle
\sum_{i=1}^{mN}}\psi _{a}(t_{i})\psi _{b}(t_{i}).
\label{Lab}
\end{equation}

Let us denote by angle brackets the averaging over an ensemble of observational data. Let us assume that the noise ensemble average $\epsilon(t)$ is equal to
zero. Then the average values of all parameters $\beta_a$ are also equal zero, i.e.
\begin{equation}
<\epsilon(t)>=0,\quad <\beta_a=0>
\end{equation}

The covariance matrix $M_{ab}=<\beta_a\beta_b>$ of parameter estimates is given by expression
\begin{equation}
M_{ab}({T})={\displaystyle \sum_{c=1}^{5}}{\displaystyle
\sum_{d=1}^{5}}L_{ac}^{-1}L_{bd}^{-1}\left[
{\displaystyle \sum_{i=1}^{mN}}{\displaystyle \sum_{j=1}^{mN}}\psi
_{c}(t_{i})\psi_{d}(t_{j})R(t_{i},t_{j})\right], \label{1.12}
\end{equation}
where $R(t_i,t_j)=<\epsilon(t_i)\epsilon(t_j)>$ autocovariance function of stochastic noise $\epsilon(t)$. Recall that the function $R(t_i,t_j)$
is related to the noise power spectrum $\epsilon(t)$ by the Wiener-Khinchin theorem
\begin{equation}
S(f)=2\int_{0}^{\infty }R(\tau )\cos (2\pi f\tau )d\tau,
\label{papa}
\end{equation}
where $R(\tau)=R(t_i-t_j)$. Specific expressions for $R(\tau)$ are given  in table 2.1.

The matrix $M_{ab}$ is symmetric ($M_{ab}=M_{ba}$), the elements in it the main diagonal represent the variances of the measured parameters
$\sigma_{\beta_a}=M_{aa}=<\beta_a^2>$, off-diagonal elements represent the correlation between parameters.

\begin{table}\centering
\label{spektr}
\begin{tabular}{|c|c|}
\hline &\\
Spectrum & Autocovariance\\
$S(f)$ & function $R(\tau)$ \\
\hline &\\
$h_0$ & $h_0\delta(\tau)$ \\&\\
$h_1/f$ & $\displaystyle -\frac{h_1}{\pi}\ln|\tau|$ \\&\\
$h_2/f^2$ & $\displaystyle-h_2|\tau|$ \\&\\
$h_2/f^3$ & $\displaystyle-\frac{1}{2\pi}h_3\tau^2\ln|\tau|$ \\&\\
$h_2/f^4$ & $\displaystyle-\frac 16 h_4|\tau|^3$ \\&\\
$h_2/f^5$ & $\displaystyle-\frac{1}{24\pi}h_5\tau^4\ln|\tau|$\\&\\
$h_2/f^6$ & $\displaystyle-\frac 1{120} h_6|\tau|^5$ \\&\\
\hline
\end{tabular}
\caption{Power spectrum and the corresponding autocovariance function white and correlated noise (Kopeikin, 1997a). The quantities $h_s$, $s=0,1,2,\ldots,6$ characterize the noise amplitude.}
\end{table}

Subtracting the accepted model from the observational data gives the residual deviations in which only random fluctuations will dominate.
Expression for root mean square residual deviations after subtracting the best-fitted model is given by the formula
\begin{equation}
<r^2(T)>=\frac1{mN}\sum\limits_{i=1}^{mN}\sum\limits_{j=1}^{mN}F(t_i,t_j)
R(t_i,t_j),
\end{equation}
where is the function
\begin{equation}
F(t_{i},t_{j})=\delta _{ij}-{\displaystyle \sum_{a=1}^{5}}
\sum_{b=1}^{5}L_{ab}^{-1}L_{cd}^{-1}\psi_a(t_{i})\psi_{b}(t_{j}),
\label{1.12b}
\end{equation}
called the filter function. The last expressions clearly demonstrate that part of the noise in the residual deviations is filtered out during fitting
parameters, and thus its value becomes smaller. At all, by choosing quite a lot of adjustable parameters, you can decrease the amplitude
residual deviations to a sufficiently small value. This is due to the fact that long-term correlated noise fluctuations with any advance
with a given degree of accuracy can be represented by a function described by a finite number of parameters. This statement applies only to
correlated noise. White noise can only be described by random function by definition.

Model information matrix (\ref{emis}), calculated by the formula (\ref{Lab}), is given by the following expression
\begin{equation}
L_{ab}=\left(
\begin{array}{ccccc}
\frac{T^3}{\nu^23} & \frac{-2x T}{\nu n^2} & \frac{- T^2}{2\nu} & 0 &
\frac{T}{\nu n} \\\\
& \frac{x^2T^3}{6} & 0 & \frac{-n^2x^2T^2}{4} & \frac{-xT}{4n} \\\\
&& T & 0 & 0 \\\\
&symm&& \frac{n^2x^2T}{2} & 0 \\\\
&&&& \frac{T}{2}
\end{array}
\right).
\end{equation}

The inverse information matrix is written as
\begin{equation}
L^{-1}_{ab}=\left(
\begin{array}{ccccc}
\frac{12\nu^2}{T^3} & \frac{432\nu}{n^2x T^5} & \frac{6\nu}{T^2} &
\frac{216\nu}{n^3xT^4} & \frac{-24\nu}{n T^3} \\\\
& \frac{24}{x^2 T^3} & \frac{216}{n^2xT^4} &
\frac{12}{nx^2T^2} & \frac{12}{nxT^3} \\\\
&& \frac{4}{T} & \frac{108}{n^3xT^3} & \frac{-12}{nT^2} \\\\
&symm&& \frac{8}{n^2x^2T} & \frac{6}{n^2xT^2} \\\\
&&&& \frac{2}{T}
\end{array}
\right).
\end{equation}

Expression
\begin{equation}
\Psi_{ab}=\sum_{i=1}^{mN}{\displaystyle \sum_{j=1}^{mN}}\psi
_{a}(t_{i})\psi_{b}(t_{j})R(t_{i},t_{j})
\end{equation}
easier to calculate after changing variables (Kopeikin, 1997a)
\begin{equation}
x=\frac{t_i-t_j}{2},\quad y=\frac{t_i+t_j}{2}
\end{equation}
\begin{equation}
\int_{0}^{T}{\displaystyle
\int_{0}^{T}}\psi_{a}(t_{i})\psi_{b}(t_{j})R(t_{i}-t_{j})dt_i\,dt_j=
\int_{0}^{T}R(x)dx \int_{0}^{x-T}A_{ab}(x,y)dy,
\end{equation}

The matrix $\Psi_{ab}$ in the case of noise $1/f$ has the form
\begin{equation}
\Psi_{ab}=\left(
\begin{array}{ccccc}
\frac{-T^4}{4\nu^2} & 0 & \frac{2T^3\ln T}{3\nu} & 0 & -\frac{2T^2}{\nu n} \ \\\
&-\frac{\pi x^2T^3}{6n} & 0 &-\frac 1{2}{\pi x^2T^2} & 0 \\\\
& & 4T^2\ln T & 0 & \frac{8T\ln T}{n} \\\\
& symm &&-{2\pi nx^2T} & 0 \\\\
&&&&-\frac{2\pi T}{n} \\\\
\end{array}\right)
\end{equation}
in the case of noise $1/f^2$ has the form
\begin{equation}
\Psi_{ab}=\left(
\begin{array}{ccccc}
\frac{T^5}{15\nu^2}& 0 &-\frac{T^4}{6\nu} & 0 & \frac{2T^3}{3\nu n} \\\\
& \frac{x^2T^3}{3n^2} & 0 &\frac{2x^2T^2}{n} & 0 \\\\
& & -\frac{4}{3}T^3 & 0 & -\frac{4T^2}{n} \\\\
&symm&& {4x^2T} & 0 \\\\
&&&&\frac{12T}{n^2} \\\\
\end{array}\right)
\end{equation}
in the case of noise $1/f^3$ has the form
\begin{equation}
\Psi_{ab}=\left(
\begin{array}{ccccc}
-\frac{T^6\ln T}{18\nu^2} & 0 &\frac{T^5\ln T}{15\nu} &0
&-\frac{2T^4\ln T}{3\nu n} \\\\
&\frac{\pi x^2T^3}{3n^3} & 0 &-\frac{2x^2T^3\ln T}{3n} & 0 \\\\
&& \frac{2}{3}T^4\ln T &0& \frac{8T^3\ln T}{3n} \\\\
&symm&& \frac{4\pi x^2T}{n} & 0 \\\\
&&&&-\frac{8T^2\ln T}{n^2} \\\\
\end{array}\right)
\end{equation}
in the case of noise $1/f^4$ has the form
\begin{equation}
\Psi_{ab}=\left(
\begin{array}{ccccc}
-\frac{T^7}{35\nu^2} & 0 & \frac{T^6}{45\nu} &0& -\frac{2T^5}{5\nu n} \\\\
&-\frac{28x^2T^3}{n^4} & 0 &-\frac{x^2T^4}{3n} & 0 \\\\
&& \frac{4T^5}{15} & 0 & \frac{4T^4}{3n} \\\\
&symm&& \frac{48x^2T}{n^2} & 0 \\\\
&&&&-\frac{16T^3}{3n^2}
\end{array}\right)
\end{equation}
in the case of noise $1/f^5$ has the form
\begin{equation}
\Psi_{ab}=\left(
\begin{array}{ccccc}
-\frac{T^8\ln T}{30\nu^2} & 0 &\frac{2T^7\ln T}{105\nu} &0
&-\frac{8T^6\ln T}{15\nu n} \\\\
&-\frac{64x^2T^4\ln T}{n^4} & 0 & -\frac{2x^2T^5\ln T}{5n} & 0 \\\\
&& \frac{4}{15}T^6\ln T & 0 & \frac{8T^5\ln T}{5n}\\\\
&symm&& \frac{96x^2T^2\ln T}{n^2} & 0 \\\\
&&&&-\frac{8T^4\ln T}{n^2} \\\\
\end{array}\right)
\end{equation}
in the case of noise $1/f^6$ has the form
\begin{equation}
\Psi_{ab}=\left(
\begin{array}{ccccc}
\frac{2T^9}{567\nu^2} & 0 &-\frac{T^8}{630\nu} &0&\frac{4T^7}{63\nu n} \\\\
&\frac{176x^2T^5}{15n^4} & 0 &\frac{2x^2T^6}{45n} & 0 \\\\
&&-\frac{8T^7}{315} & 0 & -\frac{8T^6}{45n} \\\\
&symm&&-\frac{64x^2T^3}{3n^2} & 0 \\\\
&&&& \frac{16T^5}{15n^2}
\end{array}\right)
\end{equation}

Variances of estimated parameters $\nu$, $n$, $t_0$, $T_0$, $x$ for noise with power spectra of the form $1/f$, $1/f^2$, $1/f^3$, $1/f^4$, $1/f^5$,
$1/f^6$ are presented in Table 2.2.
\begin{table}\centering
\label{shumy}
\begin{tabular}{|c|ccccccc|}
\hline&&&&&&&\\
&$h_0$ & $h_1f^{-1}$ & $h_2f^{-2}$ & $h_3f^{-3}$ &
$h_4f^{-4}$ & $h_5f^{-5}$ & $h_6f^{-6}$ \\&&&&&&&\\
\hline &&&&&&&\\
$\sigma^2_\nu$ & $\frac{12h_0}{\tau^3}$ & $\frac{240h_1\nu^2\ln \tau}{\tau^2}$ &
$\frac{312h_2\nu^2}{5\tau}$ & $\frac{128}{5}h_3\nu^2\ln \tau$ &
$ \frac{304h_4\nu^2\tau}{35}$ &$\frac{264}{35}h_5\nu^2\tau^2\ln \tau$ &
$\frac{40}{63}h_6\nu^2\tau^3$ \\&&&&&&& \\
$\sigma^2_n$&$\frac{24h_0}{x^2\tau^3}$&$\frac{672h_1\pi}{nx^2\tau^3}$ &
$\frac{1920h_2}{n^2x^2\tau^3}$ &
$\frac{384h_3\ln \tau}{n^2x^2\tau^2}$ & $\frac{192h_4}{n^2x^2\tau}$ &
$\frac{1152h_5\ln \tau}{5n^2x^2}$ & $\frac{128h_6 \tau}{5n^2x^2}$ \\&&&&&&& \\
$\sigma^2_{t_0}$&$\frac{4h_0}{\tau}$&${96h_1\ln \tau}$ &$\frac{440h_2\tau}{15}$ &
$\frac{178h_3\tau^2\ln \tau}{15}$ &
$\frac{452h_4 \tau^3}{105}$ &$\frac{418h_5\tau^4\ln \tau}{105}$ &
$\frac{16h_6\tau^5}{45}$ \\&&&&&&& \\
$\sigma^2_{T_0}$&$\frac{8h_0}{n^2x^2\tau}$&$\frac{248\pi h_1}{n^3x^2\tau}$
&$\frac{688h_2}{n^4x^2\tau}$ & $\frac{128h_3\ln \tau}{n^4x^2}$ &
$\frac{64h_4\tau}{n^4x^2}$
& $\frac{384h_5\tau^2\ln \tau}{5n^4x^2}$ & $\frac{128h_6\tau^3}{15n^4x^2}$\\&&&&&&\\
$\sigma^2_x$&$\frac{2h_0}{\tau}$&$\frac{8\pi h_1}{n\tau}$ &$\frac{368h_2}{5n^2\tau}$ &
$\frac{32h_3\ln \tau}{5n^2}$ &$\frac{256h_4\tau}{21n^2}$ &
$\frac{192h_5\tau^2\ln \tau}{7n^2}$ &
$\frac{1312h_6\tau^3}{315n^2}$\\&&&&&&&\\
\hline
\end{tabular}
\caption{Dependence of the dispersions of pulsar parameters on the type of spectrum power. Value $h_s,\; (s=0,1,\ldots,6)$ -- spectrum intensity of
noise power with spectral index $s$, $\nu$ -- rotation frequency pulsar, $n$ - orbital mean motion, $x$ -- projection large semi-axes of the pulsar's orbit per line of sight, $\tau$ is the observation interval.}
\end{table}

\section{Allan dispersion of orbital frequency}

To compare the stability of two time scales PT and BPT it is useful to introduce two parameter, as is done by metrologists (Ryutman, 1978)
\begin{equation} \label{yv}
y=\frac{\Delta \nu(t)}{\nu}\quad {\rm and} \quad v=\frac{\Delta n(t)}{n},
\end {equation}
which characterize instantaneous fractional frequency deviations. The convenience of working with such dimensionless quantities is due to the fact that they remain unchanged during frequency multiplication and division operations. In addition, one can easily compare the stability of the orbital motion of pulsars with different values of their own rotation and orbital frequencies. In practice, we deal with integral quantities defined as follows
\begin{equation}
\bar y(t)=\frac{1}{\tau}\int_{t}^{t+\tau}y(t')dt',\qquad
\bar v(t)=\frac{1}{\tau}\int_{t}^{t+\tau}v(t')dt'.
\end{equation}

Due to random fluctuations of $y(t)$ and $v(t)$, repeated measurements of $\bar y(t)$, $\bar v(t)$ give different numerical values with a scatter depending on the interval $\tau $. To statistically characterize the scatter, the $\sigma^2$ dispersion (or $\sigma$ standard deviation) of these values is used. Assuming that $y(t),\;v(t)$ have zero means, the variances will be equal to the mean square of $\bar y,\;\bar v(t)$:
\begin{equation}
\sigma^2[\bar y(t)]=<\bar y^2>,\quad \sigma^2[\bar v(t)]=<\bar v^2>.
\end{equation}
The brackets $<>$ denote either the statistical mean calculated from an infinite number of samples at a given time $t$, or the average over
infinite time interval, calculated from one sample $y(t)$ or $v(t)$ (This statement is valid only in the case of ergodicity random process). This variance is called true and is denoted
$I^2(\tau)$.

It is logical to assume that the values
$I_y^2=\sigma^2[\bar y(t)]$ and $I_v^2=\sigma^2[\bar v(t)]$ should not depend on the moment of measurements $t$, but only on the measurement interval
$\tau$. Therefore, we will write further
\begin{equation}
I_y(\tau)=<\bar y^2>^{1/2},\quad
I_v(\tau)=<\bar v^2>^{1/2}.
\end{equation}
 If we accept taking into account the formula (\ref{yv}), we can deduce
\begin{equation}
I_y(\tau)=\frac{1}{\nu}\sigma_{\nu}(\tau),\quad
I_v(\tau)=\frac{1}{n}\sigma_{n}(\tau).
\end{equation}

The true variance $I^2(\tau)$ is the theoretical idealization, since it relates to an infinite number of observational data. Practical estimates must be based on a finite number of samples $\bar y(t_k)$, $\bar v(t_k)$. The main tool used to assess the stability of time scales, this is the so-called Allan dispersion, which is introduced through the following definition
\begin{equation}
\sigma_y^2(\tau)=\frac12<(\bar y(t+\tau)-\bar y(t))^2>,\quad
\sigma_v^2(\tau)=\frac12<(\bar v(t+\tau)-\bar v(t))^2>.
\end{equation}
The formula that relates the true variance and the Allan variance has the form
\begin{equation} \label{ai}
\sigma_y^2(\tau)=2[I_y^2(\tau)-I_y^2(2\tau)],\quad
\sigma_v^2(\tau)=2[I_v^2(\tau)-I_v^2(2\tau)].
\end{equation}
Table \ref{allan} presents the values of $\sigma_y^2(\tau)$ and $\sigma_v^2(\tau)$, calculated based on the formula (\ref{ai}) and the values $\sigma_{\nu}^2$ and $\sigma_n^2 $ taken from Table \ref{shumy}. It can be seen that the value of $\sigma_y$ does not depend on the orbital parameters of the pulsar, as would be expected, since $\sigma_y$ characterizes the instability of rotation around the pulsar?s own axis, regardless of whether it is a component of a binary system or not. As for the value $\sigma_v$, it depends on the orbital period $P_b=2\pi/n$ and the projection of the semimajor axis $x$, i.e. on the parameters of the sinusoidal function in formula (\ref{1.7}).

\begin{table}\centering
\begin{tabular}{|c|cc|}
\hline&&\\
& $\sigma_y^2(\tau)$ & $\sigma_v^2(\tau)$ \\&&\\
\hline& & \\
$h_0$ & $\frac{10.5h_0}{T^3}$ & $\frac{21h_0}{n^2x^2T^3}$ \\&&\\
$\frac{h_1}{f}$ & $\frac{27h_1}{T^2}$ & $\frac{21h_1}{n^5x^2T^3}$ \\&&\\
$\frac{h_2}{f}$ & $\frac{9.6h_2}{T}$ & $\frac{24192h_2}{n^6x^4T^3}$ \\&&\\
$\frac{h_3}{f}$ & $2.5h_3 $ & $\frac{27648h_3}{n^6x^4T^2}$ \\&&\\
$\frac{h_4}{f}$ & $8.2h_4T$ & $\frac{33792h_4}{n^6x^4T}$ \\&&\\
$\frac{h_5}{f}$ & $14.4h_5T^2\ln T$ & $\frac{20503h_5}{n^6x^4}$ \\&&\\
$\frac{h_6}{f}$ & $3.6h_6T^3$ & $\frac{11366h_6}{n^6x^4}T$ \\&&\\
\hline
\end{tabular}
\caption{Dependence of the dispersions of pulsar parameters on the type of spectrum noise power of residual deviations of the pulsar TOA. Options
$h_s,\,\nu,\,n,\,x,\,T$ have the same meaning as in Table 2.2}
\label{allan}
\end{table}

\begin{figure}
\includegraphics[height=15cm]{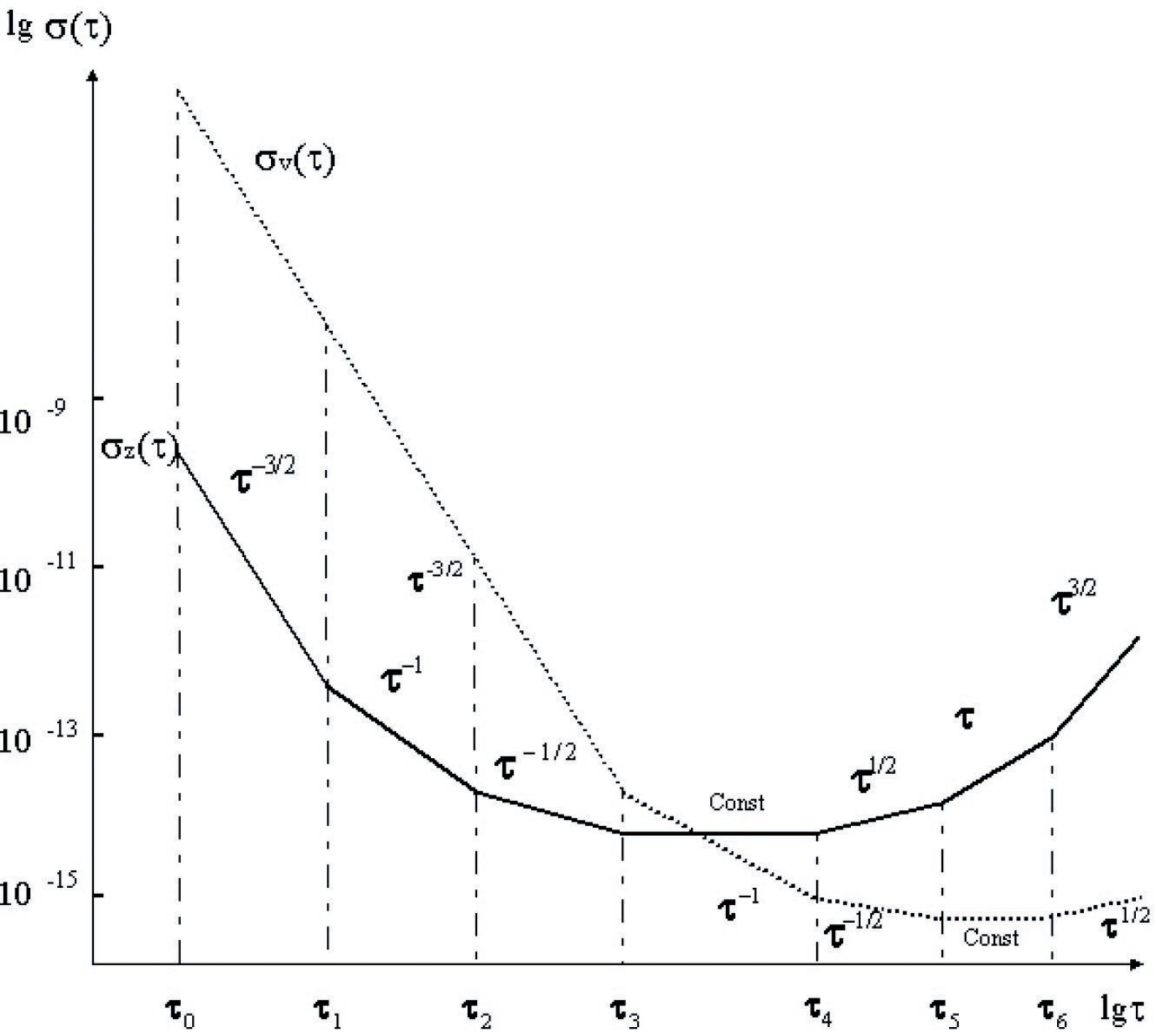}
\caption{Schematic behavior of relative rotational stability pulsar frequency, characterized by the parameter $\sigma_y$ (solid line),
and its orbital frequency, characterized by the parameter $\sigma_v$ (dotted line).}
\label{vars}
\end{figure}

Figure \ref{vars} schematically shows the behavior of quantities $\sigma_y(\tau)$ and $\sigma_v^2(\tau)$. It is assumed that the sample interval $[\tau_0,\tau_1]$ is dominated by white phase noise, mainly due to measurement errors. In the interval $\tau>\tau_1$, various components of the red noise spectrum begin to appear sequentially, the amplitudes of which $h_s$, as experience with quantum frequency standards shows, gradually decrease as the spectral noise index $s$ increases. For example, in the situation schematically depicted in Figure \ref{vars}, as the observation interval increases from $\tau_1$ to $\tau_2$, red flicker noise with spectral density $h_1/f$ dominates, and in in the interval $\tau_2<\tau<\tau_3$, red noise with spectral density $h_2/f^2$ dominates, etc. In general, the longer the observation time interval, the more significant the contribution of noise with a larger value of the spectral index $s$. This is due to the fact that, according to the assumption made above, noise with a larger $s$ has a smaller amplitude $h_s$ and therefore can begin to contribute only over longer time intervals $\tau$.

It can also be seen that the instability of the orbital phase of the pulsar $\sigma_v$ is insensitive to noise with spectral indices $s=1,2$ and, thus, does not allow $\sigma_v$ measurements to distinguish these noises from each other and from white noise. On the other hand, it is precisely this behavior of $\sigma_v(\tau)$ that makes it possible to use the orbital phase of a binary pulsar as a new, more stable over long time intervals, time standard.

As can be seen in Fig. \ref{vars}, value of $\sigma_y(\tau)$ rotational frequency pulsar begins to grow from moment $\tau_4$, while the value
$\sigma_v(\tau)$ still continues to decrease until noise with spectral index $s\ge 5$ begins to dominate. This the result does not depend on the specific numerical values of the noise amplitude. As theoretical analysis shows, the minimum of the $\sigma_v(\tau)$ curve may, under certain circumstances, be achieved much later,
than the similar minimum for $\sigma_y(\tau)$. Minimum depth for $\sigma_y(\tau)$ is due to noise with spectral index $s=3$, generated by the presence of large-scale inhomogeneities in the interstellar medium (Bandford {\it et al.} 1984). Minimum depth for $\sigma_v(\tau)$ is determined by the noise amplitude with spectral index $s\ge 5$, which arise due to the existence of a stochastic background gravitational radiation generated at the initial stage the origin of the Universe (Bertotti {\it et al.}, 1983). Noise with index $s = $5 can, in principle, also arise as a result of random fluctuations of the first derivative of the pulsar natural rotation frequency, although the occurrence of such fluctuations is extremely unlikely (Cordes, Greenstein, 1981) and will not be discussed further.
As an observational example, we can point out the minimum of the curve $\sigma_y(\tau)$ for a single pulsar PSR B1937+21, which has
the value of $10^{-14}$ over an observation interval of 2 years, and its appearance apparently due to the dominant influence of instability
rotational phase of a given pulsar (Kaspi {\it et al.}, 1994).

\section{Pulsars J1713+0747, B1913+16 and BPT scale}

\begin{figure}
\centering
\includegraphics[width=15cm]{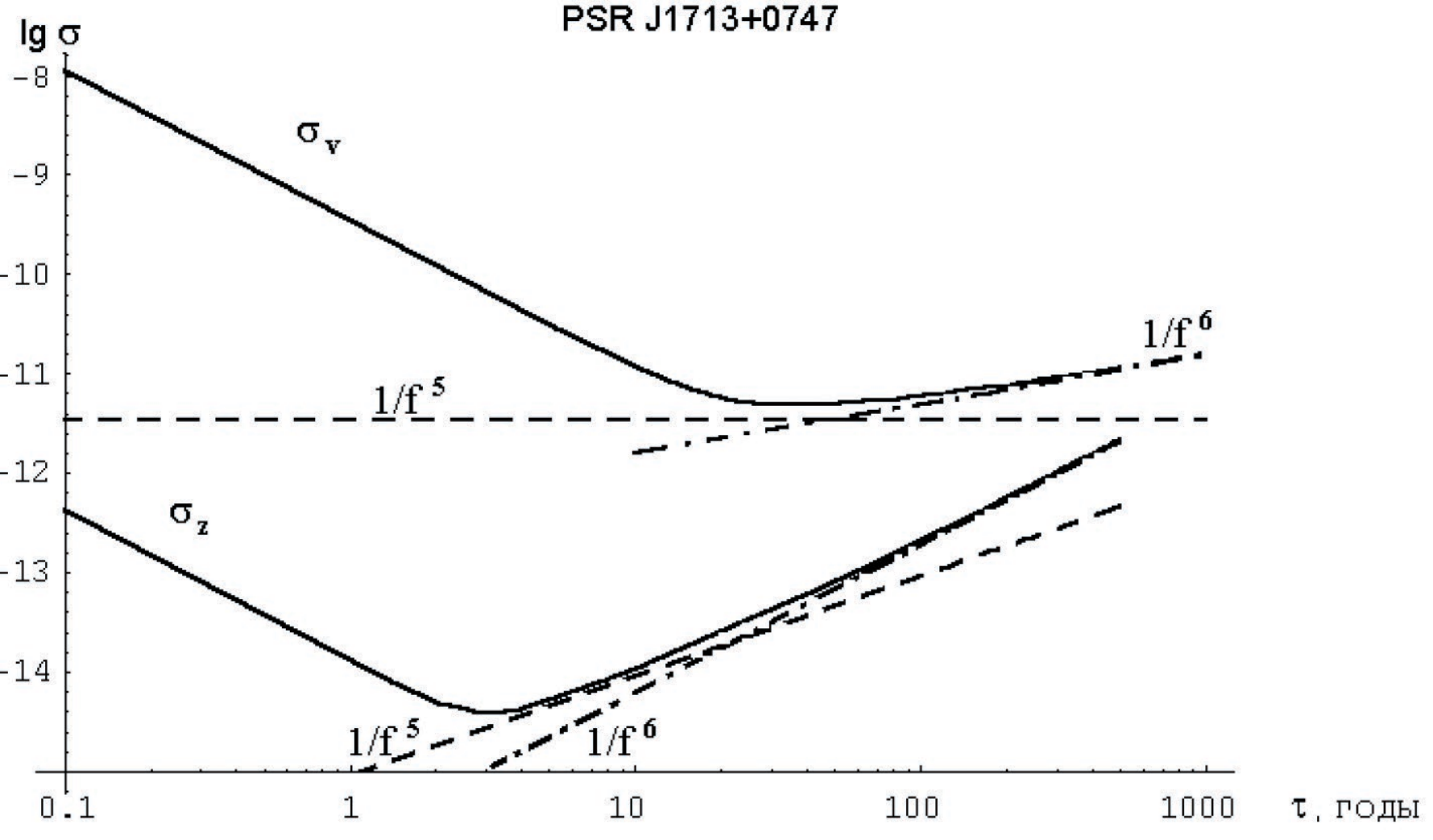}
\caption{Curves $\sigma_y$ and $\sigma_v$ calculated for the pulsar J1713+0747 in assumption that the orbital eccentricity is $e=0$. Minimum curve
$\sigma_v$ is achieved over an interval of $\tau=20$ years and is determined solely by the noise amplitude of the stochastic background gravitational
radiation (power spectrum of the form $1/f^5$). The amplitude of this noise $h_5=\Omega_g{\rm h^2}=10^{-8}$. Noise amplitude $1/f^6$ was selected
so that its contribution to $\sigma_y$ and $\sigma_v$ is noticeable only on the interval time more than 20 years. The dotted and dash-dotted lines show
noise intensity of the form $1/f^5$ and $1/f^6$ and limit the behavior curves $\sigma_y$ and $\sigma_v$ from below. This pulsar is suitable
candidate for setting an upper limit on the amplitude of the background gravitational radiation, since the time interval during which the minimums of the curves 
$\sigma_y$ and $\sigma_v$ are reached, it is sufficient short for experimental research.}
\label{fig22}
\end{figure}

\begin{figure}
\centering
\includegraphics[width=15cm]{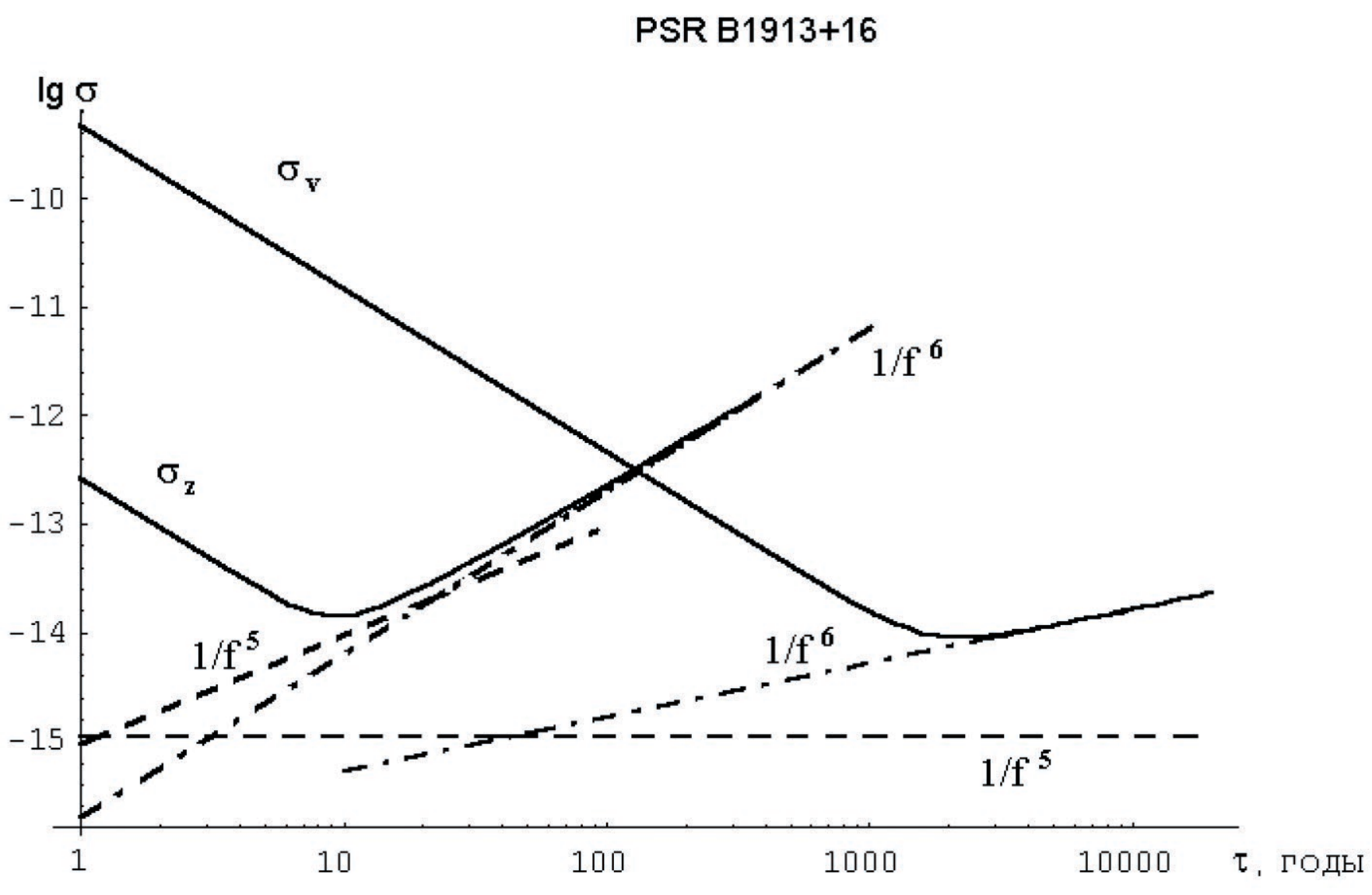}
\caption{
Curves $\sigma_y$ and $\sigma_v$ calculated for the pulsar B1913+16 without taking into account the ellipticity of its orbit. The minimum of the $\sigma_v$ curve is reached significantly later than the minimum $\sigma_y$. Dashed and dash-dotted lines show the intensity of noises of the form $1/f^5$ and $1/f^6$ for $\sigma_y$ and $\sigma_v$. The pulsar in question is not suitable for finding amplitudes of background gravitational radiation, because minimum $\sigma_v$
is determined by the noise $1/f^6$, which begins to dominate significantly before the noise of stochastic gravitational waves. This pulsar is
a good candidate as a custodian of the dynamic ephemeris scale time, because the minimum $\sigma_v$ is deep enough and is achieved over such a long time interval that all other known time and frequency standards on such an interval have significantly worse stability. Taking into account the ellipticity of the orbit does not affect significantly on the behavior of the curves $\sigma_y$ and $\sigma_v$ and not affects the results of our analytical conclusions.}
\label{fig23}
\end{figure}

Another visual illustration of using the variance $\sigma_v(\tau)$ with the point of view of fundamental applications is shown in Fig. ~\ref{fig22} and ~\ref{fig23}, on
which depict the behavior of the curves $\sigma_y(\tau)$ and $\sigma_v(\tau)$ for pulsars J1713+0747 and B1913+16, which have noticeably different
values of orbital parameters. When constructing these drawings we for clarity neglected noise with spectral indices $s=1,2,3,4$, since on the time intervals under consideration they are only have little effect on the behavior of dispersions. Fig.~\ref{fig22} demonstrates behavior of $\sigma_y(\tau)$, the initial part of which is taken from the work (Foster {\it et al.}, 1996), and reflects the influence of white noise. We we assume that $\Omega_g h^{-2}=10^{-8}$, although it is very likely that in in reality its value is significantly less. Noise amplitude value with spectral index $s=6$ was taken so that this noise starts have a noticeable effect on the behavior of $\sigma_y(\tau)$ after 20 years continuous observations. The behavior of the function depicted this way $\sigma_y(\tau)$ can be recalculated using the formulas in Table~\ref{allan} and values of orbital parameters of J1713+0747 from Table~\ref{binp} to the corresponding behavior of the function $\sigma_v(\tau)$, which is also shown in Fig.~\ref{fig22}. Comparison of curves in Fig.~\ref{fig22} shows that the double pulsar J1713+0747 is a very good detector stochastic gravitational wave radiation as measured values $\sigma_y$, and by measurements $\sigma_v$. It is important to note that using the $\sigma_v(\tau)$ curve for this purpose should help avoid uncertainty in identifying the spectral nature of noise in the residual deviations of the pulsar TOA. For example, the mere presence
minimum in the curve for $\sigma_v(\tau)$ immediately indicates the presence of noise with index $s\ge 5$. At the same time, in order to judge
the presence of similar noise along the $\sigma_y(\tau)$ curve, it is necessary fairly accurately measure the slope of this curve, which seems
problematic. Fig.~\ref{fig23} is constructed similarly. When drawing a curve $\sigma_y$ we took the error value of the parameter $y$ equal to $10^{-13}$ on
time interval of 7 years in accordance with the results of the work (Taylor, Weisberg, 1989). It was also assumed that during this time interval
the white noise of the measurements dominates. The $\sigma_v$ curve was obtained by corresponding recalculation using the formulas in Table.~\ref{allan}. Mutual behavior of curves $\sigma_y(\tau)$ and $\sigma_v(\tau)$ for the pulsar B1913+16, shown in Figure~\ref{fig23} is of a different nature compared to
J1713+0747. The minimum of the $\sigma_v(\tau)$ curve is approximately $10^{-14}$, is located below the minimum of the curve $\sigma_y(\tau)$ and
achieved after an exceptionally long period of time ($\sim$ 2000 years). This behavior of the rotational and orbital frequencies B1913+16 does not allow obtaining satisfactory limit on the amplitude of stochastic gravitational waves, but this makes pulsar a very reliable stable standard of dynamic BPT scale at time intervals when random fluctuations of residual deviations of the pulsar TOAs no longer have the character of white noise and are mainly due to the presence of red noise with spectral index $s \ge 1$.

It is very important from a practical point of view when neglecting noise with index $s=6$, the absolute value of the minimum of the curve $\sigma_v(\tau)$ can be calculated based on numerical value $h_5$, which is determined by the energy density of  background gravitational waves $\Omega_g$ (Bertotti {\it et al.}, 1983, Kopeikin, 1997).
\begin{equation} \begin{array}{rl}
S(f)=&\frac{H_0^2\Omega_g(f)}{16\pi^4}f^{-5},\\&\\
\Omega_g(f)df=&\frac{f}{\rho_c}\frac{d\rho_g(f)}{df},
\end{array}
\end{equation}
Thus, the value of $h_5$ is equal to
\begin{equation}
h_5=\frac{H_0^2\Omega_g(f)}{16\pi^4},
\end{equation}
and the minimum
$\sigma_v(\tau)$ for the case of a circular orbit is equal to
\begin{equation}
\label{minsv} \sigma_v=\frac{143H_0\sqrt{\Omega_g}}{4\pi^2n^2x}=
1.1\cdot 10^{-19}\sqrt{\Omega_g}P_b^2x^{-1}{\rm h}.
\end{equation}

Here $\rm h=H_0/100$ km/(s Mpc) is the dimensionless Hubble constant, $P_b$ - orbital period of the pulsar, $x$ - projection onto the line of sight of the
semimajor axes of the pulsar's orbit. It can be seen that the value $\sigma_v(\tau)$ does not depend only on the values of $\Omega_g$ and $\rm h$, but also on the orbital parameters of pulsar, which is quite natural, since the magnitude $\sigma_v(\tau)$ characterizes the stability of the orbital movements. In a real situation, the minimum $\sigma_v(\tau)$ also depends on noise with spectral index $s=6$. If the noise amplitude with index $s=6$ is less than the amplitude of the stochastic gravitational background, then the minimum value of $\sigma_v(\tau)$ coincides with the value determined by the formula (\ref{minsv}). Otherwise noise with $s=6$ may begin to appear before $\sigma_v(\tau)$ reaches the value
(\ref{minsv}).

It is possible to reduce the value of $\sigma_v(\tau)$ by using pulsars with small ratio $P_b^2/x$. If we take into account Kepler's third law, then $P_b^2/x$
can be expressed by the following formula
\begin{equation}\label{Px}
\frac{P_b^2}{x}=\frac{P_b^2}{a\frac{m_2}{m_1+m_2}\sin i}=
\frac{4\pi^2a^2}{Gm_2\sin i},
\end{equation}
where $m_1,\;m_2$ are the masses of the pulsar and companion, respectively, $a$ - semimajor axis of the relative orbit (in the case of a circular orbit, this is
just the distance between the companions), $i$ is the inclination of the orbital normal to line of sight, $G$ is the gravitational constant. Thus, in order to minimize 
$\sigma_v(\tau)$ to obtain the maximally stable BPT scale, it is necessary to use pulsars in close binary systems with a massive companion. This is understandable from a physical point of view: the value $Gm_2/a^2$ in the formula (\ref{Px}) is, in its meaning, nothing more than the acceleration of the pulsar. Thus, the larger it is, the more stable the binary system is to external disturbances. If we want to use a double pulsar for setting an upper limit on the amplitude of the stochastic gravitational radiation, then in this case you need a pulsar with a large ratio $P_b^2/x$, which, accordingly, will give a high minimum $\sigma_v(\tau)$, detectable over a relatively short time interval. It should be emphasized that the measurement of the minimum $\sigma_v(\tau)$, appears to provide a more accurate indication of the upper limit on amplitude stochastic gravitational waves than the measurement of the corresponding slope of the $\sigma_y(\tau)$ curve, since the measurement of the minimum can be produced with greater confidence than determining the slope of the curve $\sigma_y(\tau)$, depending on sufficiently large measurement errors $\sigma_y(\tau)$ over long time intervals (Kaspi {\it et al.}, 1994; Dewey, Thorsett, 1996).

There are currently 46 binaries in the pulsar catalog pulsars. As an illustration, we have given in table.~\ref{binp} parameters of some binary pulsars. Some of them can be taken as detectors of stochastic background of gravitational waves (J1713+0747, B1855+09) or as keepers of the dynamic pulsar time scale BPT (B1913+16).

\begin{table}\centering
\begin{tabular}{|l|clcllr|}
\hline
PSR &$P$, ms &d, kpc&$P_b$, days &$\quad x$, s&$\quad e$ &$r$, s\\
&&&&&& \\
\hline
J1713+0747& 4.5701&1.1 &67.82512988 &32.342413&$7.492\cdot 10^{-05}$ &$>$13.5\\
B1855+09 & 5.3621&0.9 &12.32717119 &9.2307802&$2.168\cdot 10^{-05}$ &1.15 \\
J0437-4715& 5.7574&0.14 &5.741042329 &3.3666787&$1.87 \cdot 10^{-05}$ &- \\
J2317+1439 & 3.4452&1.89 &2.459331464 &2.3139483&$5\cdot 10^{-07} $ &- \\
J1537+1155 &37.9044&0.68 &0.420737299 &3.729468 &$0.2736779 $ &6.70 \\
J2130+1210C&30.5292&10 &0.335282052 &2.52 &$0.68141 $ &- \\
B1913+16 &59.0299&7.13 &0.322997463 &2.3417592&$0.6171308 $ &6.83 \\
\hline
\end{tabular}
\caption{Orbital parameters of some binary pulsars. $P$ - period pulsar in milliseconds, $d$ - distance to the pulsar, $P_b$- orbital period of the pulsar in days, $x$ - projection of the semi-major axis of orbits per line of sight in light seconds, $e$ - orbital eccentricity, $r$ is the amplitude of the Shapiro effect.} \label{binp}
\end{table}

\section{Conclusions to Chapter 2}
\begin{enumerate}

\item Using the least squares method, we calculated the theoretical dependences of the behavior of the dispersion of the rotational and orbital parameters of the pulsar depending on the type of correlated noise and the length of the observation interval, which makes it possible to predict the accuracy of estimates of pulsar parameters and thereby plan pulsar observations.

\item Based on a comparison of rotational and orbital frequency dispersions as functions of the observation interval, a new astronomical ephemeris time scale BPT is proposed, which is stable over long time intervals (tens and hundreds of years).

\item It is shown that to obtain the most stable BPT scale, pulsars in close binary systems with a massive companion are needed, and to determine the amplitude of stochastic gravitational radiation (or set an upper limit on it) over a relatively short time interval (10 - 20 years), pulsars with large ratio $P_b^2/x$.

\end{enumerate}

\chapter{Long-term variations of residual deviations in time of arrivals from pulsars} \label{flight}

To date, extensive data sets have been obtained containing residual deviations of the time of pulse arrivals (TOAs) of pulsars. These
data are used in various fields of science: from relativistic astrophysics and cosmology to fundamental metrology. Character of
the behavior of the residual deviations of the pulsar TOA in time depends on many factors: for example, from various active processes occurring inside
pulsar, changes in the interstellar medium in which it propagates signal from the pulsar, planetary ephemeris errors used for
recalculation of the TOA from the observer to the barycenter, etc. In this chapter, at least part of the observed residual deviations of the pulsar TOAs
proposed to be explained by using effects due to flybys of massive bodies near the pulsar. The passage of bodies along three possible types of orbits: elliptical, parabolic and hyperbolic is considered. However, it should be noted that near the moment of periapsis passage, it is difficult to distinguish the three types of orbits from each other if their eccentricity is close to 1.

\section{Causes of long-term variations in pulsar parameters}

Various causes of disturbances in the residual deviations of the pulsar TOA are considered. On a pulsar, as an astronomical object located in and
surrounded by other massive bodies, forces are constantly acting that stellar astronomy divided into regular and irregular
(Kulikowski, 1985). Regular forces include the general gravitational force field of a quasi-stationary system, changing very slowly over time and
determining both the orbits of individual stars and the overall rotation of all subsystems in galaxies or stellar clusters. Irregular forces
which act by chance have only a short-term effect and slightly change the direction and magnitude of the velocities involved in
bringing bodies closer together The impact of irregular forces can quite significantly fluctuate and lead to strong cumulative effects in star motion
on short (compared to the lifetime of the considered systems) time intervals. If part of the mass of the star system is contained
in star clouds, clouds of diffuse matter, stellar clusters, then the role of random encounters of these objects with stars will still play
a greater role compared to the case of single stellar encounters.

\subsection{Stellar clusters}

In stellar clusters, close encounters of two or several stars occur from time to time as they move along regular orbits. In this case, the mutual accelerations experienced by the stars participating in the encounters turn out to be comparable in order of magnitude to the accelerations of these stars caused by the regular force acting on them. Change in the relative velocity vector $V$ of two bodies with one approach is equal in absolute value (Kulikovsky, 1985)
\begin{equation}\label{deltav}
\Delta V(b,V)=\frac{2V}{\sqrt{1+(b/B)^2}},
\end{equation}
and the rotation angle of the relative velocity vector $\psi$ is determined by formula (Kulikovsky, 1985)
\begin{equation}
{\rm tg}\frac{\psi}{2}=\frac{G(m_1+m_2)}{V^2 b},
\end{equation}
where $b$ is the impact distance equal to the distance from the disturbing body $m_1$ to the asymptote of the hyperbola described by the perturbed star $m_2$,
$B=\frac{G(m_1+m_2)}{V^2}$, $G$ is the gravitational constant. As a first approximation, we can assume that the modulus of the total change in velocity $|\Delta V|$ will be related to changes during individual encounters $|\Delta V_i|$ by the relation
\begin{equation}
(|\Delta V|)^2=\sum\limits_i (|\Delta V_i|)^2.
\end{equation}

Taking into account (\ref{deltav}) and summing up changes in $|\Delta V_i|$ during the time $\Delta t$ occurring during approaches with inpact distances
in the interval $(b,b+db)$ and velocities $V,V+dV$, we obtain 
\begin{equation}
\sum\limits_i\Delta V_i^2(b,V)=2\pi bf(V)\,dV\frac{4V^2}{1+(b/B)^2}V db
\,\Delta t,
\end{equation}
where $f(V)dV$ is the spatial density of stars having a speed of within $V$ and $V+dV$, and $2\pi b\,db f(V)VdV\Delta t$ is the number of stellar
encounters in a "tube" of radius $b$, wall thickness $db$ and length $V \Delta t$.

If we limit $b$ by some upper value $b_{max}$ characteristic of the problem under consideration, then, integrating the last equation over $b$, we have
\begin{equation}\label{sumdelta}
\begin{array}{rl}
\sum\limits_i[\Delta V_i(V)]^2= & \displaystyle 8\pi f(V)VdV\Delta t\,V^2
\int\limits_0^{b_{max}}\frac{b\,db}{1+\frac{b^2}{B^2}}=\\&\\
  &\displaystyle 4\pi G^2(m_1+m_2)^2\frac{\Delta t}V f(V)
\ln(1+\displaystyle\frac{b^2_{max}}{B^2})\,dV.
\end{array}
\end{equation}

It is useful to emphasize that in the last formula the result is only depends logarithmically on $b_{max}$. Therefore, in any case, it is possible
without much error, take $b_{max}$ equal to the average distance between stars in a cluster. Next we perform integration (\ref{sumdelta})
on $V$, neglecting the weak dependence of the function $B$ on $V$ under the logarithm, which allows us to consider this logarithm as a constant.
Let us further assume that $f(V)$ has a Maxwellian velocity distribution (Kholopov, 1981)

\begin{equation}\label{maxwell}
f(V)\,dV=\frac{4N}{\sqrt{\pi}\sigma^3}V^2
\exp{\left(-\frac{V^2}{\sigma^2}\right)}\,dV.
\end{equation}
Constant factor before $V^2 \exp{\left(-\frac{V^2}{\sigma^2} \right)}$ is chosen such that when integrated over all velocities
get as a result $N$ -- the number of stars per unit volume, $\sigma$ -- most likely speed. After integration we have
\begin{equation}\label{delta1} (\Delta V)^2=8\sqrt\pi G^2(m_1+m_2)^2 \frac N{\sigma}
\ln{\left[1+\frac{b^2_{max}}{B^2}\right]}\,\Delta t. \end{equation}

Next, we can introduce some simplifying assumptions. Let's assume that $b_{max}\gg B$. This assumption holds for all stellar clusters. For example, with $m_1+m_2=2m_{\odot}$, $V=10$ km/s, $B\approx 20$ a.e. The concentration of $N$ stars varies greatly depending on from the cluster. Characteristic values of $N$ are approximately equal to 
\begin{center}
\begin{tabular}{ll}
$N\approx 0.138\;{\rm pc}^{-3}$ & in the vicinity of the Sun \\
$N\approx 1\div 80\;{\rm pc}^{-3}$ & in open clusters\\
$N\approx 10^2\div 10^4\;{\rm pc}^{-3}$ & in globular clusters. \\
\end{tabular}
\end{center}

\noindent Let us take the most favorable case for us of a spherical clusters. The density of $10^3$ stars/pc$^3$ corresponds to the average distance between stars of the order of 0.1 pc. In this case, the ratio $\frac{b_{max}}{B}$ will be of the order of $10^3$. We will assume that interactions occur most often with the low-mass and, therefore, highly concentrated population of the cluster, i.e. let us assume that $m_2\ll m_1$. Based on the above assumptions, we can
write the equation (\ref{delta1}) in the following form
\begin{equation}\label{delta2}
(\Delta V)^2=16\sqrt\pi G^2 m_1^2 N\frac 1{\sigma}
\ln{\frac{b_{max}}{B}}\,\Delta t.
\end{equation}

The equation (\ref{delta2}) allows  us to estimate through what interval time, the star (or pulsar) will receive a given speed increase.
Let's substitute specific numeric values into (\ref{delta2}):
$m_1=2m_\odot=4\cdot 10^{30}$ kg, $N=10^3$ pc$^{-3}$, $\sigma=5$ km/s,
$\frac{b_{max}}{B}=10^3$. We get
\begin{equation}
(\Delta V)^2\approx 10^{-7}\Delta t.
\end{equation}
For example, in $10^8$ seconds the pulsar will receive an increase in speed $\sim\sqrt{10}\approx 3$ m/s. This value is quite sufficient for
detection by timing method, because will lead to relative a change in the pulsar period of the order of $10^{-8}$ or a change
$\dot\nu\sim 10^{-16}$ for pulsars with a period of 1 second. and $\dot\nu\sim 10^{-19}$ for millisecond pulsars.

Currently, the pulsar catalog (Taylor et al, 1993) contains 33 pulsars included in globular clusters, of which 5 have a negative
derivative of the period, which is interpreted as accelerated motion pulsar. Because in general, the derivative of the period should be
positive, this example shows that the influence of gravitational forces in globular cluster plays a significant role.

Thus, based on all of the above, we can conclude that {\it a pulsar located in a globular star cluster is affected by irregular forces, which in their magnitude are such that they can be detected in a relatively short ($\sim$ several years) period of time .}

\subsection{Remote companions of pulsars}

The presence of high-order frequency derivatives ($\ge 3$) in some pulsars can be explained (and indeed is explained) by the presence of companions revolving around the common center of mass of the system in question at elliptical orbits. A typical example is the pulsars B1620-26 and B1257+12. For the first, derivative frequencies up to and including the fourth are reliably measured, and for the second, up to $\nu^{(3)}$, inclusive (Joshi, Rasio, 1997; Rasio, 1994; Thorsett {\it et al.}, 1993). If a distant companion of a pulsar orbits in a long-period orbit with a period that greatly exceeds the available observational data, then it is almost impossible to completely restore the orbital parameters and it is only permissible to impose some (sometimes very significant) restrictions on the orbital parameters.

\subsection{Asteroid noise}

Currently, several tens of thousands of asteroids are known (Gil-Hutton, 1997). Most of them are located between the orbits of Mars and Jupiter at an average distance from the Sun of 2.8 AU. There are several dozen relatively large asteroids moving in elongated orbits that can intersect with the Earth`s orbit and, thus,
have a weak disturbing effect on the movement of the Earth. We can assume that all large asteroids (more than 50 km in diameter) have already been discovered. The latest DE405/LE405 planetary ephemerides released by the Jet Propulsion Laboratory (JPL) take into account the perturbing influence of the three large asteroids Ceres, Pallas and Vesta on the motion of the major planets (Standish et al, 1995) by explicitly describing their Keplerian orbits and refining their masses. The influence of the other 297 most influential asteroids is taken into account by dividing them into three groups according to average density and then calculating their total influence on Mars, Earth and the Moon during the day. This integral daily disturbance is then taken into account when integrating the equations of planetary motion.

The degree of influence of large asteroids directly depends on the accuracy of determining their masses, which is made from the mutual disturbances of asteroids on each other (Schubart, 1971, 1975) or from disturbances in the orbit of Mars (Williams, 1984). Since the masses of even the largest asteroids are 10 orders of magnitude less than the mass of the Sun, the accuracy of determining the masses is low and amounts to 4 - 20~\%. Periodic disturbances in the orbit of Mars have an amplitude of 0.8 and 0.2 km from Ceres and Pallas with a period of 10 years and 5 km from Vesta with a period of 52 years. About 40 more asteroids disturb Mars by about 5 m. It is clear that the disturbances in the Earth`s orbit will be less than in the case of Mars, but nevertheless they will not be negligible. These disturbances can be estimated from Euler's equations for osculating elements. It is logical to choose the semimajor axis of the orbit $a$ as the osculating element
\begin{equation}
\frac{da}{dt}=2a^2(e\sin v\, S +pr^{-1} T),
\end{equation}
where $S,\,T$ are, respectively, the radial and perpendicular to the orbital plane components of the perturbing acceleration, each of which is proportional to $\propto {m_a}/{r_{12}^2}$. Here $m_a$ is the mass of the asteroid, $r_{12}$ is the distance from the asteroid to the perturbed planet, which for estimation calculations can be assumed to be proportional to the difference between the semimajor axes of the asteroid and the planet. When integrating over a short period of time, the proportionality of the corresponding elements will remain, i.e.
\begin{equation}\label{411}
\frac{\delta a_z}{\delta a_m}=\left(\frac{a_z}{a_m}\right)^2
\frac{e_z}{e_m}\left(\frac{a_m-a_a}{a_z-a_a}\right)^2
\end{equation}
- for the $S$-component of the disturbing acceleration and
\begin{equation}\label{412}
\frac{\delta a_z}{\delta a_m}=\left(\frac{a_z}{a_m}\right)^2
\left(\frac{a_m-a_a}{a_z-a_a}\right)^2
\end{equation}
-- for the $T$-component of the disturbing acceleration. A more accurate analysis involves taking into account the ratio of orbital frequencies and various harmonics of the elliptical motion, which we neglect. Substituting the orbital elements of the Earth and Mars into formulas (\ref{411}) and (\ref{412}), we obtain $\delta a_z \approx 0.04\,\delta a_m$, $\delta a_z \approx 0.2\,\delta a_m$ for the radial and perpendicular components, respectively. Thus, the Earth disturbances from Ceres, Pallas and Vesta will be $170\pm 7$ m, $40\pm 6$ m and $1.0\pm 0.2$ km, respectively. The uncertainty in the latter value could, in principle, be detected if not for its long period.

It is possible to estimate the uncertainty in the position of the barycenter of the Solar System, caused by the inaccuracy in determining the masses of the three largest asteroids. The total uncertainty is about 30 meters, which is in units time is about 100 ns. This effect is still difficult to detect
through timing, but in the near future it may turn out essential for unambiguous interpretation of observations.

The above estimates show that disturbances caused even largest asteroids, are quite small and relatively long-period, to be confidently detected by the timing method in the present time. On the other hand, there are still a large number of small asteroids, which, during short approaches to the Earth, also cause deviations in its movement, which, in turn, leads to additional TOA noise level. Calculation of the spectrum of this noise and its observation analysis will allow us to move forward towards a more accurate determination of mass of invisible matter in the solar system.

\section{The influence of different types of orbits on the residual deviations of the TOA pulsars}

In this section, we will consider the influence of the pulsar`s encounters with other bodies and their manifestations in the structure of residual deviations of the pulsar. Approaches can occur in three types of orbits: hyperbolic, parabolic and elliptical. Flights along the first two types of orbits are the most typical, for example, in globular star clusters. Elliptical orbits can be used to analyze the disturbances of distant, invisible pulsar companions. Since motion along any of the three types of orbits cannot be expressed in terms of elementary functions of time, series expansions are used around the moment of minimal approach of the flying body to the pulsar, i.e. moment of passing through the periapsis. Thus, the projection of the pulsar shift onto the line of sight is expressed through a polynomial series, which greatly simplifies further calculations. A similar structure of the series is used further to obtain theoretical expressions for the power spectrum, calculated on the basis of the residual deviations of the TOA. Power spectra are a very important source of information about processes occurring inside or in the vicinity of a pulsar, as well as in the interstellar medium. Power spectra have already been obtained for a fairly large number of pulsars, and their correct interpretation is an urgent physical problem of modern pulsar astrometry.

\subsection{Hyperbolic orbits}

Arrival time $T$ and pulse number $N(T)$, which is also called phase of the pulsar, are related by the formula
\begin{equation}\label{phase}
N(T)=N_0+\nu T+\dot\nu T^2/2+\ddot\nu T^3/6+\ldots,
\end{equation}
where $N_0$ is an arbitrary constant, $\nu$ is the rotational frequency of the pulsar, $\dot\nu$, $\ddot\nu$ are the first and second derivatives of the rotational frequency of the pulsar, $T$ is the proper time, measured by a hypothetical clock on pulsar, the ellipsis indicates the possible presence in the rotational frequency of derivatives of higher orders than the second. In most cases, higher order derivatives are not required. It should be noted that some pulsars experience sudden changes in frequency, called glitches, as well as random changes in frequency that are not described by a time polynomial. Time $T$ is not directly measurable. For this reason, it is necessary to find a relationship that connects the pulse number $N$ and the time of arrival of this pulse to the observer. We further assume that the observer is located at the barycenter of the Solar system, as a result of which corrections associated with the movement of the observer are not considered further.

Metric in a coordinate system with the origin at the center of mass of system "the pulsar  -- flying body" (hereinafter simply "system") is described by the formula
\begin{equation}\label{metric}
ds^2=-[1+2\Phi+O({\rm v}^4)]dt^2+O({\rm v}^3)dx^i
dt+[1-2\Phi+O({\rm v}^4)](dx^2+dy^2+dz^2).
\end{equation}

Here is the Newtonian potential
\begin{equation}
\Phi({\bf r},t)=-\frac{M_1}{|{\bf r}-{\bf r}_1(t)|}-
\frac{M_2}{|{\bf r}-{\bf r}_2(t)|},
\end{equation}
where indices 1 and 2 refer to the pulsar and the body flying past it respectively. The units used are $c=G=1$ ($c$ is the speed of light,
$G$ is the gravitational constant), so the mass of the Sun is $M_{\odot}=1.477\; {\rm km}=4.925\cdot 10^{-6}\; {\rm c}$. In the formula (\ref{metric})
the symbols $O({\rm v}^3)$, $O({\rm v}^4)$ denote the degrees of expansion by $\frac{\rm v}c$ of a higher order of smallness than $\frac{\rm v^2}{c^2}$.

Let us choose a polar coordinate system $(r,\,\theta,\,v)$ associated with rectangular system $(x,\,y,\,z)$ with the usual formulas and with the beginning
coordinates at the barycenter of the binary system. Let's orient it so that the equatorial plane of the polar coordinate system coincided with
plane of the orbit, and the $x$ axis was directed to the pericenter of the orbit. Then coordinates of the pulsar and body

\begin{equation}
{\bf r}_1=(r_1,\pi/2,v),\qquad {\bf r}_2=(r_2,\pi/2,v+\pi),
\end{equation}
where $v$ is the true anomaly, $r_1$, $r_2$ are described to within first approximation of a hyperbola with focus at the center of mass of the system
\begin{equation}\label{5}
r_1=\frac{M_2}{M_1+M_2}r,\quad r_2=\frac{M_1}{M_1+M_2}r,
\quad r=\frac{a(e^2-1)}{1+e\cos v}
\end{equation}
here $a$ is the semimajor axis of the relative orbit, $e$ is the eccentricity of orbits.

The proper time $T$ is related to the coordinate time $t$ by the expression obtained from the metric (\ref{metric}):
\begin{equation}
(dT)^2=-ds^2=dt^2[1+2\Phi-{\rm v^2}+O({\rm v^4})],
\end{equation}
or in the form of a differential time equation:
\begin{equation}
\frac{dT}{dt}=1+\Phi({\bf r}_1)-\frac 12 {\rm v_1^2}+O({\rm v^4})
\end{equation}

The potential $\Phi({\bf r}_1)$ can be expanded into a series and represented in the following way

\begin{equation}
\Phi({\bf r}_1)=\frac{-M_1}{|{\bf r}_1-{\bf r}_1|}+
\frac{-M_2}{|{\bf r}_1-{\bf r}_2|}={\rm const}-\frac{M_2}r.
\end{equation}
The constant part of the potential, designated const, in further calculations we will omit it, since it leads only to the linear progression of time scales $T$ and
$t$ relative to each other. This move can be removed by overriding units of time. Next we have
\begin{equation}
v^2_1= \frac{M_2^2}{M_1+M_2}\left(\frac 2r+\frac 1a \right),
\end{equation}
from where we get, omitting the constant terms
\begin{equation} \label{dTdt}
\frac{dT}{dt}=1-\frac{M_2}r-\frac{M_2^2}{M_1+M_2}\frac 1r
\end{equation}
In order to integrate the equation (\ref{dTdt}), it is convenient to introduce new variable $H$ - an analogue of the eccentric anomaly in the case of
elliptical movement. The variable $H$ is related to coordinate time $t$ by equation
\begin{equation}\label{kepler}
t-t_{\Pi}=\frac{a^{3/2}}{\sqrt{M}}(e\sh H-H),
\end{equation}
where $t_{\Pi}$ is the moment of passage of the periapsis or the moment of maximum approaching bodies, $M=M_1+M_2$.

Let's differentiate the equation (\ref{kepler}) using the relations
\begin{equation}\label{difkep}
r=a(e\ch H-1),\quad
dt=\frac{a^{3/2}}{\sqrt{M}}(e\ch H-1)dH=\sqrt{\frac{a}{M}}r\,dH.
\end{equation}

Let's integrate the equation (\ref{dTdt})
\begin{equation} \label{intt}
\begin{array}{c}
\displaystyle
\int dT={\int}\left(1-\frac{M_2}r-\frac{M_2^2}{M_1+M_2}\frac 1r\right)\,dt=
\\ \\
\displaystyle
{\int} \left[1-\left(M_2+\frac{M_2^2}{M_1+M_2}\right)\frac 1r\right]dt=
\\\\
\displaystyle
{\int} \, dt-\int\sqrt{\frac{a}{M}}\left(M_2+\frac{M_2^2}{M_1+M_2}\right)
\,dH= \\\\
\displaystyle
t_{em}-\sqrt{\frac{a}{M}}\left(M_2+\frac{M_2^2}{M_1+M_2}\right)H.
\end{array}
\end{equation}
Let us substitute into the equation (\ref{intt}) the expression following from (\ref{difkep})
\begin{equation}
\sqrt{\frac{a}{M}}H=\sqrt{\frac{a}{M}}e \sh H-\frac ta,
\end{equation}
omitting constant terms and multiplicative constants
\begin{equation}
t_{em}-\sqrt{\frac{a}{M}}\frac{M_1 M_2+2M_2^2}{M_1+M_2}e\sh H = T
\end{equation}
or
\begin{equation}\label{gamma}
T=t_{em}-\gamma\sh H,
\end{equation}
where is a constant
\begin{equation}
\gamma=\sqrt{\frac{a}{M}}\frac{M_1 M_2+2M_2^2}{M_1+M_2}e
\end{equation}

Next, we consider the propagation of the pulse from the pulsar to the barycenter Solar system. The arrival time of the $N$th pulse is determined by the formula
\begin{equation}\label{propag}
t=t_{em}-{\bf n\cdot r}_1(t_{em})+2M_2\ln\left[\frac{2r_b}{r_1(t_{em})+
{\bf n\cdot r}_1(t_{em})}\right].
\end{equation}
Here ${\bf n}$ is the unit vector directed from the barycenter of the binary system to the barycenter of the Solar system, ${\bf r}_1$ is the radius vector of the pulsar, $r_b$ is the coordinate distance from the pulsar to the observer slowly varying with time, which will not be included in the final formula in the future, $t_{em}$ is the emission time, which is determined by the formula (\ref{gamma}). The term ${\bf n\cdot r}_1$ represents the integral Doppler effect and can be expressed in terms of orbital elements
\begin{equation}\label{remer}
{\bf n\cdot r}_1(t_{em})=-r_1\sin i\sin(v+\omega),
\end{equation}
where $i$ is the inclination of the orbital plane to the sky plane, $\omega$ is longitude of periapsis. The last term in the equation (\ref{propag}), containing
logarithm, can be converted as follows \begin{equation}\label{shap}
\ln\frac{2r_b}{r_1+{\bf n\cdot r}_1}=-\ln r_1(1-\sin
i\sin(\omega+v))= \ln\frac{1+e\cos v}{1-\sin i\sin(\omega+v)}.
\end{equation}

In the equation (\ref{shap}) only the variable part is saved, and the constant the part $\ln r_1$ and the term $\ln 2r_b$ are omitted, since the quantity $r_b$
changes slowly enough that it leads to simple redefinition of constant $N_0$ and coefficients of the time polynomial in the equation (\ref{phase}).

Taking into account the above transformations, the equation (\ref{propag}) converted to the form
\begin{equation}
t_{em}=t-r_1\sin i\sin(\omega+v)-2M_2\ln\frac{1+e\cos v}{1-\sin
i\sin(\omega+v)}.
\end{equation}
Proper time $T$ after substituting (\ref{gamma}) into (\ref{propag}) is found by the formula
\begin{equation}\label{prop1}
T=t-r_1\sin i\sin(\omega+v)-2M_2\ln\frac{1+e\cos v}{1-\sin i\sin(\omega+v)}-
\gamma\sh H
\end{equation}

Next, we express all functions containing $v$ in terms of functions $H$ (Subbotin, 1968):
\begin{equation} \label{vh}
\cos v=\frac{e-\ch H}{e\ch H-1},\qquad
\sin v=\frac{\sqrt{e^2-1}\sh H}{e\ch H-1},
\end{equation}

\begin{equation} \label{omv}
\sin(\omega+v)=\sin v\cos\omega+\cos v\sin\omega=
\frac{\sqrt{e^2-1}\sh H}{e\ch H-1}\cos\omega+\frac{e-\ch H}{e\ch
H-1}\sin\omega,
\end{equation}

\begin{equation}\label{rsiso}
\begin{array}{cl}
r_1\sin i\sin(\omega+v)&=a_1\sin i\sqrt{e^2-1}\sh H\cos\omega+
a_1\sin i(e-\ch H)\sin\omega \\
& =\alpha(e-\ch H)+\beta\sh H,
\end{array}
\end{equation}
where the constants are $\alpha=a_1\sin i\sin\omega$,
$\beta=a_1\sin i\sqrt{e^2-1}\cos\omega$,
$a_1=\frac{M_2}{M_1+M_2}a$. Besides this, we have
\begin{equation} \label{frc}
\frac{1+e\cos v}{1-\sin i\sin(\omega+v)}=
\frac{e^2-1}{e\ch H-1-[\sqrt{e^2-1}\sh H\cos\omega+(e-\ch H)\sin\omega]\sin i} .
\end{equation}
Taking into account the equations (\ref{vh}) -- (\ref{frc}) formula (\ref{prop1}) is converted to the form
\begin{equation} \label{prop2}
\begin{array}{cl}
T=&t-\alpha(e-\ch H)-(\beta+\gamma)\sh H+\\
&2M_2\ln[e\ch H-1-\sin i(\sqrt{e^2-1}\sh H\cos\omega+(e-\ch H)\sin\omega)].
\end{array}
\end{equation}

In the formula (\ref{prop2}) the value $H=H(t_{em})$. Since we are dealing with coordinate time $t$, and not with the radiation time $t_{em}$, which
depends on the position of the pulsar in orbit, then it is necessary to express all quantities through time $t$. To do this, we use the expansion of the first
order of the functions $\ch H(t_{em})$ and $\sh H(t_{em})$ around the moment $t$:
\begin{equation}
\ch H(t_{em})=\ch H(t)+\sh H(t)(H(t_{em})-H(t)).
\end{equation}
From the equation (\ref{difkep}) and (\ref{propag}) it follows that, with accuracy sufficient for observations
\begin{equation}\label{442}
H(t_{em})-H(t)=\frac{M_2}{\sqrt{Ma}}\frac 1r(t_{em}-t)
=-\frac{M_2}{\sqrt{Ma}}\sin i\sin(\omega+v).
\end{equation}

Thus, we obtain from (\ref{442})
\begin{equation}
\ch H_{em}=\ch H-\frac{M_2}{\sqrt{Ma}}\sin i\sin(\omega+v)\sh H.
\end{equation}
Likewise
\begin{equation}
\sh H_{em}=\sh H-\frac{M_2}{\sqrt{Ma}}\sin i\sin(\omega+v)\ch H.
\end{equation}
Taking into account the expressions for $\ch H_{em}$ and $\sh H_{em}$, the equation (\ref{prop2}) converted to the form
\begin{equation}\label{prop33}
\begin{array}{c}
T=t-\alpha(e-\ch
H)-(\beta+\gamma)\sh H+\\ 2M_2\ln[e\ch H-1-\sin i(\sqrt{e^2-1}\sh
H\cos\omega+(e-\ch H)\sin\omega)]-\\
\displaystyle(\alpha\sh H-(\beta+\gamma)\ch H)\frac{M_2}{\sqrt{Ma}}\sin
i\sin(\omega+v)= \\\\
t-\alpha(e-\ch H)-(\beta+\gamma)\sh H+\\
2M_2\ln[e\ch H-1-\sin i(\sqrt{e^2-1}\sh H\cos\omega+(e-\ch H)\sin\omega)]-\\
\displaystyle\frac{M_2}{\sqrt{Ma}}\sin i(\alpha\sh H-(\beta+\gamma)\ch H)\cdot
\left[\frac{\sqrt{e^2-1}\sh H}{e\ch H-1}\cos\omega+\frac{e-\ch H}{e\ch
H-1}\sin\omega \right].
\end{array}
\end{equation}

Finally we have
\begin{equation}\label{prop3}
\begin{array}{c}
T=t-\alpha(e-\ch H)-(\beta+\gamma)\sh H+\\
2M_2\ln[e\ch H-1-\sin i(\sqrt{e^2-1}\sh H\cos\omega+(e-\ch H)\sin\omega)]-\\
\displaystyle \frac{M_2}{\sqrt{Ma^3}}(\alpha\sh H-(\beta+\gamma)\ch H)
\frac{\alpha(e-\ch H)+\beta\sh H}{e\ch H-1}.
\end{array}
\end{equation}

The equation (\ref{prop3}) is the one we are looking for. It is after substitution in equation (\ref{phase}) solves the problem of finding the dependence $N=N(t)$.

\subsection{Parabolic orbits}

Formula that relates true anomaly $v$ and time $t$ (so called Barker equation (Gurzadyan, 1992))
\begin{equation}
\frac{1}{\sqrt{2q^3}}(t-T_0)=\tg\frac v2+\frac 13\tg^3\frac v2,
\end{equation}
where $q$ is the distance to the periapsis, $T_0$ is the moment of passage of the periapsis. Further calculations are similar to those already done for the hyperbolic
orbits. Here are several intermediate formulas used to derive final formula.
\begin{equation}
dv=\frac{\sqrt{2qM}}{r^2}dt,
\end{equation}
\begin{equation}
T=t-\gamma \tg \frac v2
\end{equation}
\begin{equation}
\gamma=M_2\left(1+\frac{M_2}{M}\right)\sqrt{\frac{2q}{M}},
\end{equation}
\begin{equation}
v(t_{em})-v(t)=-\frac{M_2}{\sqrt{2qM}}\sin i\sin(\omega+v)(1+\cos v).
\end{equation}
Final formula
\begin{equation}
\begin{array}{cl}
\displaystyle
T=&t-\alpha(1-\tg^2\frac v2)-(\beta+\gamma)\tg\frac v2+\\\\
&\displaystyle 2M_2\ln\left[1-\frac 12(1-\tg^2\frac v2)(1+\sin i\sin\omega)-
\sin i\cos\omega\tg\frac v2\right]-\\\\ \displaystyle
&\frac{M_2\sin i\sin(\omega+v)}{\sqrt{2qM}}(2\alpha\tg\frac v2-
(\beta+\gamma)),
\end{array}
\end{equation}
\begin{equation}
\alpha=q\sin i\sin\omega,\quad \beta=q\sin i\cos\omega.
\end{equation}

\subsection{Elliptical orbits}

Elliptical orbits have already been considered in a number of well-known works. Among them are the works of (Blandford, Teukolsky, 1976), (Hougan, 1985), (Damour, Deruelle, 1986). It is therefore possible not to repeat the derivation of the formula that connects the coordinate and proper time on an elliptical orbit. However, to complete the picture, in this work we can write the corresponding formula without intermediate calculations.

\begin{equation}
\begin{array}{c}\displaystyle
T=t-\alpha(\cos E-e)-(\beta+\gamma)\sin E\\\\ \displaystyle
+2M_2\ln[1-e\cos E-\sin i(\sqrt{1-e^2}\sin E\cos\omega+(e-\cos E)
\sin\omega)]-\\\\ \displaystyle
\frac{M_2}{\sqrt{aM}}(\alpha\sin E-(\beta+\gamma)\cos E)
\frac{\alpha(\cos E-e)+\beta\sin E}{1-e\cos E},
\end{array}
\end{equation}
where $E$ is the eccentric anomaly, $\alpha=a_1\sin i\sin\omega$,
$\beta=a_1\sqrt{1-e^2}\sin i\cos\omega$

\subsection{Analysis of the formula for the relation between pulsar and barycentric time}

In this section, the value $\Delta=T-t$ will be actively used, the use of which allows us to present the equation (\ref{prop3}) in the form:
\begin{equation}\label{tmt}
\begin{array}{c}
\Delta=-\alpha(e-\ch H)-(\beta+\gamma)\sh H+\\ \\
2M_2\ln[e\ch H-1-\sin i(\sqrt{e^2-1}\sh H\cos\omega+(e-\ch H)\sin\omega)]-\\\\
\displaystyle\frac{M_2}{\sqrt{Ma^3}}(\alpha\sh H-(\beta+\gamma)\ch H)
\frac{\alpha(e-\ch H)+\beta\sh H}{e\ch H-1}.
\end{array}
\end{equation}

The first and second terms on the right side of the formula (\ref{tmt}) describe the geometric time delay in signal propagation from the pulsar to the barycenter of the pulsar-passing body pair and are an analogue of the Roemer correction; the third term containing the logarithm is the Shapiro correction arising due to the propagation of light in the gravitational field of a flying body, the fourth term has the meaning of a correction to the first and second terms, allowing us to consider the value $H$ as a function of the coordinate barycentric time $H(t)$ .

The residual deviations of the pulsar TOA are obtained by subtracting the theoretically calculated TOA values from those obtained experimentally. In the case when the observer, as previously assumed, is located at the barycenter of the Solar System, the theoretical dependence of the TOA on time $t$ is reduced to polynomial time. To describe the motion of a pulsar along a hyperbolic polynomial trajectory, time is not enough. For this reason, the residual deviations of the TOA can have a rather complex form.

\begin{figure}\centering
\includegraphics[width=15cm]{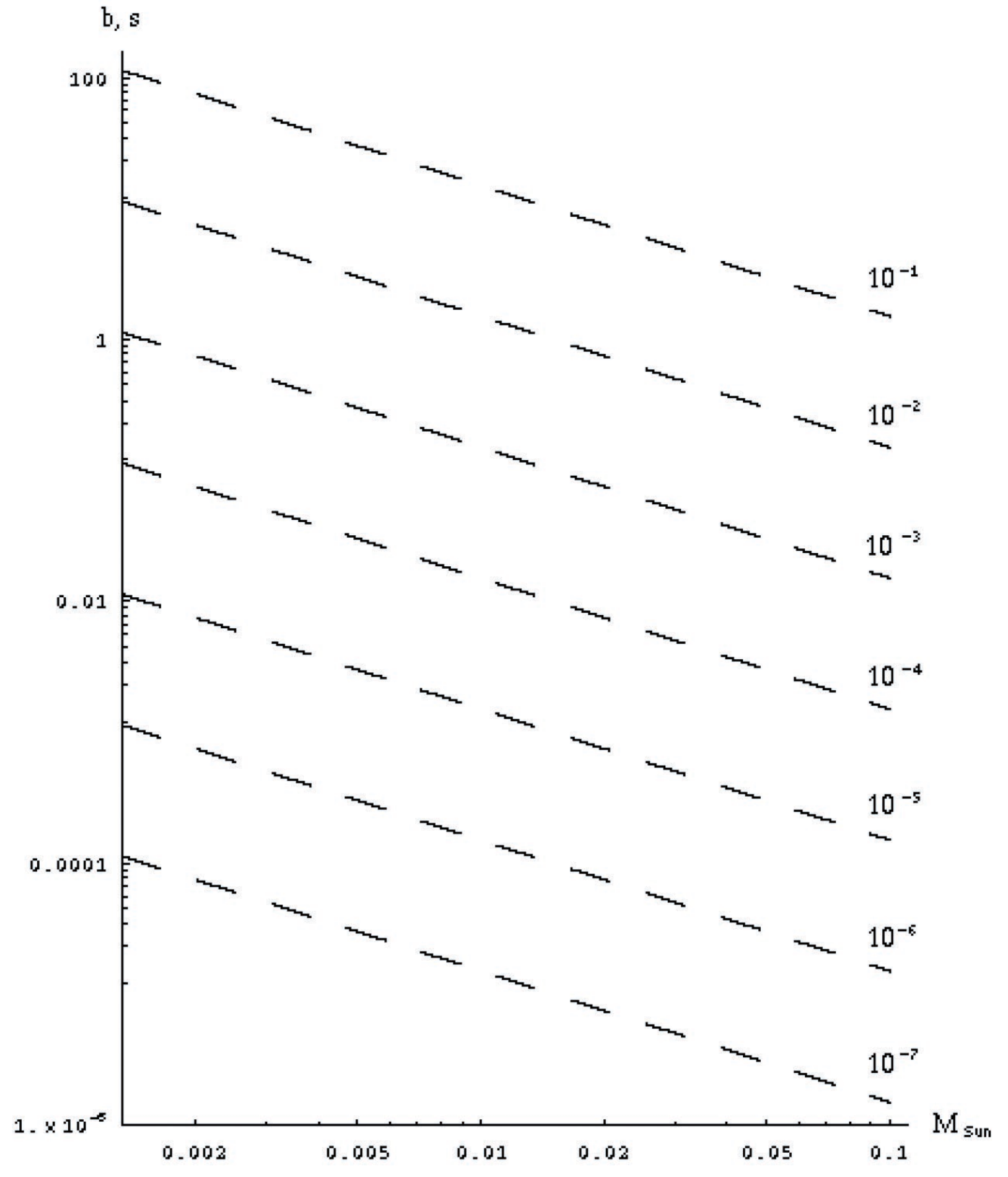}
\caption{Dependence of the amplitude of residual deviations of the pulsar TOA on the mass of the body flying past the pulsar $M$ and the impact distance $b$,
in which this body flies past the pulsar. The dotted lines show what combinations of $M$ and $b$ are needed to achieve a certain the magnitude of the residual deviations, expressed in seconds.}
\label{mb}
\end{figure}

The amplitude of residual deviations when a body passes near a pulsar depends on the mass of the given body and the impact distance. This amplitude 
can be easy estimated using formulas (\ref{5}) and (\ref{remer}):
\begin{equation}
{\bf n\cdot r}_1=-r_1\sin i\sin(\omega + v)=
-\frac{M_2}{M_1+M_2}\frac{b\sqrt{e^2-1}\sin i\sin(\omega + v)}{1+e\cos v},
\end{equation}
i.e., to the first approximation $\Delta \sim \frac{M_2}{M}\,b\,\sin i$, where $b$ is the impact distance, $M_2$ is the mass of the flying body. Let us remind you that
that the equality $b=a\sqrt{e^2-1}$ holds for a hyperbola. Dependence of amplitude on mass and impact distance is shown in Fig. \ref{mb}.

\begin{figure}\centering
\includegraphics[height=10cm]{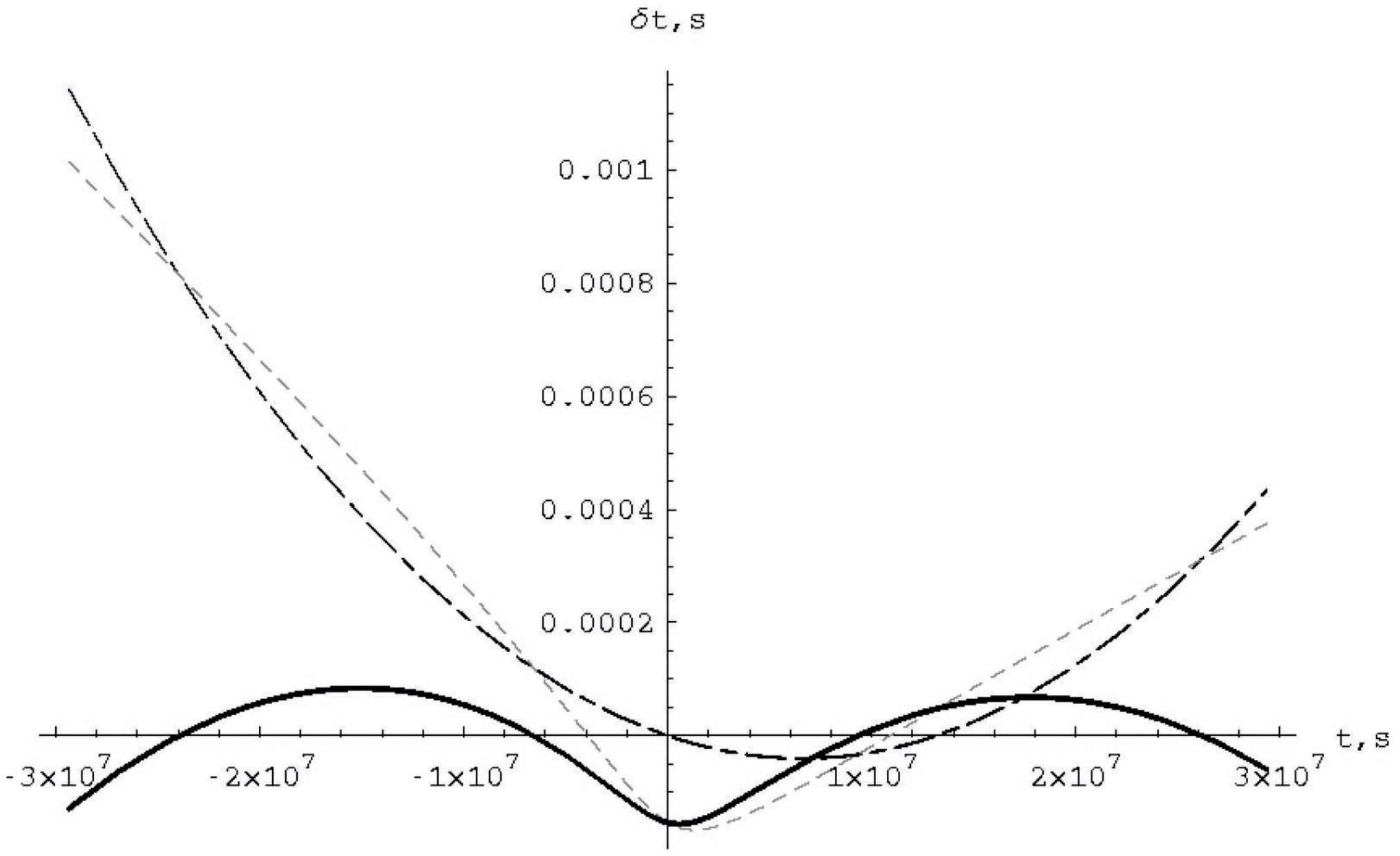}
\caption{Dependence of the shape of the residual deviation curve of the pulsar PIR on time at a certain orientation of the orbit relative to the observer.
The orbital parameters are fixed at $b=1000$ light. sec., $e=1.001$, $i=\pi/4$, masses of the pulsar $M_1=2M_{\odot}$ and the body $M_2=10^{-6}M_{\odot}$,
rotational frequency of the pulsar $\nu=1\;{\rm Hz}$ and derivative frequencies $\dot\nu=\ddot\nu=0$. The dotted line shows the relationship
(4.54). The dash-dotted line shows a quadratic polynomial time inscribed in curve (4.54) using the least squares method.
The residual deviations shown by the solid curve are the difference of the curve (4.54) and a quadratic polynomial. The abscissa axis represents time in
seconds, counted from the moment of closest approach of the pulsar and disturbing body. The ordinate axis shows the value of residual deviations in seconds.}
\label{res}
\end{figure}

\begin{figure}\centering
\includegraphics[width=15cm]{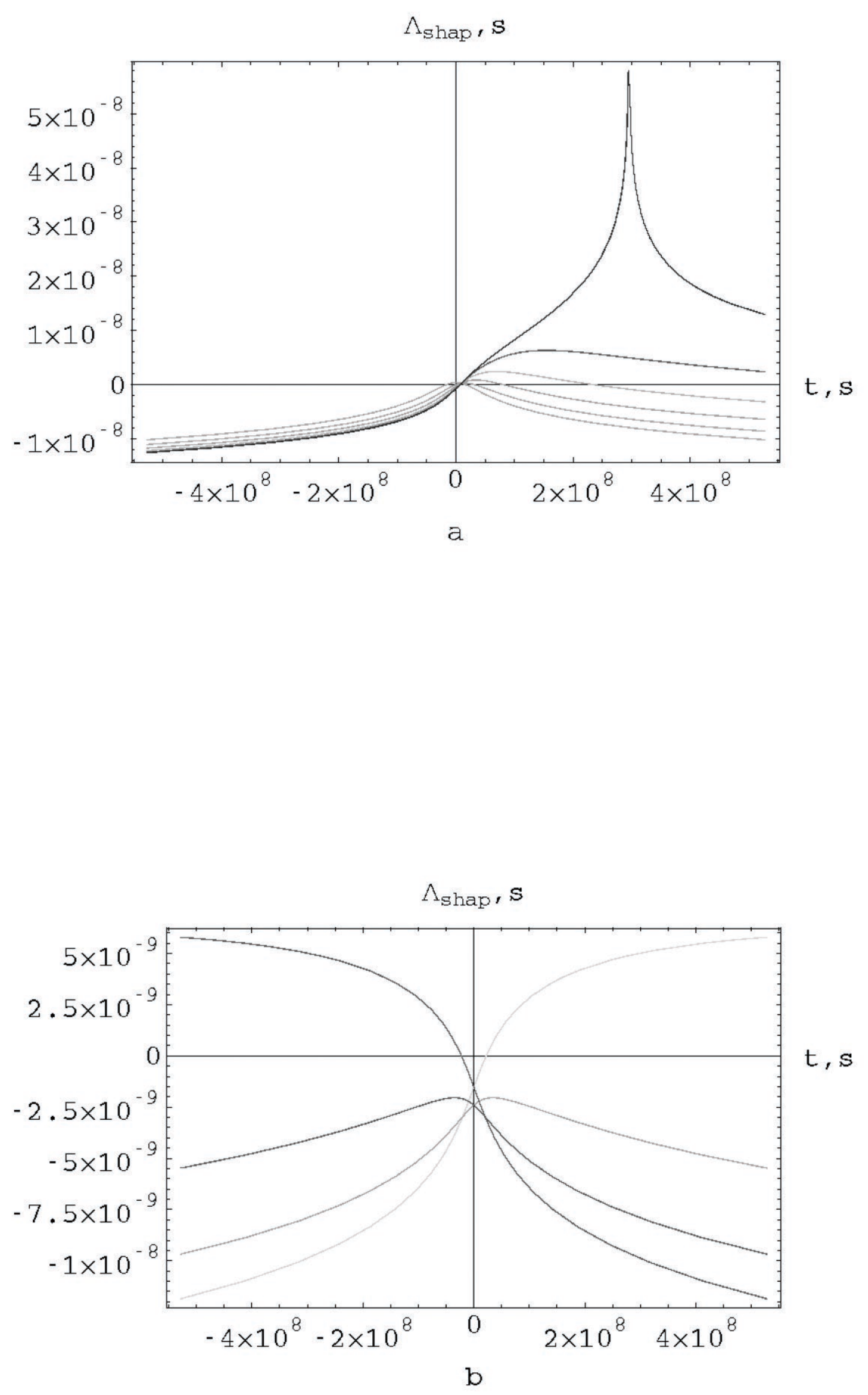}
\caption{Shapiro effect as a function of time at different inclination angles orbits $i$ and different periastron longitudes $\omega$. Parameters are fixed
$M_1=1.7M_{\odot}$, $M_2=0.00043M_{\odot}$, $b=10$ a.u., $\omega=-\pi/10$. a) The slope $i$ takes values $0^{\circ}$ (black curve),
$18^{\circ}$, $36^{\circ}$, $54^{\circ}$, $72^{\circ}$, $90^{\circ}$ (light gray curve). b) Shapiro effect as a function of time at different periastron longitudes
$\omega$ and inclination angle $i=89^{\circ}$. Periastron longitude takes values $-0.8\pi$ (dark gray curve), $-0.6\pi$, $-0.4\pi$, $-0.2\pi$ (light gray curve).}
\label{shpdel}
\end{figure}

The shape of the residual deviation curve depends primarily on the orientation of the orbit in space, as well as on the initial speed of movement of the passing body relative to the pulsar. Figure \ref{res} shows an example of what the shape of the residual deviations might be for a fixed inclination angle $i$ and periapsis longitude $\omega$. To construct this figure, it is enough to use only the first two terms of the equation (\ref{tmt}), because they are several orders of magnitude superior to the others. In Fig. \ref{res} the dashed line shows the dependence (\ref{tmt}), the dash-dotted line shows the quadratic time polynomial fitted in the curve (\ref{tmt}) using the least squares method and, finally, the solid line shows the residual deviations as the difference between the dashed line and the dash-dotted. For comparison, Figure \ref{shpdel} shows the contribution of the Shapiro effect at different orbital inclination angles $i$ and different periapsis longitudes $\omega$. The magnitude of this effect is determined primarily by the coefficient $M_2$ of the logarithm, a value quite small compared to the geometric shift $\sim \frac{M_2}{M}\,b\,\sin i$. The contribution of the Shapiro effect to the residual deviations of the TOA of pulsars has already been studied in (Sazhin, Safonova, 1993), (Hosokava, 1993), (Doroshenko, Larchenkova, 1994) therefore it is not studied further.

\begin{figure}
\centering
\includegraphics[height=16cm]{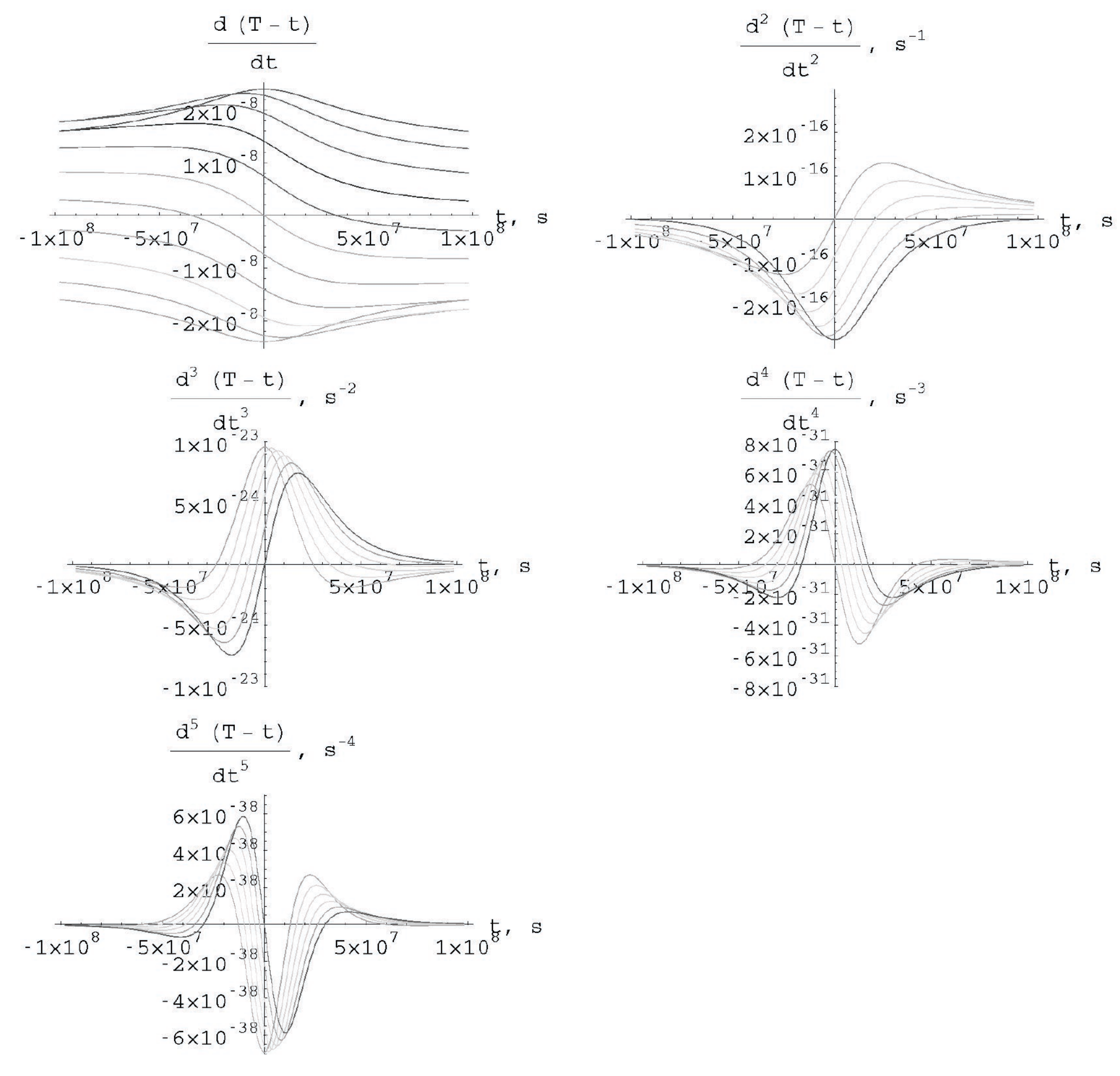}
\caption{Behavior of the Doppler factor and its derivatives in the case hyperbolic orbit. The following orbital parameters were used:
$M_{1}=1.7M_{\odot}$, $M_2=0.00043M_{\odot}$, $b=10$~a.u., $e=1.9$, $i=-\pi/2$, $\omega$ takes values from $-\pi$ (dark gray curve)
to $0$ (light gray curve) on the graph $\frac{d(T-t)}{dt}$ and from $-\pi/2$ (dark gray curve) to 0 (light gray curve) on the rest
graphs. On all graphs, time $t$ in seconds is counted from the moment closest approach between the pulsar and the disturbing body.}
\label{zz}
\end{figure}

Figure \ref{zz} shows the behavior of the derivative $z=\frac{d(T-t)}{dt}$ as functions of time at different angles $\omega$. The value of $z$ is determined by formula
\begin{equation}\label{dopf}
z=\frac{d(T-t)}{dt}=\sqrt{\frac{M}{a^3}}\frac{\alpha\sh H-(\beta+\gamma)\ch H}
{e\ch H - 1}.
\end{equation}
The quantity $(1+z)\nu$ gives the apparent frequency of the pulsar. From Figure \ref{zz} it can be seen that the observed value of the pulsar frequency is asymptotically tends to a constant value as $t\rightarrow\pm\infty$. At $\omega=0$ the pulsar frequency returns to the same value it was before the flight. This means that the direction of motion of the pulsar, although changed, but the radial velocity of the pulsar returned to its previous value. At $\omega=\pi/2$ the change in frequency as a result of the flight reaches maximum value. You can calculate this change using the formula 
\begin{equation}\label{deltaz}
\begin{array}{c}
\displaystyle
\Delta z=\lim\limits_{t\to\infty}z-\lim\limits_{t\to-\infty}z=
\sqrt{\frac{M}{a^3}}\frac{\alpha-(\beta+\gamma)}{e}-
\left(-\sqrt{\frac{M}{a^3}}\frac{\alpha+(\beta+\gamma)}{e}\right)=\\
\displaystyle =\sqrt{\frac{M}{a^3}}\frac{2\alpha}{e}=
\frac{2M_2\sin i\sin\omega}{e\sqrt{aM}}.
\end{array}
\end{equation}
You can also calculate the value
$\zeta=\frac{\lim\limits_{t\to\infty}z}{\lim\limits_{t\to -\infty}z}$:
\begin{equation}
\zeta=-\frac{\alpha-(\beta+\gamma)} {\alpha+(\beta+\gamma)}
\end{equation}
If we neglect the relativistic parameter $\gamma$, which does not play a significant role in these calculations, then the latter expression will be reduced to the formula
\begin{equation}\label{461}
\displaystyle
\left.\zeta\right|_{\gamma=0}=-\frac{\alpha-\beta}{\alpha+\beta}=
-\frac{\tg\omega-\sqrt{e^2-1}}{\tg\omega+\sqrt{e^2-1}}
\end{equation}

Unfortunately, in practice the value of $\zeta$ cannot be determined because to do this, it is necessary to know the true frequency of the pulsar, not distorted by the Doppler shift. Therefore, formula (\ref{461}) is of purely theoretical interest.

The second derivative of the quantity $\Delta=T-t$ has the form:
\begin{equation}
\frac{d^2(T-t)}{dt^2}=\frac{M}{a^3}\left[\frac{\alpha\ch H-
(\beta+\gamma)\sh H}
{(e\ch H - 1)^2}-\frac{e \sh H(\alpha\sh H-(\beta+\gamma)\ch H)}{(e\ch H - 1)^3}
\right]
\end{equation}

It is possible to give expressions for derivatives of the Doppler factor more high order, but due to their cumbersomeness we will not do this, but
we present only graphs of their behavior depending on time $t$ at different longitudes of the periapsis $\omega$.

If the body and the pulsar fly in a parabolic orbit, then the observed frequency of the pulsar will behave approximately the same as in the case of a hyperbolic orbit, with the only difference being that for any orientation of the orbit, the frequency after the flight will return to its previous value. Mathematically this is expressed as follows
\begin{equation}
\lim_{t\rightarrow\pm\infty}z=\lim_{t\rightarrow\pm\infty}
\frac{d(T-t)}{dt}=0
\end{equation}
Behavior of quantities $T-t$ and $z$ as functions of time at different angles $\omega$ is shown in Figure 4.5.
\begin{figure}\label{zp}
\centering
\includegraphics[height=15cm]{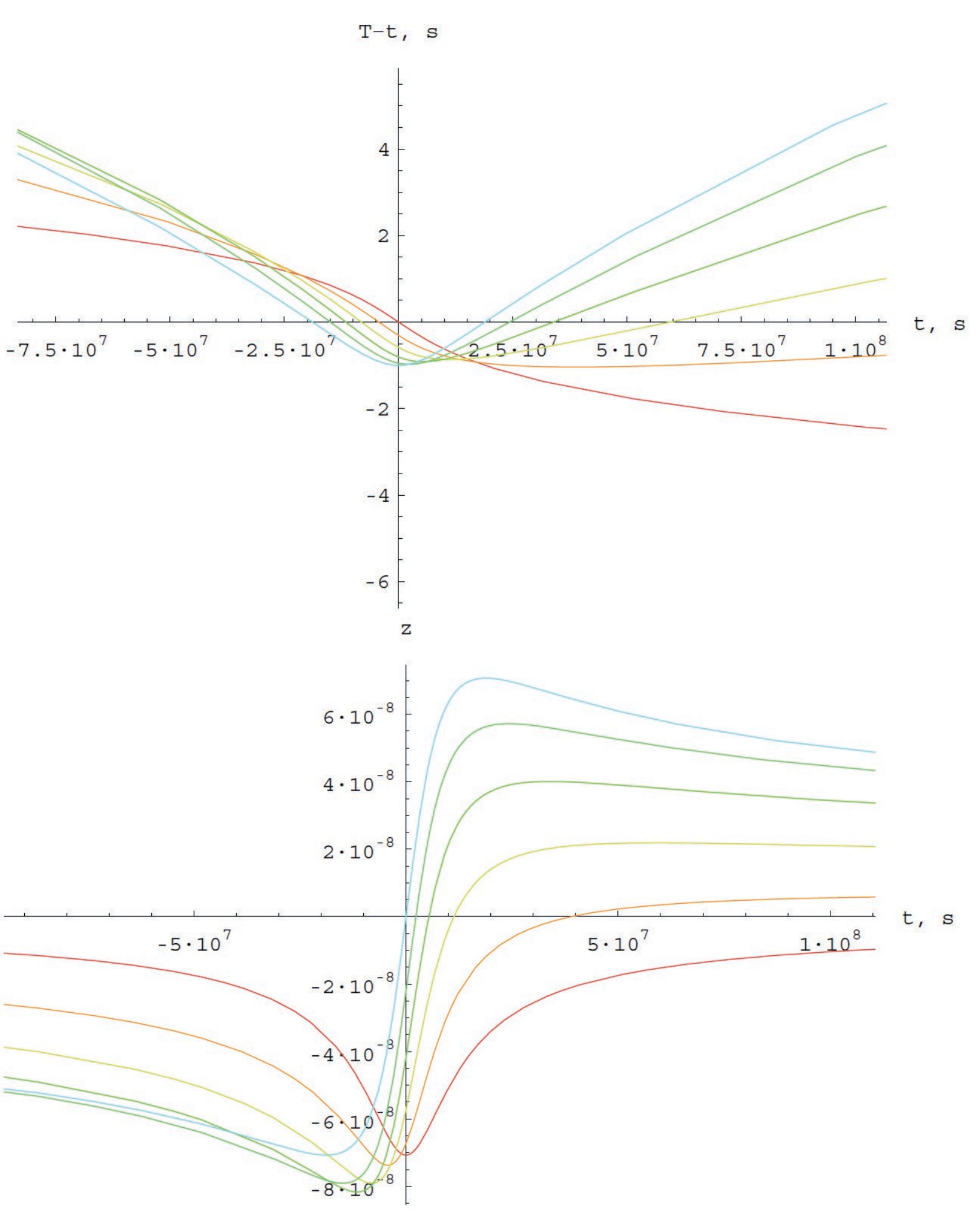}
\caption{Behavior of the value $T-t$ (top graph) and Doppler factor $z$ (bottom graph) in the case of a parabolic orbit. Orbit parameter
$q=2$~a.e., $M_1=2M_{\odot}$, $M_2=0.002M_{\odot}$. Periapsis longitude $\omega$ varies from 0 (dark gray curve) to $\pi/2$. Abscissa
the time in seconds from the moment of closest approach of the pulsar and disturbing body.} \end{figure}

Additional information can be used to numerically determine orbital parameters. As such, for example, we can use the positions of extrema of the value $z$ which are easily measured. To express these extrema in terms of orbital elements we expand the magnitude of the geometric shift of the pulsar into a series relative to the moment of passage of the pulsar through the pericenter:
\begin{equation}
\begin{array}{l}
\Delta_R(H)= -\alpha(e-\ch H)-\beta\sh H= \\\\
\displaystyle -\alpha(e-1)-\beta H+
\frac{\alpha}{2!}H^2-\frac{\beta}{3!}H^3+\frac{\alpha}{4!}H^4
-\frac{\beta}{5!}H^5+\frac{\alpha}{6!}H^6-\frac{\beta}{7!}H^7+O^8(H)
\end{array}
\end{equation}
The function $H=H(t)$ is expanded in powers of time as follows
\begin{equation}
H(t)=\frac{t}{\kappa(e-1)}-\frac{et^3}{3!\kappa(e-1)^4}+
\frac{e(1+9e)t^5}{5!\kappa^5(e-1)^7}-
\frac{(1+9e(6+25e))t^7}{7!\kappa^7(e-1)^{10}}+O^8(t)
\end{equation}
Substituting one series into another, we get
\begin{equation}
\begin{array}{c}\displaystyle
T-t=-\alpha(e-1)-\frac{\beta}{\kappa(e-1)}t+
\frac{\alpha}{2!\kappa^2(e-1)^2}t^2+\frac{\beta}{3!\kappa^3(e-1)^4}t^3-
\frac{\alpha(1+3e)}{4!\kappa^4(e-1)^5}t^4\\\\
\displaystyle
-\frac{\beta(1+9e)}{5!\kappa^5(e-1)^7}t^5+\frac{\alpha(1+24e+45e^2)}
{6!\kappa^6(e-1)^8}t^6+\frac{\beta(1+54e+225e^2)}{7!\kappa^7(e-1)^{10} }t^7
+O^8(t),
\end{array}
\end{equation}
where we limited ourselves to the 7th power of time.
Recall that in the last three formulas $\alpha=a_1\sin i\sin\omega$,
$\beta=a_1\sqrt{e^2-1}\sin i\cos\omega$,
$\kappa=\sqrt{\frac{a^3}{M_1+M_2}}$.

Let us now write down the expansion for the parabolic orbit:
\begin{equation}
\begin{array}{l}\displaystyle
\Delta_R(v)= -\alpha(1-\tg^2\frac v2)-\beta\tg\frac v2=\\\\
\displaystyle -\alpha-\frac{\beta}{2}v+\frac{\alpha}{4}v^2-
\frac{\beta}{24}v^3+\frac{\alpha}{24}v^4-
\frac{\beta}{240}v^5+\frac{17\alpha}{2880}v^6-\frac{17\beta}{40320}v^7+
O^8(v),
\end{array}
\end{equation}

\begin{equation}
v(t)=\frac{2}{\kappa}t-\frac{4}{3\kappa^3}t^3+\frac{26}{15\kappa^5}t^5-
\frac{172}{63\kappa^7}t^7+O^8(t),
\end{equation}

\begin{equation}
\begin{array}{l}\displaystyle
\Delta_R(t)=-\alpha-\frac{\beta}{\kappa}t+\frac{\alpha}{\kappa^2}t^2
-\frac{\beta}{3\kappa^3}t^3-\frac{2\alpha}{3\kappa^4}t^4
-\frac{\beta}{3\kappa^5}t^5+\frac{7\alpha}{9\kappa^6}t^6+
\frac{4\beta}{9\kappa^7}t^7+O^8(t).
\end{array}
\end{equation}
Here $\alpha=q_1\sin i\sin\omega$, $\beta=q_1\sin i\cos\omega$,
$\kappa=\sqrt{\frac{q^3}{M_1+M_2}}$, $q_1=\frac{M_2}{M_1+M_2}q$.

To complete the picture, we also present expansions for the elliptical orbit:
\begin{equation}
\begin{array}{l}
\Delta_R(E)=-\alpha(\cos E-e)-\beta\sin E=\\\\
-\alpha(1-e)-\beta E+\frac{\alpha}{2!}E^2+\frac{\beta}{3!}E^3-
\frac{\alpha}{4!}E^4-\frac{\beta}{5!}E^5+\frac{\alpha}{6!}E^6+
\frac{\beta}{7!}E^7+O^8(E),
\end{array}
\end{equation}

\begin{equation}
E(t)=-\frac{t}{\kappa(1-e)}+\frac{et^3}{3!\kappa^3(1-e)^4}+
\frac{e(1+9e)t^5}{5!\kappa^5(1-e)^7}+\frac{e(1+9e(6+25e))t^7}
{7!\kappa^7(1-e)^{10}}+O^8(t),
\end{equation}

\begin{equation}\begin{array}{c}\displaystyle
\Delta_R(t)=-\alpha(1-e)+\frac{\beta}{\kappa(1-e)}t+
\frac{\alpha}{2!\kappa^2(1-e)^2}t^2-\frac{\beta}{3!(1-e)^4\kappa^3}t^3-
\frac{\alpha(1+3e)}{4!\kappa^4(1-e)^5}t^4+\\\\ \displaystyle
\frac{\beta(1+9e)}{5!(1-e)^7\kappa^5}t^5+
\frac{\alpha(1+24e+45e^2)}{6!\kappa^6(1-e)^8}t^6-
\frac{\beta(1+54e+225)}{7!(1-e)^{10}\kappa^7}t^7+O^8(t).
\end{array}
\end{equation}
Here the parameters are $\alpha=a_1\sin i\sin\omega$,
$\beta=a_1\sqrt{1-e^2}\sin i\cos\omega$,
$\kappa=\sqrt{\frac{a^3}{M_1+M_2}}$.

To find the extrema of the value $z$, it is necessary to solve the equation $\frac{d^2\Delta_R}{dt^2}=0$. Since in the general case there are two extrema (this can be seen from Figure \ref{zz}), it is enough to leave the terms of the 4th power of time $t$ in the expansion of $\Delta_R(t)$, and the rest can be neglected. Then, as a result of differentiation, a quadratic equation will be obtained, which will give two {\it approximate} solutions. For a hyperbolic orbit, the required equation is
\begin{equation}
\frac{\alpha(1+3e)}{\kappa^2(e-1)^3}t^2-\frac{2\beta}{\kappa(e-1)^2}t-
2\alpha=0
\end{equation}
has a solution
\begin{equation}\label{474}
t_{1,2}=\frac{(e-1)(\beta\pm\sqrt{(6e^2-4e-2)\alpha^2+\beta^2})}
{\alpha(1+3e)}\kappa.
\end{equation}
For a parabolic orbit, an equation similar in meaning
\begin{equation}
\alpha+\frac{\beta}{\kappa}t-\frac{4\alpha}{\kappa^2}t^2=0
\end{equation}
has a solution
\begin{equation}\label{476}
t_{1,2}=\frac{\beta\pm\sqrt{16\alpha^2+\beta^2}}{8\alpha}\kappa.
\end{equation}
The moments $t_1$ and $t_2$ are easily determined from observational data, therefore equations (\ref{474}) and (\ref{476}) along with other equations
(which will be derived below) can be used to find numerical values of the orbital parameters of the pulsar.

Additional information for finding orbital parameters arises if we involve derivatives of the pulsar rotation frequency more than
high orders (for example, above the 3rd). This method was proposed in works (Thorsett et al, 1993, Rasio, 1994, Rasio et al. 1995) and applied
in works (Joshi \& Rasio, 1997; Rodin, 1999; Rodin, 1999b). Let's describe this method in more detail.

Consider the phase of the pulsar
\begin{equation}\label{477}
N(T)=\nu T + \frac 12\dot\nu T^2
\end{equation}
which we will count from the moment of passing the periapsis. This does not reduce the generality of the reasoning, but it significantly simplifies the calculations. Pulsar time $T$ is related to barycentric time $t$ by the following formula
\begin{equation}
T=t-\Delta,
\end{equation}
Without significantly losing the accuracy of the reasoning, let us replace for simplicity $\Delta$ by $\Delta_R$ and expand $\Delta_R$ into a series in powers of time
near the moment of passing the periapsis
\begin{equation}\label{479}
\Delta_R=\sum\limits_{i=0}^n c_i t^i
\end{equation}

Let us substitute the expansion (\ref{479}) into the expression for the pulsar phase (\ref{477}) and group the coefficients at the same powers of $t$. As a result, we obtain a series in powers of time, the coefficients of which are presented in Table 3.1. Substituting their numerical values for the coefficients $c_i$ shows that in the overwhelming majority of cases, terms with the factor $\dot\nu$ can be neglected, except for the factor at $t^2$, where $\dot\nu$ is comparable to the value $\alpha \nu/\kappa^2$. Table 3.2 contains the expansion coefficients at the corresponding powers of $t$ for the hyperbolic, parabolic and elliptical orbits, respectively, after the simplification described above.

\begin{table}[t]\centering
\begin{tabular}{|c|c|}
\hline&\\
$t$ & $c_i$ \\&\\
\hline &\\
1 & $-c_0+\frac{1}{2}\frac{\dot\nu}{\nu}c_0$ \\&\\
$t$ & $(\nu-\dot\nu c_0)(1-c_1)$ \\&\\
$t^2$ & $-\nu c_2+\frac{\dot\nu}{2}((c_1-1)^2+2c_0 c_1)$ \\&\\
$t^3$ & $-\nu c_3+\dot\nu((c_1-1)c_2+c_0 c_3)$ \\&\\
$t^4$ & $-\nu c_4 +\frac{\dot\nu}{2}(c_2^2+2(c_1-1)c_3+2c_0c_4)$ \\&\\
$t^5$ & $-\nu c_5 +\dot\nu(c_2c_3+(c_1-1)c_4+c_0 c_5)$ \\&\\
$t^6$ & $-\nu c_6 +\frac{\dot\nu}{2}(c_3^2+2(c_2c_4+(c_1-1)c_5+c_0c_6))$ \\&\\
$t^7$ & $-\nu c_7 +\dot\nu(c_3c_4+c_2c_5+(c_1-1)c_6+c_0c_7)$\\&\\
\hline
\end{tabular}
\caption{Coefficients of expansion of the pulsar phase in powers of time.}
\label{sss}
\end{table}

\begin{table}\centering
\begin{tabular}{|c|c|c|c|}
\hline &&& \\
$t$ & hyperbola & parabola & ellipse \\&&&\\
\hline &&&\\
1 & $\alpha(e-1)$ & $\alpha$ & $\alpha(1-e)$ \\&&&\\
$t$ & $\nu+\frac{\beta\nu}{\kappa(e-1)}$ &
$\nu+\frac{\beta\nu}{\kappa}$ & $\nu-\frac{\beta\nu}{\kappa(1-e)}$ \\&&&\\
$t^2$ & $-\frac{\alpha\nu}{2!\kappa^2(e-1)^2}+\frac{\dot\nu}{2}$ &
$-\frac{\alpha\nu}{\kappa^2}+\frac{\dot\nu}{2}$ &
$-\frac{\alpha\nu}{2!\kappa^2(1-e)^2}+\frac{\dot\nu}{2}$ \\&&&\\
$t^3$ & $-\frac{\beta\nu}{3!\kappa^3(e-1)^4}$ &
$-\frac{\beta\nu}{3\kappa^3}$ &
$\frac{\beta\nu}{3!\kappa^3(1-e)^4}$ \\&&&\\
$t^4$ & $\frac{(1+3e)\alpha\nu}{4!\kappa^4(e-1)^5}$ &
$\frac{2\alpha\nu}{3\kappa^4}$ &
$\frac{(1+3e)\alpha\nu}{4!\kappa^4(1-e)^5}$ \\&&&\\
$t^5$ & $\frac{(1+9e)\beta\nu}{5!\kappa^5(e-1)^7}$ &
$\frac{\beta\nu}{3\kappa^5}$ &
$-\frac{(1+9e)\beta\nu}{5!\kappa^5(1-e)^7}$ \\&&&\\
$t^6$ & $-\frac{(1+24e+45e^2)\alpha\nu}{6!\kappa^6(e-1)^8}$ &
$-\frac{7\alpha\nu}{9\kappa^6}$ &
$-\frac{(1+24e+45e^2)\alpha\nu}{6!\kappa^6(1-e)^8}$ \\&&&\\
$t^7$ & $-\frac{(1+54e+225e^2)\beta\nu}{7!\kappa^7(e-1)^{10}}$ &
$-\frac{4\beta\nu}{9\kappa^7}$ &
$\frac{(1+54e+225e^2)\beta\nu}{7!\kappa^7(1-e)^{10}}$\\&&&\\
\hline
\end{tabular}
\caption{Coefficients of expansion of the pulsar phase in powers of time at movement along a hyperbolic, parabolic and elliptical orbit.}
\label{different}
\end{table}

Based on the expansion terms from Table \ref{different}, the orbital parameters can be determined. As noted above, the observed pulsar frequency and its derivative are distorted by the influence of the perturbing mass, while higher-order derivative frequencies are almost entirely caused by disturbance from a flying body. Under certain physical assumptions (for example, comparing the ratio of $\nu$ and $\dot\nu$ for other pulsars), the frequency and frequency derivative can be reconstructed as if there were no disturbances.

Let us write down the solution for the orbital parameters $\alpha$, $\beta$, $e$,
$\kappa$, neglecting for now the presence of the derivative frequency $\dot\nu$.
\begin{equation}
\alpha=\frac{3\dot f^2\ddot f}{\nu(5\ddot f f^{(3)}-2\dot f f^{(4)})},
\end{equation}
\begin{equation}
\beta=\frac{3\sqrt{3}\dot f^{3/2}\ddot f^{5/2}}{\nu(5\ddot ff^{(3)}-
2\dot f f^{(4)})(-4\ddot f f^{(3)}+\dot f f^{(4)})},
\end{equation}
\begin{equation}
e=-\frac{\ddot f f^{(3)}-\dot f f^{(4)}}{3\dot f f^{(4)}-9\ddot f f^{(3)}},
\end{equation}
\begin{equation}
\kappa=\frac 32 \frac{\sqrt{3\dot f\ddot f}(-3\ddot f f^{(3)}+
\dot f f^{(4)})}{-5\ddot f f^{(3)}+2\dot f f^{(4)}}.
\end{equation}

These relations can be used directly only after $\dot\nu$ has been subtracted from the value $\dot f$. It should also be noted that these relationships were obtained from expansions near the moment of passage of the periapsis, which was assumed to be known. For some pulsars, this moment is determined quite reliably from the extremum of the second derivative of the rotational phase of the pulsar. It should also be noted that the epoch of the rotational parameters of the pulsar (phase and its derivatives) must coincide with the epoch of the periapsis. Calculations become more complicated if the observation interval does not contain the moment of passage of the periapsis, since in this case one more additional orbital parameter appears.

\section{Power spectra of TOA variations}

Gravitational disturbances acting on a pulsar in a globular cluster, and disturbances in the Earth's motion from passing asteroids can be
viewed from a completely different perspective. To do this you need to consider disturbances such as shot noise, i.e. sum of a large number
short-term impulses (Korn \& Korn, 1984)
\begin{equation}
x(t)=\sum\limits_{k=1}^\infty a_k v(t-t_k),
\end{equation}
whose form is given by the function $v=v(t)$, which has a Fourier transform
\begin{equation}
V_F(i\omega)=\int\limits_{-\infty}^{\infty}v(t)e^{-i\omega t}\,dt,
\end{equation}
while the pulse amplitude $a_k$ is a random variable with finite variance, and the sequence of random moments $t_k$ represents a Poisson process with an average sample rate $\theta$. Such a process is stationary and ergodic if it starts from $t=-\infty$; it is approximated by a Gaussian random process if the pulses overlap each other quite often. For this process the first and the second moments are given by the expressions:
\begin{equation}\label{kemp1}
{\bf M}x(t)=\xi=\theta{\bf M}a_k\int\limits_{-\infty}^{\infty} v(t)\,dt,
\end{equation}
\begin{equation}\label{kemp2}
{\bf M}x^2(t)=\xi^2+\theta{\bf M}a_k^2\int\limits_{-\infty}^{\infty}
v^2(t)\,dt,
\end{equation}
autocovariance function
\begin{equation}\label{kemp3}
R_{xx}(\tau)=\xi^2+\theta{\bf M}a_k^2\int\limits_{-\infty}^{\infty}
v(t)v(t+\tau)\,dt
\end{equation}
and power spectrum
\begin{equation}\label{kemp4}
S_{xx}(\omega)=2\pi\xi^2\delta(\omega)+\theta{\bf M}a_k^2\left|
V_F(i\omega)\right|^2.
\end{equation}
In formulas (\ref{kemp1})-(\ref{kemp4}) ${\bf M}$ is the average value, $R_{xx}$ - autocorrelation function, $S_{xx}$ - spectral function
density. In the special case when $a_k$ are fixed constants, formulas (\ref{kemp1})-(\ref{kemp4}) are known as Campbell's formulas (Terebizh, 1992)

In the calculations presented in this chapter, $v(t)$ is a time polynomial of the $n$th degree, which, however, cannot be characterized by any one scale factor $a_k$. In the value ${\bf M} a_k^2$, which is a linear combination of the polynomial coefficients, we can identify a common factor that is equal to $\frac{aM_2\sin i}{M_1+M_2}$. The values of $t_k$ correspond to the moments when the pulsar's companion passes the pericenter of the orbit.

So, let's consider the Fourier transform of a 5th degree polynomial from which, as is usually done when processing timing data a quadratic polynomial is subtracted. (In the next chapter, using a specific example, it will become clear why a polynomial of the 5th degree is considered). Since in practical calculations we always deal with finite series, we consider the Fourier transform on a finite interval, which gives
\begin{equation}
\begin{array}{rcl}
V_F(i\omega) & = &\int\limits_{-T/2}^{T/2}(\ddot\nu t^3+\nu^{(3)}t^4+
\nu^{(4)}t^5)e^{-i\omega t}\,dt=\\&&\\
&& \displaystyle \frac{T^3(10T\nu^{(3)}\sin\frac{\omega T}{2}+
i(80\ddot\nu+T^2\nu^{(4)})\cos\frac{\omega T}{2})}{1920\omega}+ \\&&\\
&& \displaystyle \frac{T^2(8T\nu^{(3)}\cos\frac{\omega T}{2}-
i(48\ddot\nu+T^2\nu^{(4)})\sin\frac{\omega T}{2})}{192\omega^2}+ \\&&\\
&& \displaystyle \frac{T(-6T\nu^{(3)}\sin\frac{\omega T}{2}-
i(24\ddot\nu+T^2\nu^{(4)})\cos\frac{\omega T}{2})}{24\omega^3}- \\&&\\
&& \displaystyle \frac{4T\nu^{(3)}\cos\frac{\omega T}{2}+
i(8\ddot\nu+T^2\nu^{(4)})\sin\frac{\omega T}{2}}{4\omega^4}+ \\&&\\
&& \displaystyle \frac{2\nu^{(3)}\sin\frac{\omega T}{2}+
iT\nu^{(4)}\cos\frac{\omega T}{2}}{\omega^5}- \\&&\\
&& \displaystyle \frac{2i\nu^{(4)})\sin\frac{\omega T}{2})}{\omega^6}
\end{array}
\end{equation}

In the squared Fourier transform, replacements were made $\sin^2\frac{\omega T}{2}\rightarrow\frac 12$, $\cos^2\frac{\omega T}{2} \rightarrow\frac 12$ . These replacements are possible, since these terms are periodic functions of the observation interval and, thus, they introduce additional fluctuations into the power-law part of the spectrum, complicating its analysis. For the same reason, periodic terms were excluded from the even components of the spectrum. The squared Fourier transform after the above simplifications looks like this:
\begin{equation}\label{kvf}
\begin{array}{rcl}
\left| V_F(i\omega) \right|^2 = V_F\cdot V_F^* & =& \displaystyle
\frac{T^6}{\omega^2}\left(\frac{\nu_2^2}{2304}+\frac{T^2\nu_3^2}{73728}+
\frac{T^2\nu_2\nu_4}{46080}+\frac{T^4\nu_4^2}{14745600}\right)-\\&&\\
&&\displaystyle\frac{T^5\sin\omega T}{\omega^3}
\left(\frac{(48\nu_2+T^2\nu_4)(80\nu_2+T^2\nu_4)}{368640}-
\frac{T^2\nu_3^2}{4608}\right)-\\&&\\
&&\displaystyle\frac{T^4}{\omega^4}\left(\frac{\nu_2^2}{96}+
\frac{T^2\nu_3^2}{2304}+\frac{11T^2\nu_2\nu_4}{11520}+
\frac{T^4\nu_4^2}{122880}\right)+\\&&\\
&&\displaystyle\frac{T^3\sin\omega T}{\omega^5}
\left(\frac{\nu_2^2}{3}-\frac{T^2\nu_3^2}{64}+
\frac{13T^2\nu_2\nu_4}{480}+\frac{T^4\nu_4^2}{2880}\right)+\\&&\\
&&\displaystyle\frac{T^4}{\omega^6}\left(
\frac{\nu_2\nu_4}{96}+\frac{T^2\nu_4^2}{11520}\right)-\\&&\\
&&\displaystyle\frac{T\sin\omega T}{\omega^7}
\left(2\nu_2^2-\frac{T^2\nu_3^2}{3}+
\frac{2T^2\nu_2\nu_4}{3}+\frac{T^4\nu_4^2}{60}\right)+\\&&\\
&&\displaystyle\frac{2\nu_2^2}{\omega^8}+\frac{T\sin\omega T}{\omega^9}
\left(-2\nu_3^2+\frac{\nu_4(12\nu_2+T^2\nu_4^2)}{3}\right)+\\&&\\
&&\displaystyle\frac{2\nu_3^2-4\nu_2\nu_4}{\omega^{10}}-
\frac{2T\nu_4^2\sin\omega T}{\omega^{11}}+\frac{2\nu_4^2}{\omega^{12}}
\end{array}
\end{equation}

The resulting theoretical power spectrum (\ref{kvf}) has a maximum slope of -12. In general, the power spectrum of a polynomial of degree $n$ has a maximum slope of $-2n-2$. Thus, the higher the degree of polynomial that describes the residual deviations, the greater the slope of the power spectrum. When deriving the spectrum (\ref{kvf}), no words were said about the magnitude and origin of the derivatives of the rotational frequency. If we make the natural assumption that various physical causes will lead to the appearance of derivatives of the rotational speed having different relative values, then this difference will immediately show up in the power spectrum by how parts of the spectrum with different slopes are related to each other. In the next section, this power spectrum property will be used to determine some of the orbital parameters of the pulsar PSR B1620-26.

\section{Experimental data and their interpretation}

This section is devoted to observational data from two pulsars PSR B1620-26 and B1822-09. For the pulsar PSR B1620-26, only the rotational frequency and its derivatives up to and including the fourth are known, and only a limit is imposed on the fifth derivative, which does not coincide in sign with the value $f^{(5)}$, derived from the theoretical relations of Table 3.2. The sign of the derivatives of the rotational frequency from $\dot\nu$ to $\nu^{(5)}$ behaves as $--++-$, which for positive $\alpha$ and $\beta$ gives a negative sign for $ \nu^{(5)}$. Due to this discrepancy, the value $f^{(5)}$ will not take part in further calculations. Based on the numerical values of the rotational parameters of PSR B1620-26, the orbital parameters of the pulsar will be derived.

For the pulsar PSR B1822-09, there are observational data in the form of residual deviations, which were provided to the author of this work by T.~V.~Shabanova (1999). This data will be used to derive numerical values for derivatives and relative changes in rotational speed. Based on the obtained numerical parameters, two
possible orbits will be found: hyperbolic and elliptical.

\subsection{Pulsar PSR B1620-26}

As initial data, we can use the rotational parameters of the pulsar PSR B1620-26 for the epoch ${\rm JD}=2448725.5$, taken from (Thorsett et al, 1999):
\begin{center}
\begin{tabular}{l}
$f=90.287332005426(14) {\rm s}^{-1}$,\\
$\dot f=-5.4693(3)\times 10^{-15}{\rm s}^{-2}$,\\
$\ddot f=1.9283(14)\times 10^{-23}{\rm s}^{-3}$,\\
$f^{(3)}=6.39(25)\times 10^{-33}{\rm s}^{-4}$,\\
$f^{(4)}=-2.1(2)\times 10^{-40}{\rm s}^{-5}$,\\
$f^{(5)}=3(3)\times 10^{-49}{\rm s}^{-6}$.
\end{tabular}
\end{center}

When substituting these rotational parameters into the formulas of Table \ref{different} it turns out that $e\approx-0.5$ which is meaningless by definition
eccentricity. To get a more reasonable result you can solve the inverse problem: assume that the orbit is elliptical, i.e. $0\leq e<1$, and then calculate the unperturbed quantity $\dot \nu$:
\begin{equation}
\dot\nu=\dot f-\frac{1+9e}{1+3e}\frac{\dot f f^{(3)}}{f^{(4)}}.
\end{equation}

As $e$ changes from 0 to 1, the frequency derivative $\dot\nu$ changes from $-4.7353\cdot 10^{-15}$ to $-3.6331\cdot 10^{-15}$. To calculate the values $\alpha$, $\beta$, $\kappa$ using the formulas of Table  \ref{different}, it is necessary to substitute $\dot f-\dot\nu$ instead of the value $\dot f$, i.e. that part of the derivative of the rotational frequency that is caused precisely by gravitational disturbance. Assuming that the eccentricity $e=0$ and the total mass of the system $M=1.7 M_{\odot}$, we obtain $\alpha=0.75$ s, $\beta=5.95$ s, period $P_b=2\pi/n =60$ years, semimajor axis projection $x=a_1\sin i =6.0$ s, semimajor axis $a=18$ a.~u., companion mass $m_2\sin i=\frac{a_1}{a}M \approx {10^{-3}}M_{\odot}$.

\begin{figure}
\centering
\includegraphics[width=15cm]{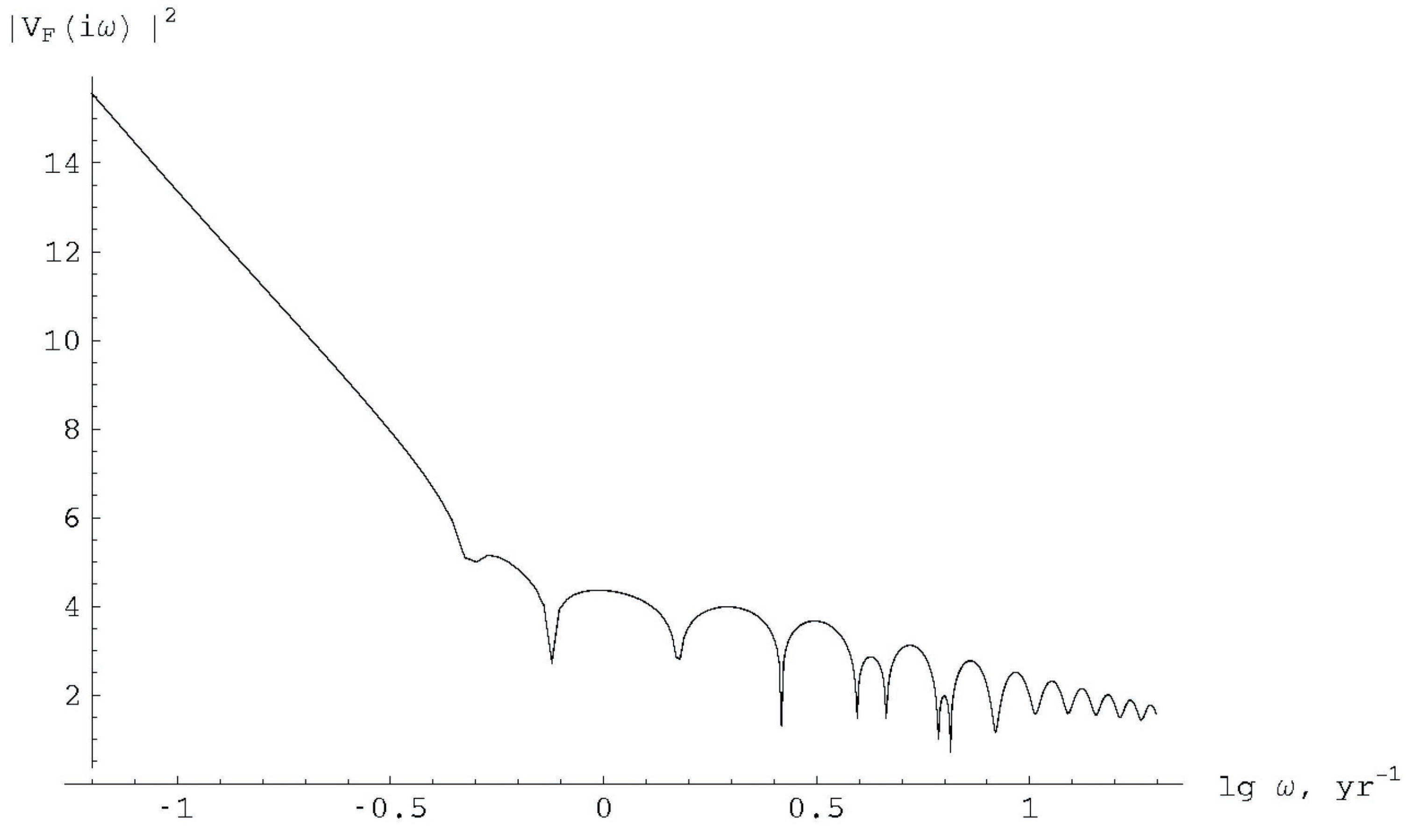}
\caption{Power spectrum of residual deviations of the pulsar TOA, calculated based on the measured rotational parameters of the pulsar B1620-26 and
formulas (4.89). The spectrum was calculated only taking into account the parameters $\ddot\nu$, $\nu^{(3)}$, $\nu^{(4)}$, because~. it was assumed that from
residual deviations as a result of fitting have already been subtracted quadratic time polynomial. The spectrum is plotted over the observation interval $T=10^8$ s.}
\label{1620}
\end{figure}

Figure \ref{1620} shows the theoretical power spectrum of the residual deviations of the pulsar PSR 1620-26. To calculate it, the rotational parameters $\ddot\nu$, $\nu^{(3)}$, $\nu^{(4)}$ were used, since it was assumed that, as a result of the standard procedure for adjusting the parameters, a polynomial of the 2nd degree was subtracted from the residual deviations (this was partly due to the goal of simplifying subsequent calculations). Only two components are visible: with a slope of -2 and with a slope of -12. All spectral components with intermediate slopes were absorbed by these two. Further, knowing from the formula (\ref{kvf}) the expressions of amplitudes for the components $\omega^{-2}$ and $\omega^{-12}$, and also knowing how $\ddot\nu$, $ \nu^{(3)}$, $\nu^{(4)}$ are expressed through orbital parameters, information about orbital parameters can be calculated from the spectrum. It should be noted that in this case the orbital parameters determined from the power spectrum must be interpreted as statistically average, i.e. as if the pulsar had a system of a large number of (small) planets each of which would disturb the motion of the pulsar.

Equating spectral components with slopes -2 and -12 at the point where their intensities are equal, gives the following equation (assuming that
$e=0$)
\begin{equation}
\frac{2\nu^2_4}{\omega^{12}}=\frac{\nu_2^2T^6}{2304\omega},
\end{equation}
from where after elementary transformations
\begin{equation}
\langle\kappa^2\rangle=\frac{\nu_2}{\nu_4}=\frac{\sqrt{4608}}
{\omega^5 T^3},
\end{equation}
\begin{equation}\label{pbkappa}
\left<P_b\right>=2\pi \langle\kappa\rangle=\frac{8.24}{\omega^{5/2}
T^{3/2}}.
\end{equation}
The average statistical period determined from the equation (\ref{pbkappa}) $P_b\approx 70$ years agrees well with the period of 60 years calculated previously.

\subsection{Pulsar PSR B1822-09}

Observations of this pulsar are the subject of an article by T.V. Shabanova (Shabanova, 1998), who kindly made the residual deviations available to the author.

This pulsar has activity, which is expressed in the fact that its period suddenly begins to change spontaneously, which then stabilizes. This behavior does not fit into the usual understanding of glitches occurring in the inner regions of pulsars, not to mention the fact that such behavior simply does not fit the definition of a glitch, which should be a fairly short-lived event.

The gradual change in period finds its natural explanation (although not the only one) if we remember that pulsars are moving objects that can interact with other bodies, as a result of which their radial velocity changes. This immediately leads to the observer noticing a change in the observed rotation frequency of the pulsar.

Before starting to calculate the derivatives of the rotation frequency of this pulsar, a cubic spline approximation of the data was carried out, on the basis of which a new uniform series was obtained with an interval between data points of 10 days. This interval was chosen because the average distance between the points of the original series was also about 10 days. The operation of uniformizing a series distorts only its high-frequency part, while low-frequency oscillations are completely preserved. The shape of the power spectrum is also preserved in the low frequency domain. The change in rotation frequency $\Delta\nu/\nu$ was found by numerically taking the derivative of a uniform series (finite differences were calculated). Since taking the derivative numerically is an unstable operation, before calculating the frequency derivatives $\dot\nu/\nu$, $\ddot\nu/\nu$, $\nu^{(3)}/\nu$ and $ \nu^{(4)}/\nu$ each new series was pre-smoothed by 21 points
with a rectangular window.

\begin{figure}
\centering
\includegraphics[width=15cm]{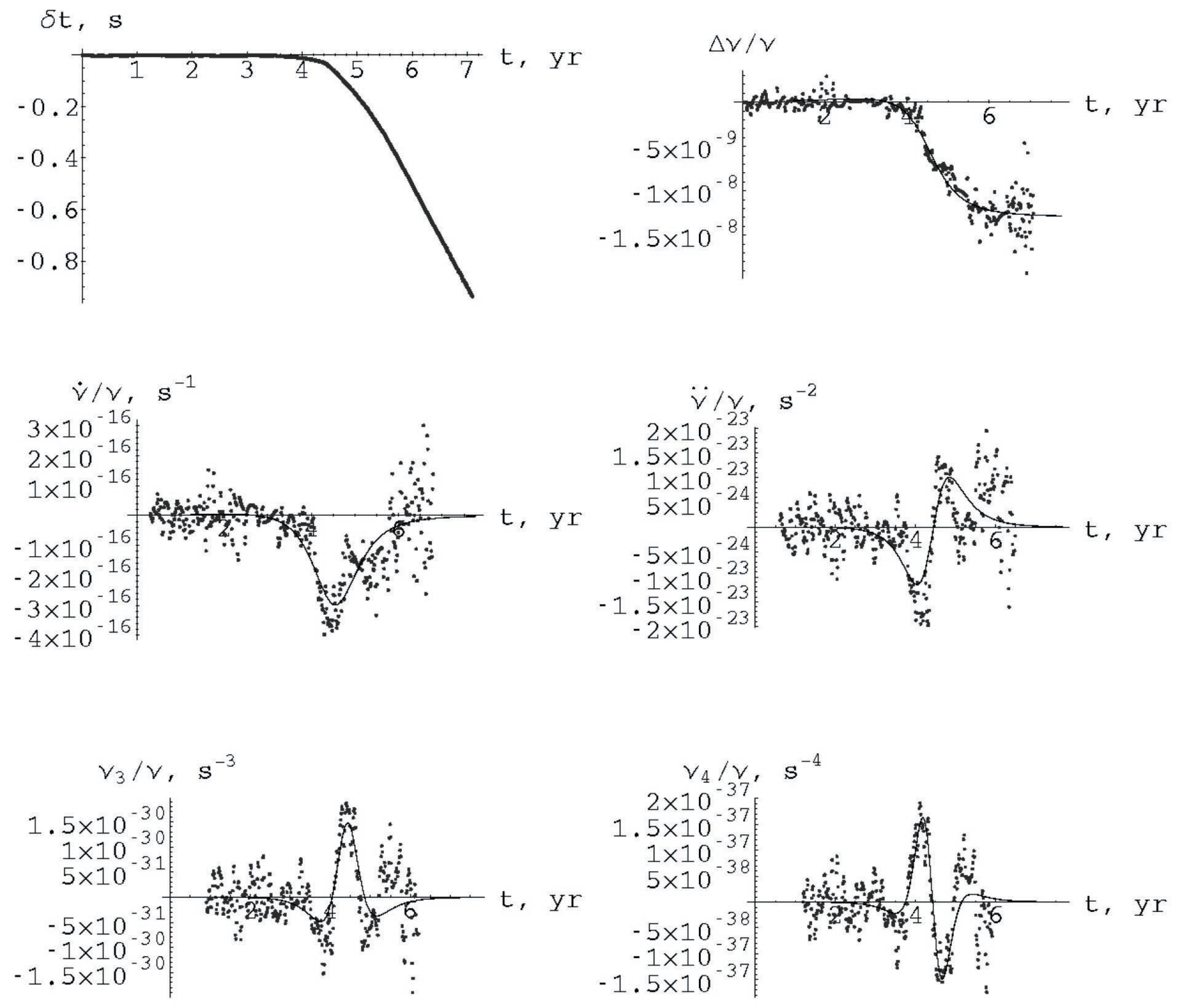}
\caption{Residual deviations of the pulsar PSR 1822-09 and frequency derivatives rotation. The calculation was carried out under the assumption of a hyperbolic orbit.
The dots show observational data, the solid curve -- theoretical calculations. For brevity, the notation is introduced $\nu^{(3)}=\nu_3$, $\nu^{(4)}=\nu_4$}.
\label{1822}
\end{figure}

The moment of ${\rm MJD}\,(T_0)=49940$ was taken as the moment of maximum approach of the pulsar to the supposed body (Shabanova, 1998). To find the numerical values of the observed parameters of the pulsar's orbit $\alpha$, $\beta$, $\kappa$, $e$, equations valid for hyperbolic and elliptical motion were used. We do not consider parabolic movement due to its improbability. Let us write down the equations of hyperbolic motion
\begin{equation}
\begin{array}{rl}
\Delta z &= \displaystyle
\frac{2\alpha}{\kappa e}, \\&\\
\displaystyle\frac{\ddot f}{f} &=
\displaystyle -\frac{\beta}{\kappa^3(e-1)^4},\\&\\
\displaystyle\frac{f^{(3)}}{f} &= \displaystyle
\frac{(1+3e)\alpha}{\kappa^4(e-1)^5},\\&\\
\displaystyle\frac{f^{(4)}}{f}
&= \displaystyle \frac{(1+9e)\beta}{\kappa^5(e-1)^7},
\end{array}
\end{equation}
where the following values were taken for the left-hand sides of the equations
$\frac{\dot f}{\nu}=-3.1(7)\times 10^{-16}$, $\frac{\ddot
f}{\nu}=-4(5)\times 10^{-24}$, $\frac{f^{(3)}}{\nu}=1.7(5)\times
10^{-30}$, $\frac{f^{(4)}}{\nu}=4(5)\times 10^{-38}$,
$\frac{f^{(5)}}{\nu}=-2.0(4)\times 10^{-44}$, $|\Delta z|=1.26\times
10^{-8}$. Observational data coupled with theoretical behavior curves rotational phase, frequency and its derivatives depending on time shown in Figure \ref{1822}.

Assuming hyperbolic motion, the following were obtained orbital parameters: $\alpha=-0.27$ s, $\beta=-0.038$ s, $n=7.2\times 10^{-8}$ s$^{-1}$, eccentricity $e=3.15$, impact distance $b=7$ a.~u., companion mass $m_2\sin i=3.7\times 10^{-4}M_{\odot}$, speed at infinity $V_\infty=17$ km/s.

\begin{figure}
\centering
\includegraphics[width=15cm]{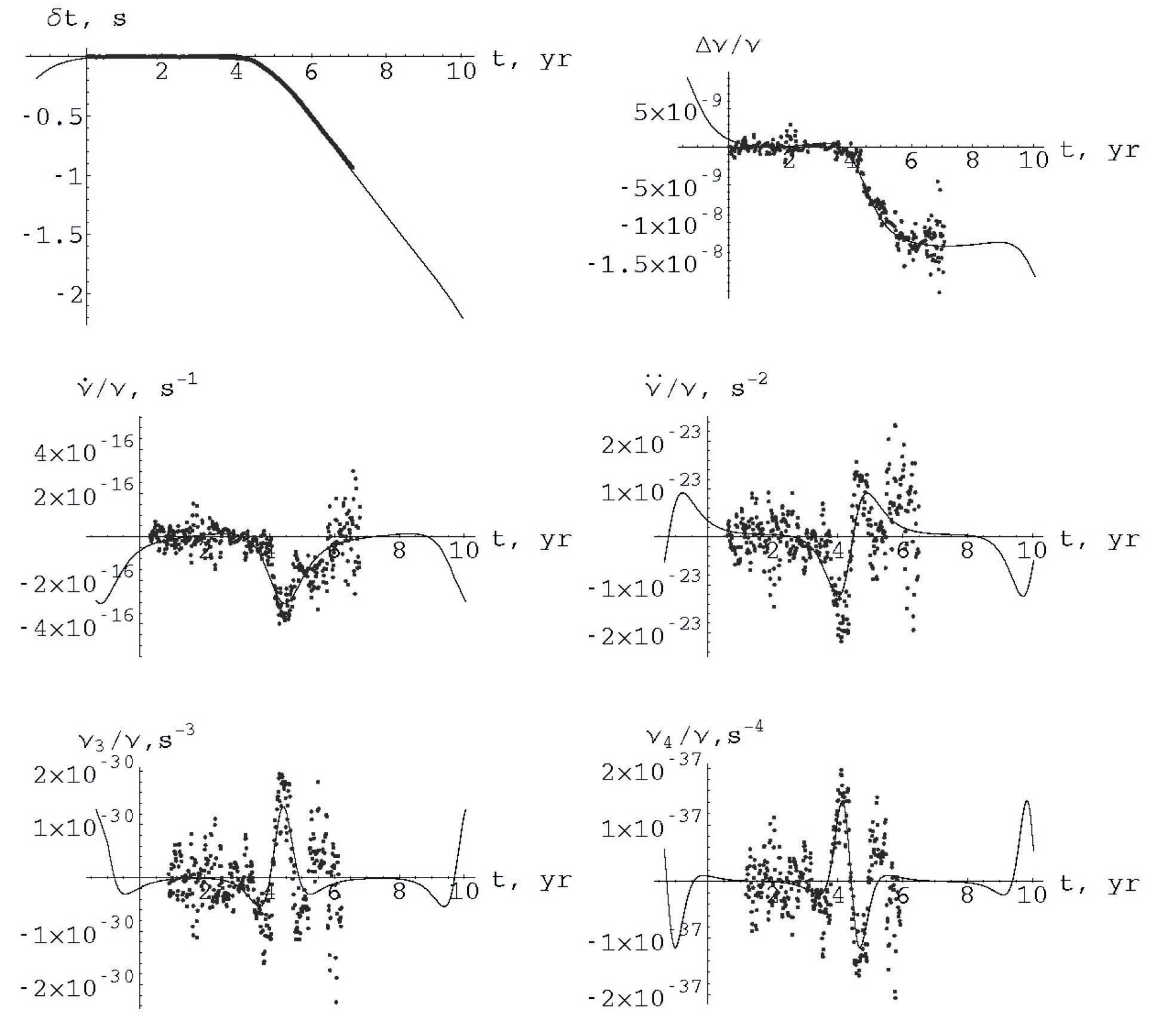}
\caption{Residual deviations of the pulsar PSR 1822-09 and derivatives of the rotation frequency for an elliptical orbit. The dots show observational data, the solid curve shows theoretical calculations. For clarity, two periods are shown. It is clearly seen that within the framework of the elliptical motion there is a smooth change in the period followed by its stabilization. The change in period corresponds to the passage of the pulsar near the periapsis, while the constant period corresponds to the portion of the orbit near the apocenter. The same notations are used as in the previous figure.}
\label{1822ell}
\end{figure}

To explain the observed residual deviations of the pulsar PSR B1822-09, it is also possible to use the assumption that the pulsar's companion is moving in an elliptical orbit. If the orbital period of the companion around the pulsar is long enough, then using a quadratic polynomial it is, in principle, possible to fit part of the orbital motion around the apocenter of the orbit, where the pulsar's motion is relatively slow, and low-order derivatives are sufficient to approximate the rotational phase of the pulsar. The situation changes radically when the pulsar approaches the pericenter of the orbit, and several frequency derivatives are no longer sufficient to adequately adjust the rotational phase of the pulsar. Calculations similar to those given above for a hyperbolic orbit were also made for an elliptical orbit. The calculation results are shown in Figure \ref{1822ell}.

Within the framework of elliptical motion, the following orbital values were obtained parameters: $\alpha=0.14$ s, $\beta=-0.027$ s, $n=3.5\times 10^{-8}$
s$^{-1}$, eccentricity $e=0.27$, semimajor axis $a=3.6$ a.~u., orbital period $P_b=5.7$ years, companion mass $m_2\sin i=1.2\times
10^{-4}M_{\odot}$.

\begin{figure}
\includegraphics[width=16cm]{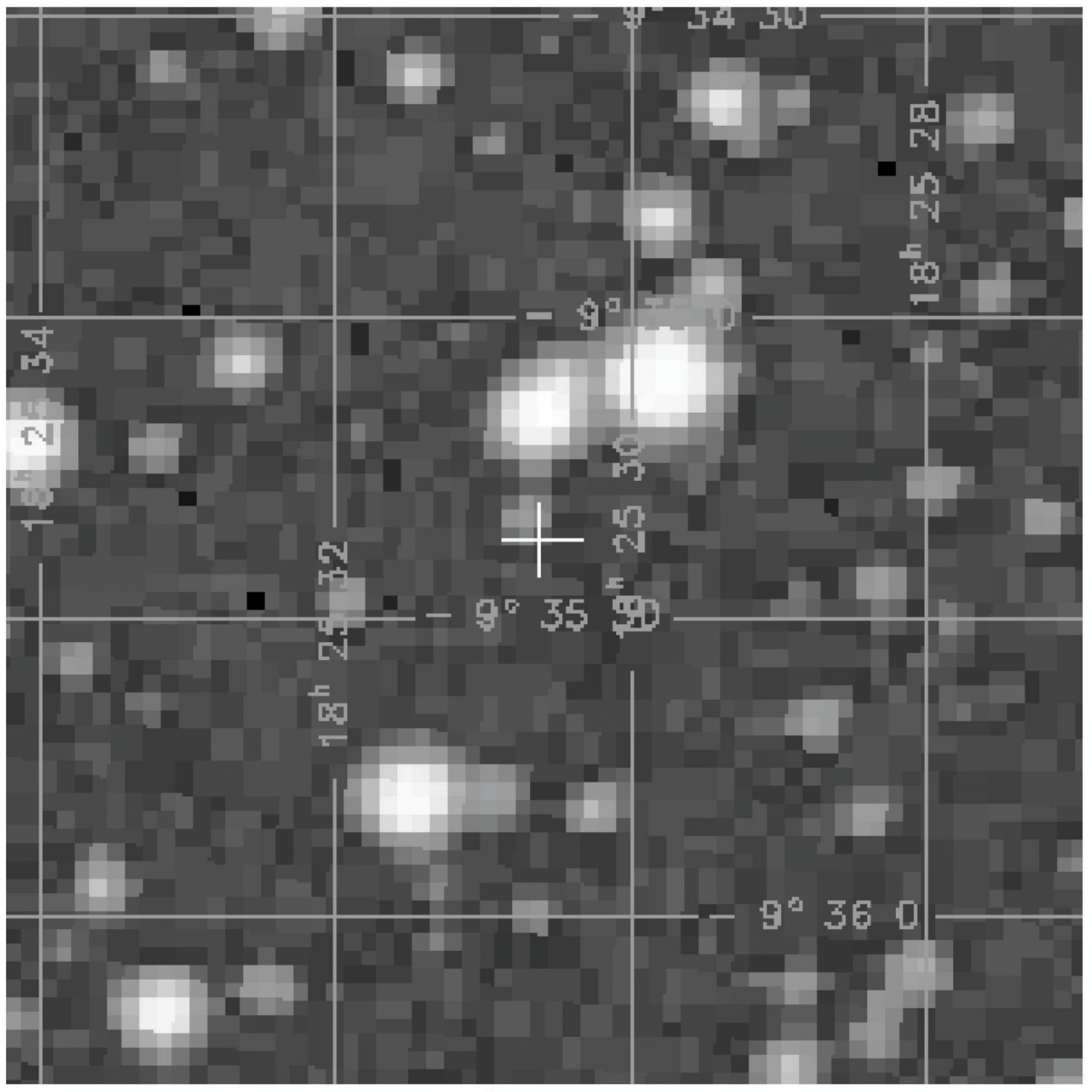}
\caption{Map of the surroundings of the pulsar PSR B1822-09. The cross indicates the position of the pulsar.}
\label{around}
\end{figure}

The previous part examined the practical application of the formulas (\ref{deltaz}), (\ref{kvf}) and those given in Table \ref{different} using the example of the pulsars PSR B1620-26 and B1822-09. The solution obtained for the pulsar B1620-26 completely coincides with the solution from (Joshi, Rasio, 1997, Thorset et. al., 1999). The solution within the framework of the theory of gravitational perturbations acting on the pulsar B1822-09 is new, and it explains much better the observed behavior of the residual deviations of the TOA. In this case, a hyperbolic solution is possible if the pulsar is located, for example, in a globular cluster consisting of low-luminosity stars and therefore not observable from Earth. In this regard, the surroundings of this pulsar were explored to search for such a cluster. Figure \ref{around} shows a map of the pulsar's surroundings, taken from the Digitized Sky View (http://skyview.gsfc.nasa.gov). At an angular distance of $\sim 3 "$ from the pulsar there is a star-shaped object of $18^m$ magnitude. In a linear measure, this corresponds to $\sim 1.5\times 10^{-2}$ pc. Thus, if this is not a random coincidence along line of sight, then we can assume a gravitational connection between the pulsar and the object. Unfortunately no other information about the vicinity of this pulsar is available.

\newpage
\section{Conclusions to Chapter 3}
\begin{enumerate}
\item Formulas for timing a pulsar perturbed by a massive body flying in a hyperbolic, parabolic and elliptical orbit are derived. This was done for the first time for hyperbolic and parabolic orbits.

\item The derived formulas are used to explain the long-period variations in the gravitational field and spontaneous changes in the rotational frequency observed in the pulsars PSR 1620-26 and PSR 1822-09.

\item The orbital parameters, masses, impact distance and approach velocity (for the pulsar PSR 1822-09) of the disturbing bodies are estimated.

\item Based on the derived timing formulas, the power spectrum of residual MPI deviations was calculated, which turned out to coincide with the spectrum of colored noise and has the form $1/f^n$, where $n=2,3,4,...$.

\item It is shown that by analyzing the slope and amplitude of the power spectrum, it is also possible to derive the average orbital parameters of the disturbing bodies.
\end{enumerate}

\chapter{Prospects for development}

It seems extremely desirable in the future to carry out regular astronomical observations of selected pulsars from various points of view: astrophysical, astrometric, metrological. In this case, the choice of pulsars should, of course, depend on the purpose of the study. To a large extent, regular timing of pulsars is carried out by a number of radio astronomy observatories. In this regard, it would be desirable as a prospect to be able to analyze data from various observatories presented in some unified format. This has already begun in the work of (Lorimer, 1998).

I would like to note the possibility of using pulsars for navigation purposes in space, similar to how this is done with the help of navigation satellites. Pulsars are created by nature itself as high-precision frequency standards. Having on board the spacecraft at least three antennas aimed at navigation pulsars and a pulse counter, it is possible to measure the position of the spacecraft relative to the barycenter of the Solar system using the integral Doppler effect with very good accuracy for cosmic scales.

\section{VLBI observations of a network of reference pulsars}

VLBI - observations of pulsars is a high-precision method for determining the coordinates of pulsars, independent of timing. The potential accuracy of VLBI measurements is quite high and reaches $10^{-4}$ arcsec, subject to an optimal observation program, base configuration and taking into account the influence of the ionosphere and troposphere. This dissertation describes measurements carried out on a single-base Kalyazin-Kashima interferometer. Adding one more point to the existing two will allow you to manage your observation time more economically. In this case, for sources with high declination it will no longer be necessary to rotate the base by $90^\circ$, but a relatively short period of time will be sufficient.

Periodic observations of selected pulsars, distributed as uniformly as possible across the celestial sphere, establish a kinematic reference frame as described in the introduction to this work. Since its accuracy depends on the accuracy of determining the proper motions of pulsars, repeated observations are required to better refine this value.

\section{Timing of double pulsars}

As shown in chapter \ref{dvs} double pulsars are suitable indicators of the presence or absence of correlated noise in the spectra power of residual deviations of the TOA. One of the sources of such noise are gravitational waves with a power spectrum of $\sim 1/f^5$, where $f$ -- frequency. As already mentioned, to detect gravitational waves we need pulsars with a large $P_b^2/x$ ratio, which corresponds to wide systems with a low-mass companion. In addition to the presence of such systems their regular observation is necessary, since only having a rich statistical material can really try to measure subtle effects of gravitational waves.

As for the BPT time scale, its stability can indeed exceed the stability of the conventional atomic scale over very long periods time intervals of $\sim 10^2 \div 10^3$ years. Watching at the same time several double pulsars, a "weighted average" scale can be derived as this is done in metrology.

\section{Observations of pulsars in globular clusters}

Pulsars in globular clusters are of interest primarily because astrophysical point of view, since the acceleration of the pulsar in the gravitational the cluster field does not allow us to consider such a pulsar as highly stable with a well-predicted rotational phase. However, the presence of derivatives of the rotation frequency allows, in turn, explore the structure of the cluster's gravitational field, distribution mass inside the cluster and thus allows us to consider pulsars in globular clusters as a kind of probes.

\chapter*{Conclusion}
\addcontentsline{toc}{chapter}{Conclusion}

High-precision astrometric and radiophysical observations of pulsars are a powerful means of studying the physics of the magnetosphere of pulsars, their
kinematics, the vicinity of pulsars from the point of view of their gravitational interactions with other bodies, physics of the interstellar medium,
gravitational background in the Universe, link between various celestial systems coordinates, construction of long-term highly stable astronomical
time scales and, ultimately, the construction of space-time reference systems. Pulsars are therefore unique objects that allow them to be used in completely different
fields of science.

Since observations of pulsars are carried out over long periods time, then the experimental data are highly susceptible to distortion influence of long-term factors, such as stochastic background gravitational waves, variations in the dispersion measure, gravitational perturbations from closely passing bodies, imperfection of planetary ephemerides, fluctuations earth's ionosphere, troposphere, etc. For this reason, it is very important becomes the task of adequately taking into account this kind of low-frequency disturbances.

In single-frequency VLBI observations, the main distorting factor is ionosphere. Variations in total electron content (TEC) in the ionosphere occur with a period of one day. Error in the coordinates of the radio source leads to modulation of the group delay with a period of also one day. Thus, there is a strong correlation between the coordinate correction of radio sources and TEC value. If we do not take into account the impact on group delay of the ionosphere, which can be considered in this
example as correlated interference, then get the correct coordinates radio sources seems difficult.

In observations of the arrival times of pulsar pulses, correlated noise manifest themselves to their fullest extent, significantly making the TOA of pulsars noisy.
However, in this situation with the harmful effects of low-frequency noise can be largely overcome by using more adequate model when fitting pulsar parameters using the method of least squares. For example, a Fourier series can be used as a model. When fitting the residual deviations of the TOA of pulsars with such a series, it is necessary include a member for a period of one year. Then knowing the cosine and sine components of this term, we can derive the correction of the coordinates of the studied pulsar.

If we consider pulsars as astronomical clocks, i.e. from the point from the point of view of studying the stability of their period, then here too it is necessary to
carefully investigate the influence of correlated low-frequency noise on accuracy of determination of pulsar parameters. In this consideration it should be taken into account that low-frequency noise has a power spectrum that is described by the law $f^{-n}$, $n=1,\,2,\,3,\,\ldots$. Sometimes you have to
apply a linear combination of these quantities. Correlation function which corresponds to the power spectra of a given type and represents is a function of time increasing in modulus (see table 2.1), included in expressions for the variances of the estimated parameters and thereby leads to to the fact that as the time interval increases, the dispersion of estimates begins to grow. This conclusion is especially important for predicting how the dispersion of the rotational and orbital frequency of the pulsar with increasing time interval of observations. A detailed analysis of the behavior of these variances depending on the observation interval showed (see Chapter 2), that the fractional stability of the orbital frequency $\sigma_v$ may be less compared to the fractional stability value rotational frequency $\sigma_y$ over a relatively long interval observations (several hundred or thousand years).

A study of the properties of low-frequency noise allows us to conclude that to adequately take into account their impact on the estimated parameters, it is necessary
use more advanced computational algorithms that use a more complete reduction model and take into account correlation properties of noise.

As a concrete example of how astrophysical information can be extracted from low-frequency noise, one can cite calculations of gravitational disturbances acting on a pulsar from a closely passing body. Such an effect will, in particular, manifest itself in the presence of high-order frequency derivatives in the rotational phase of the pulsar. For some pulsars, derivatives of the rotation frequency up to the fifth inclusive have already been measured. This makes it possible to reconstruct some orbital parameters by observing the pulsar for a fraction of its orbital period. If we consider gravitational disturbances of this kind as shot noise, then it seems possible to theoretically derive the power spectrum in order to subsequently compare it with the observed spectra of pulsar noise and draw a conclusion about their nature.

If we consider pulsars from an observational point of view then pulsars are also useful objects here. Possessing very weak fluxes and pulsed radiation, pulsars impose very strict requirements on the level of recording equipment which automatically implies a high level of specialists creating pulsar equipment and the quality of the element base. When creating modern VLBI equipment, in particular VLBI correlators, the presence of a correlation mode with gating is certainly noted as mandatory processing attributes, which makes it possible to increase the signal-to-noise ratio for pulsed radio sources. The presence of such a mode implies high technology of equipment and advanced software.

Summarizing all of the above, it must be said that it is necessary to continue observing and studying pulsars with all possible efforts.

\end{document}